\PassOptionsToPackage{dvipsnames,svgnames,x11names}{xcolor}
\documentclass[final,3p,fleqn,sort]{elsarticle}
\usepackage[labelfont=bf, singlelinecheck=false,
            justification=justified]{caption}
\usepackage{xcolor}
\usepackage{amsmath,amssymb}
\usepackage{subcaption}
\usepackage{mathrsfs}
\usepackage{stmaryrd}
\usepackage{booktabs}
\usepackage{microtype}
\usepackage{mathtools}
\usepackage{multirow}
\usepackage{enumitem}
\usepackage{setspace}
\usepackage{placeins}
\usepackage{etoolbox,siunitx}
\usepackage{bm,cancel}
\usepackage{float}
\robustify\bfseries

\usepackage{amsthm}

\theoremstyle{remark}

\makeatletter
\g@addto@macro\normalsize{%
	\setlength\abovedisplayskip{5pt}
	\setlength\belowdisplayskip{5pt}
	\setlength\abovedisplayshortskip{6pt}
	\setlength\belowdisplayshortskip{6pt}
}
\makeatother

\usepackage[T1]{fontenc}
\SetSymbolFont{stmry}{bold}{U}{stmry}{m}{n}

\let\orho\rho
\let\rho\varrho

\usepackage{lineno,hyperref}
\modulolinenumbers[5]

\renewcommand{\vec}[1]{\ensuremath{\boldsymbol #1}}
\newcommand{\derivative}[2]{\frac{\mathrm{d} #1}{\mathrm{d} #2}}
\newcommand{\pderivative}[2]{\frac{\partial #1}{\partial #2}}
\newcommand{\eg}{\textit{e.g.}~}
\newcommand{\ie}{\textit{i.e.}~}
\newcommand{\He}{\ensuremath{\bm{\mathcal{H}}}}
\newcommand{\Tee}{\ensuremath{\bm{\mathcal{Z}}}}
\newcommand{\Ree}{\ensuremath{\bm{\mathcal{R}}}}
\newcommand{\ree}{\ensuremath{\vec{r}}}
\renewcommand{\H}{\ensuremath{\bm{\hat{\mathcal{H}}}}}
\newcommand{\mat}[1]{\ensuremath{\mathbf{#1}}}
\newcommand{\avg}[1]{\left\{\hspace*{-3pt}\left\{#1\right\}\hspace*{-3pt}\right\}}
\newcommand{\avgln}[1]{\ensuremath{#1^\mathrm{ln}}}

\newcommand{\jump}[1]{\ensuremath{\left\llbracket #1 \right\rrbracket}}
\newcommand{\rholn}{\ensuremath{\rho\textsuperscript{ln}}}
\newcommand{\betaln}{\ensuremath{\beta\textsuperscript{ln}}}
\newcommand{\pln}{\ensuremath{p\textsuperscript{ln}}}

\newcommand{\avgp}{\ensuremath{\tilde{p}}}
\newcommand{\uavg}{\overline{\lVert \vec{u} \rVert^2}}
\newcommand{\betaavg}{\ensuremath{\overline{\beta^2}}}

\newcommand{\Eline}{\ensuremath{\overline{E}}}

\newcommand{\half}{\frac{1}{2}}

\allowdisplaybreaks
\renewcommand{\L}{\mathrm{L}}
\newcommand{\R}{\mathrm{R}}
\DeclareMathOperator{\diag}{diag}

\usepackage[framemethod=tikz]{mdframed}

\makeatletter
\def\mathcolor#1#{\@mathcolor{#1}}
\def\@mathcolor#1#2#3{%
  \protect\leavevmode
  \begingroup
    \color#1{#2}#3%
  \endgroup
}
\makeatother
\begin{document}
\begin{frontmatter}
\title{Ideal GLM-MHD: About the entropy consistent nine-wave magnetic field divergence diminishing ideal magnetohydrodynamics equations}
\author[physik]{Dominik Derigs\corref{mycorrespondingauthor}}
\cortext[mycorrespondingauthor]{Corresponding author}
\ead{derigs@ph1.uni-koeln.de}
\author[mathematik]{Andrew R.~Winters}
\author[mathematik]{Gregor J.~Gassner}
\author[physik]{Stefanie Walch}
\author[mathematik]{Marvin Bohm}
\address[physik]{I.\,Physikalisches Institut, Universit\"at zu K\"oln, Z\"ulpicher Stra\ss{}e~77, 50937 K\"oln}
\address[mathematik]{Mathematisches Institut, Universit\"at zu K\"oln, Weyertal 86-90, 50931 K\"oln}
\numberwithin{equation}{section}
\begin{keyword}
magnetohydrodynamics \sep entropy stability \sep divergence-free magnetic field \sep divergence cleaning
\end{keyword}
\begin{abstract}
The paper presents two contributions in the context of the numerical simulation of magnetized fluid dynamics. First, we show how to extend the ideal magnetohydrodynamics (MHD) equations with an inbuilt magnetic field divergence cleaning mechanism in such a way that the resulting model is consistent with the second law of thermodynamics. As a byproduct of these derivations, we show that not all of the commonly used divergence cleaning extensions of the ideal MHD equations are thermodynamically consistent. Secondly, we present a numerical scheme obtained by constructing a specific finite volume discretization that is consistent with the discrete thermodynamic entropy. It includes a mechanism to control the discrete divergence error of the magnetic field by construction and is Galilean invariant. We implement the new high-order MHD solver in the adaptive mesh refinement code \texttt{FLASH} where we compare the divergence cleaning efficiency to the constrained transport solver available in \texttt{FLASH} (unsplit staggered mesh scheme).
\end{abstract}
\end{frontmatter}

\section{Introduction}
Widespread applications of the ideal magnetohydrodynamic (MHD) equations emerged in many disciplines, including, but not limited to, astrophysical and magnetically confined fusion plasma applications.
The mentioned applications, usually deal with extreme conditions like near vacuum up to ultra-high density environments where shocks of varying type and strength are the rule rather than the exception. Hence, the accuracy and robustness of numerical simulation codes are very important.

Yet, the mathematical model of ideal MHD has some shortcomings that can cause the results to become unphysical.
As neither the full set of universally valid thermodynamics laws nor the divergence-free condition of the magnetic field is directly coupled into the classic mathematical model of ideal MHD, it is possible to obtain physically invalid solutions.

In order to single out physically relevant solutions, we have to augment the system with additional admissibility criteria.
One admissibility criterion in the case of ideal MHD is the divergence-free condition of the magnetic field that is expressed by Gau\ss's law for magnetism
\begin{equation}\label{eq:divB}
	\nabla\cdot\vec{B} = 0.
\end{equation}
The implementation of \eqref{eq:divB} into a numerical approximation is a major difficulty.
As detailed in the early 1980s by Brackbill and Barnes \cite[eq.~(2)]{Brackbill1980}, numerical discretization errors always have a noticeable impact on the temporal evolution of the magnetic field divergence:
\begin{equation}\label{eq:BB}
\pderivative{}{t} (\nabla \cdot \vec{B}) = 0 +  \mathcal{O}\big( \Delta x^m,\,\Delta t^n \big),
\end{equation}
where $\Delta x$ and $\Delta t$ are the space and time discretization steps, and $m,n \ge 0$ are scheme dependent constants that also depend on the smoothness of the problem.
Since the advent of sufficiently powerful computers, many approaches have been proposed to address this issue and ``clean'' such errors, including projection methods, constrained transport, and hyperbolic divergence cleaning. We give a brief overview over these methods with special focus on the hyperbolic divergence cleaning as we find it particularly useful in the context of highly efficient, highly parallelizable numerical schemes for large-scale applications.

Another natural admissibility criterion is given by the universally valid laws of thermodynamics.
Most numerical schemes do not take the second law of thermodynamics directly into account but rather equip the system with a minute amount of diffusion.
We seek to derive a numerical scheme that complies with the laws of thermodynamics and especially the second law, \ie the entropy inequality.
Following the work of \eg \cite{Derigs2016,Winters2016,Fjordholm2011,Tadmor1987}, we use this law as an additional admissibility criterion to construct discretizations that agree with the laws of thermodynamics.
In this way, we enforce that the solution does not converge towards physically irrelevant solutions which increases their numerical robustness, as thermodynamically impossible physical processes that could \eg lead to negative thermal pressures are forbidden.

In this work, we derive a new mathematical model that is built to explicitly take into account all the mentioned relevant physics in order to impede unphysical results of numerical simulations.
We start from first principles to avoid commonly done simplifications that, as we explain herein, make the typically used classic mathematical model of ideal MHD unsuitable in describing the behavior of physical flows in certain regimes of numerically computed, discrete solutions.

For this, we derive a {new} system of equations, which we deem the ideal GLM-MHD equations, that allows the construction of a novel magnetic field divergence diminishing scheme that naturally complies with thermodynamics.
Furthermore, we investigate our new model in great detail and present both the methods we use as well as the details to allow a straightforward implementation of our scheme into the reader's own simulation code.

We shortly summarize the new contributions presented in this work:
\begin{itemize}
	\item We derive the ideal MHD equations from the compressible Euler equations where we explicitly allow non-zero magnetic field divergence in Maxwell's equations. We physically motivate and highlight important findings (Sec.~\ref{Sec:idealMHDdivB})
	\item We derive a {new} physically motivated mathematical model, the \textit{ideal GLM-MHD equations} which we thoroughly investigate from both mathematical and physical perspectives (Sec.~\ref{Sec:GLM})
	\item We derive entropy stable numerical flux functions. We give all necessary details to code the {high-order} scheme we are describing in a straightforward manner (Secs.~\ref{Sec:EC} and \ref{Sec:ES})
\end{itemize}

\section{The ideal MHD equations in the case of $\nabla\cdot B \ne 0$}\label{Sec:idealMHDdivB}
We begin with the three-dimensional compressible Euler equations describing non-magnetized single-fluid flows. We then couple the effect of magnetic fields to the model.
We find that the classic model of ideal MHD in the form of conservation laws is not valid for arbitrary flows, but only for fluids where \eqref{eq:divB} holds \emph{exactly}. Interestingly, our derivations reveal results that are known from previous publications \cite{Godunov1972,Powell1999} while our independent and fundamentally different approach motivates these findings with a solely physically motivated mindset which eases the interpretation.

An important first step is, due to the findings of Brackbill and Barnes \cite[eq.~(2)]{Brackbill1980}, see also \eqref{eq:BB}, that we shouldn't assume the divergence-free condition of the magnetic field \eqref{eq:divB} is fulfilled \emph{exactly} at \emph{all} times in numerical simulations. In fact, Dirac \cite{Dirac31} showed that quantum mechanics does not preclude the existence of isolated magnetic monopoles although physicists often reason that magnetic monopoles are unlikely to exist. Their reasoning mainly comes from the fact that quantum mechanics, as it is usually established, seems possible only when there are no isolated magnetic monopoles. Dirac, however, disproved this view.
One of the most important theoretical motivations of introducing magnetic monopoles is to obtain Maxwell's equations in symmetric form with regard to charges.
Note that the Maxwell equations retain all their properties, such as invariance under a global duality transformation that mixes electric and magnetic fields \cite{Jackson06}, when magnetic monopoles are included, whether or not they exist.

During the derivation of the classic ideal MHD equations in conservative form, \eqref{eq:divB} is used to simplify the computations. However, this ultimately destroys the validity of the system of ideal MHD equations for flows where the divergence-free condition is not fulfilled to a certain extent, which is generally the case in discrete approximations. This is commonly found in simulations due to errors caused by any given numerical approximation. Note that this more general form of the ideal MHD equations is, of course, perfectly valid in the case of $\nabla\cdot\vec{B} = 0$, as the only difference is that certain terms aren't neglected early on in the derivation.

We start from the three-dimensional compressible Euler equations written compactly as a system of conservation laws,
\refstepcounter{equation}\label{eq:3DEuler}
\begin{align}\tag{\theequation a-c}
\pderivative{}{t}\, \vec{q} + \nabla\cdot\vec{f} =
\pderivative{}{t}
\begin{bmatrix}
\rho \\ \vphantom{\big(}\rho\vec{u} \\ \vphantom{\big(}E \end{bmatrix}
+
&\nabla\cdot
\begin{bmatrix}
\rho\vec{u} \\
\rho(\vec{u}\otimes\vec{u}) + p\mat{I} \\
\vec{u}\big(\frac{1}{2}\rho\|\vec{u}\|^2 + \frac{\gamma p}{\gamma - 1} \big) \\
\end{bmatrix} = \begin{bmatrix}
0\\\vec{F}\\E_\mathrm{a}
\end{bmatrix},
\end{align}
where $\rho$, $\rho\vec{u}$, and $E$ are the density, volume specific momenta, and {total energy density} of the plasma system, $p$ is the thermal pressure, $\mat{I}$ is the $3 \times 3$ identity matrix. $\vec{F}$ and $E_\mathrm{a}$ denote the sum of all (external) forces and {energy source terms} which are important for the fluid.
The multidimensional flux function is denoted by $\vec{f}$. The flux Jacobian has only real eigenvalues and the right eigenvectors are linearly independent, \ie the compressible Euler equations are hyperbolic. We assume that the total energy and the thermal pressure are related through the ideal gas law
\begin{equation}
{p} = \rho R T = (\gamma-1)\left(E - \frac{1}{2}\rho\lVert\vec{u}\rVert^2\right),
\end{equation}
with the ratio of specific heats, $\gamma = \frac{c_p}{c_v}$.

For coupling electromagnetic fields to the fluid, we must examine the equations which describe their behavior.
The generalized Maxwell's equations for non-vanishing magnetic charge densities are
\refstepcounter{equation}\label{eq:generalizedMaxwell}
\begin{equation}\tag{\theequation a-d}
\nabla\cdot\vec{E} = \frac{\orho_e}{\varepsilon_0}, \qquad \nabla\cdot\vec{B} = \mu_0\orho_m, \qquad \nabla\times\vec{B} = \frac{1}{c^2}\pderivative{\vec{E}}{t} + \mu_0 \vec{j}_e, \qquad -\nabla\times\vec{E} = \pderivative{\vec{B}}{t} + \mu_0\vec{j}_m,
\end{equation}
where $\vec{E}$ and $\vec{B}$ are the electric and magnetic fields \cite[Sec.~6.11]{Jackson06}. The charge densities are $\orho_{e,m}$, where the subscript $e$ refers to electric charges and $m$ refers to magnetic charges. A similar distinction is made for the current densities, {$\vec{j}_{e,m} := \orho_{e,m}\vec{u}$}. The equations \eqref{eq:generalizedMaxwell} are invariant under a global duality transformation that mixes electric and magnetic fields \cite{Jackson06}. This underlines that the limitation $\orho_m = 0$ is only a convention.

By investigating how the magnetic field influences the fluid we can integrate the effects of magnetic fields into the compressible Euler equations. For this, we consider the Lorentz force caused by the electric and magnetic fields, denoted by $\vec{F}_L$. In our derivations, we assume that the magnetic field in the only cause of external forces, \ie $\vec{F} \equiv \vec{F}_L$ and use it on the right-hand side momentum equation (\ref{eq:3DEuler}b). If the system contains additional forces, \eg gravitational acceleration or radiation, these forces need to be added to the momentum equation as well.
The most general form of the Lorentz force is
\begin{equation}
\vec{F}_L = q_e \left(\vec{E} + \vec{u}\times\vec{B}\right) + q_m \left(\vec{B} - \frac{\vec{u} \times \vec{E}}{c^2}\right),
\end{equation}
where $q_{e,m}$ are the electric and magnetic charges. Note that the second term on the right-hand side vanishes for $q_m \propto \nabla\cdot\vec{B}=0$, see \cite{Moulin01}.
The total Lorentz force per unit volume is then
\begin{equation}
\vec{F}_L = e_e(n_i - n_e)\vec{E} + e_e(n_i \vec{u}_i - n_e \vec{u}_e)\times\vec{B} + e_m(n_i - n_e)\vec{B} - {e_m}\frac{n_i \vec{u}_i - n_e \vec{u}_e}{c^2}\times\vec{E}
\end{equation}
where the ions and electrons are denoted by subscripts $i$, and $e$, respectively. Their number densities are $n_{i,e}$. The unit charges of electric and magnetic monopoles are denoted by $e_e$ and $e_m$, respectively.
The quasi-neutrality assumption, $n := n_i = n_e$, that is due to the single-fluid model leads to
\begin{align}
\vec{F}_L &= n e_e (\vec{u}_I - \vec{u}_E)\times\vec{B} - \frac{n e_m}{c^2}(\vec{u}_I - \vec{u}_E)\vec{E}\notag\\
&= \vec{j}_e \times \vec{B} - \frac{1}{c^2} \vec{j}_m \times \vec{E},
\end{align}
where we can use the magnetic current density given by (\ref{eq:generalizedMaxwell}b)
\begin{equation}\label{eq:jm}
	\vec{j}_m = \orho_m \vec{u} = \mu_0^{-1}(\nabla\cdot\vec{B})\vec{u}.
\end{equation}
We use the ideal Ohm's law for ionized fluids in motion,
\begin{equation}\label{eq:Ohm}
\vec{E} + \vec{u}\times\vec{B} = \eta\vec{j}_e = 0,
\end{equation}
with the assumption of vanishing resistivity, $\eta = 0$, \ie infinite conductivity of the plasma which is an essential assumption of ideal MHD, to obtain
\begin{align}
\vec{F}_L &= \vec{j}_e \times \vec{B} + \frac{1}{c^2} \,\vec{j}_m \times (\vec{u}\times\vec{B}) \notag\\
&= \frac{1}{\mu_0}\left(\nabla\times\vec{B} - \frac{1}{c^2}\pderivative{\vec{E}}{t}\right) \times \vec{B} + \frac{1}{c^2} (\nabla\cdot\vec{B})\vec{u} \times (\vec{u}\times\vec{B}).
\end{align}
With this result, we add the magnetic forces into the momentum equation to obtain the non-divergence-free form of the ideal MHD momentum equation
\begin{align}
\pderivative{}{t}(\rho\vec{u})+\nabla\cdot\left(
\rho(\vec{u}\otimes\vec{u}) + \big(p+\frac{\|\vec{B}\|^2}{2\mu_0}\big)\mat{I} - \frac{\vec{B}\otimes\vec{B}}{\mu_0}\right)
&=
\frac{\nabla\cdot\vec{B}}{\mu_0} \left(\frac{1}{c^2} \vec{u} \times (\vec{u}\times\vec{B}) - \vec{B}\right) - \frac{1}{\mu_0 c^2}\pderivative{\vec{E}}{t}\times \vec{B}.
\label{eq:quiteugly}\\[-1\baselineskip]
\shortintertext{In the non-relativistic limit, $\|\vec{u}\| \ll c$, \eqref{eq:quiteugly} simplifies to become}
\label{eq:lessugly}
\pderivative{}{t}(\rho\vec{u})+\nabla\cdot\left(
\rho(\vec{u}\otimes\vec{u}) + \big(p+\frac{\|\vec{B}\|^2}{2\mu_0}\big)\mat{I} - \frac{\vec{B}\otimes\vec{B}}{\mu_0}\right)
&=
- \frac{\nabla\cdot\vec{B}}{\mu_0}\, \vec{B}
\end{align}
as the displacement current, where the rightmost term in \eqref{eq:quiteugly} can be neglected for a Newtonian theory of MHD \cite[Sec.~3.1.4]{Ogilvie2016}. We observe that \eqref{eq:lessugly} reduces to the standard form of the momentum equation in the ideal MHD equation system for $\nabla\cdot\vec{B} \rightarrow 0$.
Note that the non-relativistic limit is not a restriction in our derivations, but a natural assumption as we chose to start from the compressible Euler equations \eqref{eq:3DEuler} which are themselves derived for the non-relativistic case.

Now that we accounted for the influence of the magnetic field on the fluid in the momentum equation, we must add a new evolution equation for the magnetic field components to the system of equations. From \eqref{eq:generalizedMaxwell}, we obtain the generalized induction equation
\begin{equation}
	\pderivative{\vec{B}}{t} = - \nabla \times \vec{E} - \mu_0\vec{j}_m .
\end{equation}
Using \eqref{eq:Ohm} we get
\begin{equation}\label{eq:induction}
	\pderivative{\vec{B}}{t} - \nabla \times (\vec{u}\times\vec{B}) = -\mu_0\vec{j}_m = -(\nabla\cdot\vec{B})\vec{u}.
\end{equation}
The obtained induction equation \eqref{eq:induction} is added in the system of compressible Euler equations to model the evolution of the magnetic field.

To close the system of the generalized ideal MHD equations, we need to compute the total energy equation including the effects of the aforementioned modifications. The total energy update equations is
\begin{equation}\label{eq:Et}
	\pderivative{E}{t} = \pderivative{}{t}\Bigg(\frac{1}{2}\rho\|\vec{u}\|^2 + \epsilon + \frac{1}{2}\|\vec{B}\|^2\Bigg).
\end{equation}
After many manipulations{, that can be found in \ref{app:thegorydetails},} we find
\begin{equation}\label{eq:Etresult}
	\pderivative{E}{t} + \nabla \cdot \left(\vec{u}\left(\frac{1}{2}\rho\|\vec{u}\|^2 + \frac{\gamma p}{\gamma - 1} + \frac{\|\vec{B}\|^2}{2\mu_0} \right) - \frac{\vec{B}(\vec{u}\cdot\vec{B})}{\mu_0} \right) = -\mu_0^{-1}(\nabla\cdot\vec{B})(\vec{u}\cdot\vec{B}),
\end{equation}
which is the commonly known form of the ideal MHD total energy conservation law equipped with a non-conservative part on the right hand side. Note that, for the sake of convenience, we set $\mu_0=1$ hereafter to express the ideal MHD equations in dimensionless units. See \ref{App:DimidealMHD} for a full presentation of the ideal MHD equations in physical units.

We summarize the ideal MHD equations in their general form to be
\refstepcounter{equation}\label{eq:3DgeneralizedidealMHD}
\begin{align}\tag{\theequation a-d}
\pderivative{}{t}\, \vec{q} + \nabla\cdot\vec{f} =
\pderivative{}{t}
\begin{bmatrix}
\rho \\ \vphantom{\big(}\rho\vec{u} \\ \vphantom{\big(}E \\ \vec{B} \end{bmatrix}
+
&\nabla\cdot
\begin{bmatrix}
\rho\vec{u} \\
\rho(\vec{u}\otimes\vec{u}) + \big(p+\frac{1}{2}\|\vec{B}\|^2\big)\mat{I} - \vec{B}\otimes\vec{B} \\
\vec{u}\big(\frac{1}{2}\rho\|\vec{u}\|^2 + \frac{\gamma p}{\gamma - 1} + \|\vec{B}\|^2 \big) - \vec{B}(\vec{u}\cdot\vec{B}) \\
\vec{u}\otimes\vec{B} - \vec{B}\otimes\vec{u}
\end{bmatrix} = -(\nabla\cdot\vec{B})
\begin{bmatrix}
0\\
\vec{B}\\
\vec{u}\cdot\vec{B}\vphantom{\big(}\\
\vec{u}
\end{bmatrix}
\end{align}
with the new pressure equation that includes the magnetic energy
\begin{equation}\label{eq:idealMHDpressure}
{p} = (\gamma-1)\left(E - \frac{1}{2}\rho\lVert\vec{u}\rVert^2-\frac{1}{2}\lVert\vec{B}\rVert^2\right)
\end{equation}
defined in the domain $\Omega \subset \mathbb{R}^3$. 

A remarkable outcome of the physically motivated derivation is that we obtain a set of equations which is known to have a number of desirable properties lacking in the classical ideal MHD equations. The system (\ref{eq:3DgeneralizedidealMHD}a-d) is not only symmetrizable \cite{Barth1999} but also Galilean invariant \cite{Powell1999, Janhunen2000, Winters2016}. The non-conservative terms found on the right hand side were first mentioned by Godunov \cite{Godunov1972} who investigated whether the equations of ideal MHD can be put into symmetric hyperbolic form. Interestingly, he found the same additional non-conservative terms were required to obtain a PDE system that is symmetrizable, he took an altogether different approach.
The formulation of ideal MHD system considered here is often referred to as the \emph{eight-wave} formulation, because it supports eight traveling plane wave solutions. As the non-conservative term on the right hand side is proportional to the divergence of the magnetic field, it is, on the continuous level, nothing but adding zero in a clever way.
Further advantages are that the flux Jacobian has only real eigenvalues and the right eigenvectors are linearly independent, \ie the ideal MHD equations in form (\ref{eq:3DgeneralizedidealMHD}a-d) are hyperbolic. We will further investigate on the importance of these consequences later in this work.

Numerical simulations of this system are known to be more stable than the same numerical methods applied to the original ideal MHD equations.
This has been demonstrated by Powell \cite{Powell1999} and numerically confirmed by others (see \eg \cite[Sec.~6.1]{Yee2017}). From the derivations allowing for $\nabla\cdot\vec{B}\ne0$ a new physical understanding for the addition of the non-conservative terms emerge and it is clear that their appearance are essential parts of the system. We conclude that the classical ideal MHD equations are invalid for regions where $\nabla\cdot\vec{B}\ne0$ even if this deviation is only minor when the divergence of the magnetic field is controlled to a sufficient degree. This is due to the fact that classical ideal MHD models contain the divergence-free condition as a decoupled partial differential equation and hence assume that $\nabla\cdot\vec{B}\ne0$ can never happen.
However, our fundamentally physically motivated derivation reveals that classical numerical schemes which neglect the magnetic field divergence terms on the right hand side of (\ref{eq:3DgeneralizedidealMHD}a-d) may discretely describe the wrong physics if they cannot assure \eqref{eq:divB} pointwise.

Without the non-conservative terms, a modeled magnetized fluid may not behave in a physically correct way if the magnetic field divergence is not negligible.
To highlight this, we investigate what effect the Lorentz force has on the fluid with and without the derived non-conservative terms:
\begin{enumerate}
	\item Lorentz force \textbf{with} non-conservative terms
	\begin{align}
		\vec{F}_L &= \vec{j}_e \times \vec{B} = (\nabla \times \vec{B})\times \vec{B} \notag\\&= -\nabla\cdot\left(\frac{1}{2}\|\vec{B}\|^2 - \vec{B}\otimes\vec{B}\right) - (\nabla\cdot\vec{B})\vec{B}
	\end{align}
	The projection of the Lorentz force onto the magnetic field is
	\begin{equation}
		\vec{F}_L \cdot \frac{\vec{B}}{\|\vec{B}\|} = 0,
	\end{equation}
	so $\vec{F}_L \perp \vec{B}$ as expected and the fluid does not feel a force parallel to the magnetic field lines even in the presence of non-vanishing magnetic field divergence.
	\item Lorentz force \textbf{without} non-conservative terms
	\begin{equation}\label{eq:divMaxwellStressTensor}
		\hat{\vec{F}_L} = -\nabla\cdot\left(\frac{1}{2}\|\vec{B}\|^2 - \vec{B}\otimes\vec{B}\right)
	\end{equation}
	The projection of this form of the Lorentz force onto the magnetic field is
	\begin{equation}
		\hat{\vec{F}_L} \cdot \frac{\vec{B}}{\|\vec{B}\|} = - (\nabla\cdot\vec{B})\|\vec{B}\|.
	\end{equation}
	We see that a modeled magnetized fluid only behaves correctly if the magnetic field divergence is zero or at least negligible. In case of any notable non-zero magnetic field divergence, the fluid feels an artificial force parallel to the magnetic field lines. This leads to physically wrong behavior and makes it clear that the ideal MHD system without the correct choice of non-conservative terms is invalid in the case of $\nabla\cdot\vec{B}\ne0$. {Note that $\hat{\vec{F}_L}$ is identical to the divergence of the Maxwell stress tensor for vanishing electric fields.}
\end{enumerate}

In the eight-wave formulation of ideal MHD, the magnetic field divergence is an advected quantity with the fluid.
This can easily be seen by {taking the divergence of} the induction equation (\ref{eq:3DgeneralizedidealMHD}d),
\begin{align}
&\pderivative{}{t}(\nabla\cdot\vec{B}) = -\nabla \cdot\big( \nabla \times (\vec{u}\times\vec{B})\big) -\mu_0\nabla\cdot\vec{j}_m) = -\nabla \cdot(\mu_0\vec{j}_m) = -\nabla\cdot\big(\vec{u}(\nabla\cdot\vec{B})\big),\notag\\
\Rightarrow\quad&\pderivative{}{t}(\nabla\cdot\vec{B}) + \nabla\cdot(\vec{u}(\nabla\cdot\vec{B})) = 0.\label{eq:divBadvection}\end{align}

The appearance of the non-conservative term in the total energy equations can be understood using similar reasoning.
Assume a positive magnetic field divergence which may also be expressed as a \emph{source} of magnetic field. Such a source may generate additional magnetic and/or kinetic energy when moving through the fluid. If, however, we artificially enforce total energy conservation by neglecting the non-conservative term on the total energy, this \emph{increase} in energy can inevitably lead to a \emph{loss} of internal energy. This is due to the fact that internal energy is the reminder of the subtraction of the other energies from the total energy \eqref{eq:idealMHDpressure}. Hence, errors in the computation of the energies will always be shifted into the computed internal energy.
It is clear that in a region with sufficiently strong magnetic sources, the pressure could easily become negative if the total energy is not corrected accordingly from the magnetic fields. If, however, the non-conservative term is included, then the gain in total energy is accounted by the non-conservative term in the total energy evolution equation. In other words, as we do not strictly enforce total energy conservation.
Thus, the thermal energy is not artificially modified and the positivity of a numerical scheme is improved.

Furthermore, the non-conservative terms are necessary to ensure Galilean invariance of the system for any $\nabla\cdot\vec{B}\ne 0$. Note that Galilean invariance is a necessary property of any well-posed theory in non-relativistic physics.

Although we derive the ideal MHD equations for the general case of arbitrary $\nabla\cdot\vec{B}$, we want to
minimize the magnetic field divergence everywhere in numerical simulations to match the evolution simulation results to the observational constraint \eqref{eq:divB}. Hence, the remainder of this paper is concerned with the derivation of an entropy stable scheme that starts from the equations derived in this section. The scheme we build will discretely satisfy the second law of thermodynamics as well as minimize the divergence of the magnetic field by construction.

\section{Incorporating the divergence-free constraint into the model}\label{Sec:divB}
In this section, we investigate the coupling of the magnetic field divergence into the ideal MHD equations.
The investigations in this section are, in principle, self-sufficient and independent from any non-conservative parts being present in the system of equations.
However, we will merge new findings with our results from Sec.~\ref{Sec:idealMHDdivB} wherever appropriate to construct a mathematical model that is valid in regions of non-vanishing magnetic field divergence.

There exist different ways of enforcing \eqref{eq:divB} discretely, commonly called \emph{divergence cleaning} techniques as they are designed to ``clean up" divergence errors made by the numerical algorithms. Many schemes are designed to ``treat'' the divergence errors in the magnetic field, but never get rid of them entirely.
The conventional divergence cleaning methods are shortly described in the following.

\subsection{Non-conservative term approach}
With the non-conservative term approach (also known as source term approach, \eg \cite{Powell1999}), a \emph{non-conservative term} is added to the system of conservation laws that acts to minimize magnetic field divergence. Source terms generally lead to a loss of conservation in the quantities they affect which may be undesirable. The non-conservative term we use in this work can be understood as advecting non-zero magnetic field divergence with the fluid speed \eqref{eq:divBadvection}. As the authors found in previous work (\eg \cite{Derigs2016}) such a divergence advection is especially problematic at stagnation points of the flow where magnetic field divergence can build up due to the dependence of the divergence cleaning on the local fluid velocity. Hence, the non-conservative term approach is typically insufficient to ensure the numerical fulfillment of \eqref{eq:divB} on its own.

\subsection{Projection method}
An alternative approach is the \emph{projection method} {described} by Brackbill and Barnes \cite{Brackbill1980} and Marder \cite{Marder1987}.
The projection method has successfully been applied by, \eg Zachary et al.~\cite{Zachary1994}, Balsara \cite{Balsara1998}, and more recently by Crockett et al.~\cite{Crockett2005} as well as the authors \cite{Derigs2016}. The projection method is implemented for divergence cleaning as a completely separate post-processing step, \ie the original scheme remains unchanged. It has been extensively described by the authors in \cite[Sec.~3.10]{Derigs2016}.
In essence, the projection method acts as a source term and affects the conservation of the magnetic field, but it changes the magnetic field components in an unpredictable way. Therefore, it is unclear if cleaning the divergence errors with a projection method can build a numerical scheme that is thermodynamically consistent.

\subsection{Constrained transport}
Another approach is the \emph{constrained transport method} developed by Evans and Hawley \cite{Evans1988} or Balsara and Spicer \cite{Balsara1999} (reviewed in \cite{Toth2000}). This method originally comes from the staggered-mesh scheme of Yee \cite{Yee66} for electromagnetism in a vacuum. Technically, the divergence-free constraint is satisfied by representing the magnetic field as cell face averaged
quantities (as opposed to the usual choice of cell center volume averages). On such a grid, the MHD equations can be approximated such that they preserve numerical solenoidality of the magnetic field by construction through Stokes' theorem. Note that Balsara and Kim \cite{Balsara2004b} found advantages for the staggered-mesh in their comparison between divergence-\emph{cleaning} and divergence-\emph{free} methods for stringent test cases.
However, it is not clear if provably entropy stable schemes can be constructed for staggered-meshes \cite{Waagan2009}.

\subsection{Generalized Lagrangian multipliers}\label{Sec:GLM}
As detailed by Munz et al.~\cite{Munz2000}, the divergence constraint for the electric field can be coupled with the induction equation by introducing a new real scalar field $\psi$ also known as \emph{generalized Lagrangian multiplier} (GLM). Dedner et al.~\cite{Dedner2002} applied this technique to ideal MHD in order to incorporate the divergence-free condition \eqref{eq:divB} into the ideal MHD equations. However, as we will show later, the GLM modification of the ideal MHD equations  as presented by Dedner et al.~is inconsistent with the second law of thermodynamics as we show in Sec.~\ref{Sec:Dedner}.

In this work, we describe a novel entropy consistent formulation involving generalized Lagrangian multipliers. We call the resulting scheme the ideal GLM-MHD system. Similar to Dedner et al., the idea is not to enforce the divergence-free condition \eqref{eq:divB} exactly, but rather to construct a scheme that is designed to evolve towards a divergence-free state.

We couple the divergence of the magnetic field to Faraday's equation and add a new evolution equation for $\psi$ using a hyperbolic ansatz. The new equations read:
\begin{equation}\label{eq:mod_maxwell}
\derivative{}{t}\vec{B} \vphantom{\frac{\partial\psi}{\partial t}} = \nabla\times\left(\vec{u}\times\vec{B}\right)
\quad \Longrightarrow \quad
\derivative{}{t}\vec{B} = \nabla\times\left(\vec{u}\times\vec{B}\right) \mathcolor{red}{-  c_h \nabla\psi}\, ,\quad \mathcolor{red}{\derivative{}{t}\psi := -c_h(\nabla\cdot\vec{B})}
\end{equation}
where we highlight the modifications in red. {Note that we specified \eqref{eq:mod_maxwell} using Lagrangian derivatives, also known as convective derivatives. The advantage is that this directly leads to a Galilean invariant formulation.}

The newly introduced \emph{divergence-correcting field} is denoted by $\psi$, where $c_h$ is the \emph{hyperbolic divergence cleaning speed}.
Our definition of the generalized Lagrangian multiplier $\psi$, compared to the definition of Dedner et al.~results in a nicer set of entropy variables reducing the complexity of the forthcoming thermodynamic analysis.
It is easily seen that for vanishing magnetic field divergence the correcting field $\psi\rightarrow 0$ and the highlighted contributions in \eqref{eq:mod_maxwell} vanish, returning the model to the ideal MHD equations as derived in Sec.~\ref{Sec:idealMHDdivB}. Thus, the GLM modifications to the ideal MHD model are consistent and correctly restore the continuous limit.

Before we continue, we re-write \eqref{eq:mod_maxwell} into a form similar to \eqref{eq:3DgeneralizedidealMHD} for the sake of convenience:
\begin{subequations}\label{eq:GLMpsi}
	\begin{align}\label{eq:GLM}
	\pderivative{}{t} \vec{B}  + \nabla\cdot (\vec{u}\otimes\vec{B} - \vec{B}\otimes\vec{u} \mathcolor{red}{+ c_h \psi\mat{I}}) + \vec{u}(\nabla\cdot\vec{B}) &= 0,\\
	\label{eq:psi}
	\mathcolor{red}{\pderivative{}{t} \psi + c_h (\nabla\cdot\vec{B}) + \vec{u}\cdot\nabla\psi} &\mathcolor{red}{= 0},
	\end{align}
\end{subequations}
where we again highlight the modification with respect to (\ref{eq:3DgeneralizedidealMHD}d) in red.

If we assume that the solution is sufficiently smooth, such that all derivatives are well defined, we differentiate with respect to time and space to obtain the following relations from (\ref{eq:GLMpsi}):
\begin{align}
	&\frac{\partial}{\partial t} (\nabla \cdot \vec{B}) = - \nabla\cdot\big(\vec{u}(\nabla\cdot\vec{B})\big) - c_h \nabla^2 \psi = - \frac{1}{c_h} \pderivative{{}^2}{t^2} \psi - \frac{1}{c_h}\pderivative{}{t}(\vec{u}\cdot\nabla\psi),\ \text{and}\\
	&\frac{\partial}{\partial t} (\nabla^2 \psi) = - c_h \nabla^2 (\nabla\cdot\vec{B}) - \nabla^2(\vec{u}\cdot\nabla\psi) = -  c_h^{-1} \pderivative{}{t}\nabla\cdot\big(\vec{u}(\nabla\cdot\vec{B})\big) - c_h^{-1} \pderivative{{}^2}{t^2} (\nabla \cdot \vec{B}).
\end{align}
From these relations, it is straightforward to derive wave equations for both the magnetic field divergence as well as the correcting field $\psi$
\begin{alignat}{5}\label{eq:divB_wave}
	&\frac{\partial^2}{\partial t^2} (\nabla \cdot \vec{B}) &&- c_h^2 \nabla^2 (\nabla\cdot\vec{B}) &&- c_h \nabla^2(\vec{u}\cdot\nabla\psi) &&+ \pderivative{}{t}\nabla\cdot(\vec{u}(\nabla\cdot\vec{B})) &&= 0,
	\shortintertext{and}\label{eq:psi_wave}
	&\frac{\partial^2}{\partial t^2} \psi &&- c_h^2 \nabla^2 \psi&& - c_h \nabla\cdot(\vec{u}\cdot\nabla\psi) &&+ \pderivative{}{t}\big(\vec{u}\cdot\nabla\psi\big) &&= 0.
\end{alignat}
We see that the two wave equations \eqref{eq:divB_wave} and \eqref{eq:psi_wave} are coupled through the term $c_h \nabla\cdot(\vec{u}\cdot\nabla\psi)$ and we look at a combined wave equation,
\begin{equation}\label{eq:complexwaveequation}
	\pderivative{{}^2}{t^2}(\nabla\cdot\vec{B}) - c_h^2 \nabla^2 (\nabla\cdot\vec{B}) + \pderivative{}{t}\nabla\big(\vec{u}(\nabla\cdot\vec{B})\big) = \pderivative{{}^2}{t^2}(\hat{\nabla}\psi) - c_h^2 \nabla^2 (\hat{\nabla}\psi) + \pderivative{}{t}\nabla(\vec{u}\cdot(\hat{\nabla}\psi)),
\end{equation}
where we used the notation $\hat{\nabla}\psi := \sum_{i=x,y,z} \pderivative{\psi}{i}$.

Here, we see a complex interaction between the divergence treatment due to the advection of the magnetic monopoles with the fluid velocity, $\vec{u}$, and the newly introduced GLM correction field, $\psi$, as well as the hyperbolic cleaning speed, $c_h$. This complex interaction is expected as $\psi$ and $\vec{B}$ are not independent quantities, but a gradient in $\psi$ is created by a non-zero divergence in the magnetic field.
As such, we investigate the behavior of the magnetic field divergence based on \eqref{eq:complexwaveequation}. It is important to note that the effects described in the following take place simultaneously. However, for the sake of simplicity, we split the analysis into separate parts and discuss the effect as if they are independent.
\begin{description}
	\item[Wave components in \eqref{eq:complexwaveequation}]:
	\begin{equation}
		\frac{\partial^2}{\partial t^2} (\nabla\cdot\vec{B}) - c_h^2 \nabla^2 (\nabla\cdot\vec{B}) = 0
	\end{equation}
	This part is a wave equation describes the isotropic propagation of $\nabla\cdot\vec{B}$ with constant speed $c_h$. So, local divergence errors by this part of \eqref{eq:complexwaveequation} are isotropically propagated away from where they have been created with finite speed $c_h$. Looking back at the initially discussed source term approach, it becomes clear that the GLM approach does not suffer from the problem of accumulating magnetic field divergence at stagnation points of the fluid as this term is independent of the fluid velocity, $\vec{u}$, and we always have $c_h > 0$.
	\begin{equation}
		\frac{\partial^2}{\partial t^2} (\hat{\nabla}\psi) - c_h^2 \nabla^2 (\hat{\nabla}\psi) = 0
	\end{equation}
	We see that $\hat{\nabla}\psi$ propagates isotropically with constant speed $c_h$, just as the magnetic field divergence.
	\item[Advective components in \eqref{eq:complexwaveequation}]:
	\begin{equation}\label{eq:wave_divBadvection}
		\frac{\partial}{\partial t} (\nabla\cdot\vec{B}) + \nabla\cdot(\vec{u}(\nabla\cdot\vec{B})) = 0
	\end{equation}
	This equation is a standard advection equation describing the transport of $\nabla\cdot\vec{B}$ by bulk motion. It is clear that \eqref{eq:wave_divBadvection}, as a continuity equation, conserves $\nabla\cdot\vec{B}$ by construction. This part is a direct consequence of the non-conservative terms we derived in Section \ref{Sec:idealMHDdivB} and corresponds to the ``divergence wave'' of the well-known eight-wave formulation.
	\begin{equation}\label{eq:wave_psiadvection}
	\frac{\partial}{\partial t} (\hat{\nabla}\psi) + \nabla\cdot(\vec{u}(\hat{\nabla}\psi)) = 0
	\end{equation}
	As before, we see that $\hat{\nabla}\psi$ behaves identically to the magnetic field divergence in that it is advected with the flow. Just like \eqref{eq:wave_divBadvection}, this passive advection equation is an expected result for a Galilean invariant formulation.
\end{description}

In \eqref{eq:mod_maxwell} we introduce the possibility to advect the divergence error with the correcting field $\psi$. Also, the correcting field couples into the induction and therefore can alter the magnetic field. This transfer of information between the magnetic field components and the correcting field is important for divergence cleaning, but raises the question of how can $\psi$ affect the magnetic energy $E_\mathrm{mag} = \half\|\vec{B}\|^2$. It stands to reason that the correcting field contains some form of ``energy'' for which we should account. As the thermal energy is computed by subtracting the kinetic and magnetic energies from the total energy, any information regarding loss/gain of magnetic energy would be falsely attributed into thermal energy.

Tricco and Price~\cite{Tricco2012} investigated the effect of Dedner et al.'s GLM modification of the ideal MHD equations in the framework of smoothed particles hydrodynamics (SPH). They pointed out that the energy contained within the $\psi$ field should be defined such that, in the absence of damping terms, any change in magnetic energy should be balanced by a corresponding change in the energy stored in the correcting field.
Hence, we introduce a new kind of energy stored in the $\psi$ field, $E_\psi$, which becomes a new component of the total fluid energy, $E$, for ensuring total energy conservation.
As $\vec{B}$ and $\psi$ both have units of magnetic fields, and because we observe a surprising symmetry between $B_1$ and $\psi$ in the one-dimensional form of \eqref{eq:mod_maxwell} for vanishing fluid velocities,
\begin{equation*}
	(B_1)_t + c_h (\psi)_x + u (B_1)_x = 0,\quad\text{and}\quad (\psi)_t + c_h (B_1)_x + u (\psi)_x = 0,
\end{equation*}
we make the ansatz
\begin{equation}\label{eq:Epsi}
E_\psi := \frac{1}{2} \psi^2.
\end{equation}

Since we introduce a new form of energy into the system, we must account for its temporal evolution. Therefore, we investigate the effect which the modifications \eqref{eq:GLM} and \eqref{eq:psi} have on the conservation law for the total energy, $E=\frac{p}{\gamma -1}+\frac{\rho}{2}\|\vec{u}\|^2+\half\|\vec{B}\|^2+\half\psi^2$. To do so we examine the evolution of the magnetic energy
\begin{equation}
\pderivative{}{t}\left(\mathcolor{red}{\frac{1}{2}\|\vec{B}\|^2}\right) = \vec{B} \cdot \underbrace{\pderivative{\vec{B}}{t}}_{\eqref{eq:GLM}} = - \vec{B} \cdot \nabla \left(\vec{u}\otimes\vec{B} - \vec{B}\otimes\vec{u}\right) \mathcolor{red}{- c_h \vec{B} \cdot \nabla \psi} + (\vec{B} \cdot \vec{u})(\nabla\cdot\vec{B}),
\end{equation}
as well as a contribution from the new $\psi$ field energy \eqref{eq:Epsi}
\begin{equation}
\pderivative{}{t}\left(\mathcolor{red}{\frac{1}{2}\psi^2}\right) = \psi \cdot \underbrace{\pderivative{\psi}{t}}_{\eqref{eq:psi}} = \mathcolor{red}{- c_h \psi (\nabla \cdot  \vec{B})+ \vec{u}\psi(\nabla\psi)}.
\end{equation}
Since
\begin{equation*}
\mathcolor{red}{c_h \vec{B} \cdot \nabla \psi} + \mathcolor {red}{c_h \psi (\nabla \cdot \vec{B})} = \mathcolor {red}{c_h \nabla\cdot (\psi \vec{B})},
\end{equation*}
we find that the correct form of the total energy conservation equation for ideal GLM-MHD is given by
\begin{multline}
\pderivative{}{t} \left( \frac{1}{2} \rho \|\vec{u}\|^2 + \frac{p}{\gamma-1} + \mathcolor{red}{\frac{1}{2} \|\vec{B}\|^2} + \mathcolor{red}{\frac{1}{2} \psi^2} \right) = - \nabla \cdot \Bigg( \vec{u}\left(\frac{1}{2}\rho\|\vec{u}\|^2 + \frac{p}{\gamma-1} + p + \|\vec{B}\|^2 \right) - \vec{B}(\vec{u}\cdot\vec{B})\\
\mathcolor{red}{ + c_h \psi \vec{B}} \Bigg) - (\vec{B} \cdot \vec{u})(\nabla\cdot\vec{B}) \mathcolor{red}{+ \vec{u}\psi(\nabla\psi)}.
\end{multline}

Hence, the new \textbf{ideal GLM-MHD system} reads
\refstepcounter{equation}\label{eq:3DIDEALGLMMHD}
\begin{equation}\tag{\theequation a-e}
\boxed{%
	\pderivative{}{t}\, \vec{q} + \nabla\cdot\vec{f} =
	\pderivative{}{t}\begin{bmatrix} \rho \\ \vphantom{\big(}\rho\vec{u} \\ \vphantom{\big(}E \\ \vec{B}\\ \psi \end{bmatrix}
	+
	\nabla\cdot\begin{bmatrix} \rho\vec{u} \\
	\rho(\vec{u}\otimes\vec{u}) + \big(p+\frac{1}{2}\|\vec{B}\|^2\big)\mat{I}-\vec{B}\otimes\vec{B} \\
	\vec{u}\big(\frac{1}{2}\rho\|\vec{u}\|^2 + \frac{\gamma p}{\gamma - 1} + \|\vec{B}\|^2 \big) - \vec{B}(\vec{u}\cdot\vec{B}) + c_h \psi \vec{B} \\
	\vec{u}\otimes\vec{B} - \vec{B}\otimes\vec{u} + c_h \psi\mat{I} \\
	c_h \vec{B}
	\end{bmatrix}
	=
	-\vec{\Upsilon}_\mathrm{GLM},}
\end{equation}
with
\begin{equation}\label{eq:GLM-source}
\vec{\Upsilon}_\mathrm{GLM} := (\nabla \cdot \vec{B})
\begin{bmatrix}
0 \\ \vphantom{\big(}\vec{B} \\ \vphantom{\big(}\vec{u}\cdot\vec{B} \\ \vec{u} \\ 0
\end{bmatrix}
+ (\nabla\psi)\cdot
\begin{bmatrix}
\vec{0} \\ 0 \\ \vec{u}\psi \\ 0 \\ \vec{u}
\end{bmatrix}
\end{equation}
and the new thermodynamic pressure
\begin{equation}
p = (\gamma-1)\,\epsilon \qquad \mathrm{and}\qquad \epsilon = E - \frac{1}{2}\rho\lVert\vec{u}\rVert^2-\frac{1}{2}\lVert\vec{B}\rVert^2 - \frac{1}{2} \psi^2.\label{eq:pressure}
\end{equation}

If the divergence of the magnetic field is zero, the new system is identical to the original ideal MHD system (\ref{eq:3DgeneralizedidealMHD}). If, however, the initial solution does not satisfy the divergence-free constraint then the deviations will decay. This new set of equations ensures that any magnetic divergence caused by inaccuracies of a numerical method for the ideal GLM-MHD system remains small.

{The ideal GLM-MHD system is invariant under a Galilean transformation, \ie invariant to a transformation into a frame of reference moving with a constant relative velocity $\vec{u}_0$ ($\vec{x}' = \vec{x} - \vec{u}_0 t$, $\vec{u}' = \vec{u} - \vec{u}_0 $, $t'=t$). It shows the correct transformation behavior of $\derivative{}{t'} = \derivative{}{t}$ and $\pderivative{}{t'} = \pderivative{}{t} + \vec{u}_0\cdot\nabla$.}

Next, we investigate the structure of the obtained ideal GLM-MHD system and discuss crucial properties such as the hyperbolicity as well as the partially altered eigenstructure of the new system. We then compare the new set of equations to existing formulations in the beginning of the next section.

\subsection{Multi-dimensional structure of the ideal GLM-MHD equations}
To simplify the discussion of the new system we write \eqref{eq:3DIDEALGLMMHD} in one-dimensional form,
\begin{equation}
\frac{\partial}{\partial t} \vec{q} + \frac{\partial}{\partial x} \vec{f}^x + \vec{\Upsilon}^x = \vec{0},\label{eq:1DIDEALGLMMHD}
\end{equation}
where $\vec q = \vec q(\vec{x},t)$ is the vector of conservative variables, $\vec{f}^{x}(\vec{q})$ is the flux vector in $x$-direction, and $\vec{\Upsilon}$ is the non-conservative term.
\begin{align}
\vec{q} &=
\begin{bmatrix}\rho & \rho u & \rho v & \rho w & E & B_1 & B_2 & B_3 & \psi \end{bmatrix}^\intercal,\\
\vec{f}^x &=
\begin{bmatrix}\rho \, u \\ \rho u^2 + p + \frac{1}{2} \lVert \vec{B} \rVert^2 -B_1 B_1 \\ \rho \, u \, v - B_1  B_2 \\ \rho \, u \, w - B_1  B_3 \\ u \hat{E} - B_1 \big(\vec{u} \cdot \vec{B}\big) + c_h \psi B_1 \\ c_h \psi \\ u\, B_2 - v \, B_1 \\ u \, B_3 - w\, B_1 \\ c_h B_1 \end{bmatrix}, \qquad
\vec{\Upsilon}^x = \pderivative{B_1}{x} \begin{bmatrix} 0 \\ B_1 \\ B_2 \\ B_3 \\ \vec{u}\cdot\vec{B} \\ u \\ v \\ w \\ 0 \end{bmatrix}
+ \pderivative{\psi}{x} \begin{bmatrix} 0 \\ 0 \\ 0 \\ 0 \\ u\psi \\ 0 \\ 0 \\ 0 \\ u \end{bmatrix}\label{eq:GodunovSource},
\end{align}
with
\begin{equation}
\hat{E} := \frac{1}{2}\rho\|\vec{u}\|^2 + \frac{\gamma p}{\gamma - 1} + \|\vec{B}\|^2.
\end{equation}

We limit the analysis of the ideal GLM-MHD system \eqref{eq:3DIDEALGLMMHD} to one spatial dimension in the following. 
The main motivation for this restriction is because the analysis of the eigenstructure as well as the derivation of the numerical fluxes described later in this work proved to be quite intense. However, this restriction is done without loss of generality because the spatial dimensions are decoupled. Thus, for completeness, we summarize the results of the derivations in the $y$ and $z$-direction in \ref{app:twoandthreedims}.

\subsection{Eigenvalues of the ideal GLM-MHD system}
An important step in the investigation of the properties of the system is to compute the eigenvalues of the ideal GLM-MHD system, which are the speeds of the different waves involved in the solution. In doing so we find that the new ideal GLM-MHD system does not show a degeneracy of the eigenvalues like in the {eight-wave formulation of ideal MHD} where the entropy and divergence waves travel with the same speed, and hence have the same eigenvalue \cite[Sec.~3.5.1]{Powell1999}.

First, we compute the flux Jacobian of the ideal GLM-MHD system
\begin{multline}\label{eq:A_matrix}
\mat{A}^x := \pderivative{\vec{f}^x}{\vec{q}} =\\
\resizebox{.9\hsize}{!}{$ %
	\begin{bmatrix} 0 & 1 & 0 & 0 & 0 & 0 & 0 & 0 & 0
	\\ \frac{1}{2}\left(\gamma-1\right)\|\vec{u}\|^2-u^2 &
	u\,\left(3-\gamma\right) & v\,\left(1-\gamma\right) & w\,\left(1-
	\gamma\right) & \gamma-1 & -{B_1}\,\gamma & {B_2}\,\left(2-
	\gamma\right) & {B_3}\,\left(2-\gamma\right) & \psi\,\left(1-
	\gamma\right) \\ -u\,v & v & u & 0 & 0 & -{B_2} & -{B_1} & 0
	& 0 \\ -u\,w & w & 0 & u & 0 & -{B_3} & 0 & -{B_1} & 0 \\ A_{5,1} & A_{5,2} & u\,v\,\left(1-
	\gamma\right)-\frac{{B_1}\,{B_2}}{\rho} & u\,w\,\left(1-
	\gamma\right)-\frac{{B_1}\,{B_3}}{\rho} & u\,\gamma &
	{c_h}\,\psi- u \gamma B_1 - v B_2 - w B_3 & {B_2}\,u\,\left(2-\gamma
	\right)-{B_1}\,v & {B_3}\,u\,\left(2-\gamma\right)-{B_1}
	\,w & {B_1}\,{c_h}-\psi\,u\,\gamma \\
	0 & 0 & 0 & 0 & 0 & 0 & 0 & 0 & {c_h} \\
	\frac{{B_1}\,v-{B_2}\,u}{\rho} & \frac{{B_2}}{\rho} & -\frac{{B_1}}{\rho} & 0 & 0 & -v & u & 0 & 0\\
	\frac{{B_1}\,w-{B_3}\,u}{\rho} & \frac{{B_3}}{\rho} & 0 & -\frac{{B_1}}{\rho} & 0 & -w & 0 & u & 0 \\ 0 & 0 & 0
	& 0 & 0 & {c_h} & 0 & 0 & 0 \\ \end{bmatrix},$}
\end{multline}
with
\begin{align}
A_{5,1}^x &= -2 \gamma u \left(E-\frac{1}{2}\psi^2\right) + 2 B_1 \left(\vec{u}\cdot\vec{B} + uB_1\right) + 2 u (\gamma-1) \left( \rho \|\vec{u}\|^2\right) + u (\gamma-2) \left(\frac{1}{2}\|\vec{B}\|^2 \right)
, \\
A_{5,2}^x &= -\frac{1}{2}(\gamma-1)\left(\|\vec{u}\|^2 + 2 u^2\right) + \frac{\gamma}{\rho} \left(E - \frac{1}{2} \|\vec{B}\|^2 - \frac{1}{2} \psi^2 \right) + \frac{B_2^2 + B_3^2}{\rho}.
\end{align}
We then add the non-conservative term written in matrix form
\begin{equation}
\vec{\Upsilon}^x =
\pderivative{B_1}{x}
\begin{bmatrix}
0\\ B_1 \\ B_2 \\ B_3 \\ \vec{u}\cdot\vec{B} \\ u \\ v \\ w \\ 0
\end{bmatrix}
+ \pderivative{\psi}{x} \begin{bmatrix} 0 \\ 0 \\ 0 \\ 0 \\ u\psi \\ 0 \\ 0 \\ 0 \\ u \end{bmatrix}
=
\begin{bmatrix} 0 & 0 & 0 & 0 & 0 & 0 & 0 & 0 & 0
\\ 0 & 0 & 0 & 0 & 0 & B_1 & 0 & 0 & 0 \\ 0 & 0 & 0 & 0 & 0 & B_2 & 0
& 0 & 0 \\ 0 & 0 & 0 & 0 & 0 & B_3 & 0 & 0 & 0 \\ 0 & 0 & 0 & 0 & 0
& \vec{u}\cdot\vec{B} & 0 & 0 & u\psi \\ 0 & 0 & 0 & 0 & 0 & u & 0 & 0 & 0 \\ 0 & 0 & 0
& 0 & 0 & v & 0 & 0 & 0 \\ 0 & 0 & 0 & 0 & 0 & w & 0 & 0 & 0 \\ 0
& 0 & 0 & 0 & 0 & 0 & 0 & 0 & u \\ \end{bmatrix}
\pderivative{}{x}
\begin{bmatrix}\rho \\ \rho u \\ \rho v \\ \rho w \\ E \\ B_1 \\ B_2 \\ B_3 \\ \psi \end{bmatrix}
:= \mat{\hat{\Upsilon}}^x\, \pderivative{\vec{q}}{x},
\end{equation}
to \eqref{eq:A_matrix} and obtain
\begin{multline}\label{eq:AP_matrix}
	\hspace*{-2\parindent}
	\mat{A}_\mat{\Upsilon}^x := \mat{A}^x + \mat{\hat{\Upsilon}}^x =\\
	\hspace*{-2\parindent}
	\resizebox{\hsize}{!}{$ %
	\begin{bmatrix} 0 & 1 & 0 & 0 & 0 & 0 & 0 & 0 & 0
	\\ \frac{1}{2}\left(\gamma-1\right)\|\vec{u}\|^2-u^2 &
	u\,\left(3-\gamma\right) & v\,\left(1-\gamma\right) & w\,\left(1-
	\gamma\right) & \gamma-1 & {B_1}\,(1-\gamma) & {B_2}\,\left(2-
	\gamma\right) & {B_3}\,\left(2-\gamma\right) & \psi\,\left(1-
	\gamma\right) \\ -u\,v & v & u & 0 & 0 & 0 & -{B_1} & 0
	& 0 \\ -u\,w & w & 0 & u & 0 & 0 & 0 & -{B_1} & 0 \\ A_{5,1} & A_{5,2} & u\,v\,\left(1-
	\gamma\right)-\frac{{B_1}\,{B_2}}{\rho} & u\,w\,\left(1-
	\gamma\right)-\frac{{B_1}\,{B_3}}{\rho} & u\,\gamma &
	{c_h}\,\psi- u B_1 (1-\gamma) & {B_2}\,u\,\left(2-\gamma
	\right)-{B_1}\,v & {B_3}\,u\,\left(2-\gamma\right)-{B_1}
	\,w &c_h B_1 + u\psi(1-\gamma) \\ 0 & 0 & 0 & 0 & 0 & u
	& 0 & 0 & {c_h} \\ \frac{{B_1}\,v-{B_2}\,u}{\rho} &
	\frac{{B_2}}{\rho} & -\frac{{B_1}}{\rho} & 0 & 0 & 0 & u & 0
	& 0 \\ \frac{{B_1}\,w-{B_3}\,u}{\rho} & \frac{{B_3}}{
		\rho} & 0 & -\frac{{B_1}}{\rho} & 0 & 0 & 0 & u & 0 \\ 0 & 0 & 0
	& 0 & 0 & {c_h} & 0 & 0 & u \\ \end{bmatrix}.$}
\end{multline}

\begin{figure}[h]
	\centering
	\includegraphics[scale=1]{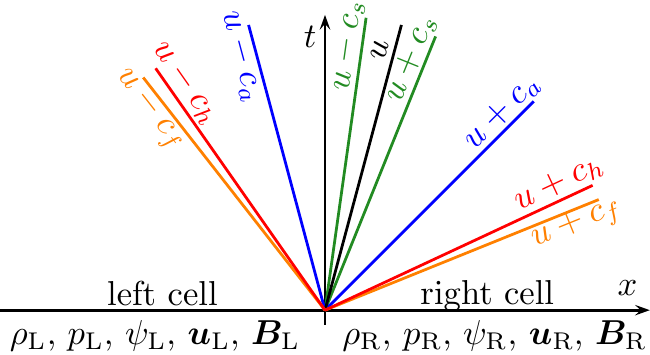}
	\caption{Spacetime sketch of a typical Riemann fan spanned by the eigenvalues \eqref{eq:eigenvalues} with some global $c_h$ that equals the fastest magnetoacoustic speed present in the entire numerical simulation. We do this to not affect the global time step size, see Section~\ref{Sec:ch}.}
	\label{fig:Riemann}
\end{figure}

From \eqref{eq:AP_matrix} we compute the eigenvalues of the ideal GLM-MHD system in $x$-direction:
\newcommand{\f}{\ensuremath{\mathrm{f}}}
\renewcommand{\s}{\ensuremath{\mathrm{s}}}
\renewcommand{\a}{\ensuremath{\mathrm{a}}}
\newcommand{\E}{\ensuremath{\mathrm{E}}}
\begin{align}\label{eq:eigenvalues}
	\lambda_{\pm \f}^x &= u \pm c_\f, \qquad \lambda_{\pm \s}^x = u \pm c_\s, \qquad \lambda_{\pm \a}^x = u \pm c_\a, \qquad \lambda_\E^x = u, \qquad \lambda_{\pm \psi}^x = u \pm c_h,
\shortintertext{with}
	\label{eq:alotofequations}
	c_\a^2& = b_1^2, \quad
	c_\mathrm{f,s}^2 = \frac{1}{2}\left(a^2+\|\vec{b}\|^2 \pm \sqrt{(a^2+\|\vec{b}\|^2)^2 - 4a^2 b_1^2}\right), \ a^2 = \gamma \, \frac{p}{\rho}, \quad \vec{b} = \frac{\vec{B}}{\sqrt{\rho}}, 
\end{align}
where $c_\f$ and $c_\s$ are the fast and slow magnetoacoustic wave speeds, respectively, and $c_\a$ is the Alfv\'en wave speed. In \eqref{eq:alotofequations}, the plus sign corresponds to the fast magnetoacoustic speed, $c_\f$, and the minus sign corresponds to the slow magnetoacoustic speed, $c_\s$. We find that all eigenvalues have multiplicity one. They are depicted in Fig.~\ref{fig:Riemann} Note that, however, some eigenvalues may coincide depending on the magnetic field strength and direction. Hence, it is not straightforward to compute the complete set of eigenvectors \cite{Brio1988,Cargo}.

We see that the divergence wave, commonly found when analyzing the original ideal MHD system in conjunction with the source term, splits into a left and a right going $\psi$-wave. If we set $\psi = c_h = 0$, $\mat{A}_\mat{\Upsilon}^x$ becomes identical to the flux Jacobian matrix of the eight-wave formulation of the ideal MHD equations and we recover a single divergence wave with eigenvalue $\lambda_\mathrm{D}^x = u$.

We note that due to $\lambda^x_{\pm \psi} \ne c_h$, we find that the commonly chosen approximation to set $c_h = \max(\lambda)$ (see \eg \cite[Section 3.4]{Feng2016}) may lead to a violation of the CFL criterion, which may cause robustness issues when the CFL number is not adapted accordingly.

\subsection{The hyperbolic propagation speed $c_h$}\label{Sec:ch}
An important issue is how to select the cleaning wave speed $c_h$. We implement the new numerical scheme for the ideal GLM-MHD system in the multi-physics code \texttt{FLASH} (see \cite{Derigs2016}) where we often experience that the MHD solver accounts for 10\% of the overall CPU time in real applications (see \eg \cite{Walch2014}). As such, it is important to determine a cleaning speed $c_h$ such that the propagation of the $\psi$ field is most effective, but does not influence the size of the time step compared to standard ideal MHD implementations.
We immediately see that this requirement is fulfilled by choosing $c_h$ to be the difference between the maximum eigenvalue, $\lambda_\mathrm{max}$, and the maximum fluid speed,
\begin{equation}\label{eq:ch}
	c_h = \lambda_\mathrm{max} - u_\mathrm{max,\Omega},
\end{equation}
where $u_\mathrm{max,\Omega} = \max\limits_\Omega(|u|,|v|,|w|)$ is the largest (physical) speed found in the entire simulation domain.

If the fluid speed is zero (\ie at stagnation points) the divergence correction is most effective with $c_h = \lambda_\mathrm{max}$. If $u_\mathrm{max,\Omega} \ne 0$, then $c_h$ is reduced such that neither of the $\psi$ eigenvalues exceed the maximum of the remaining wave speeds to guarantee that the GLM modification does not negatively affect the time step of the simulation.
Note that \eqref{eq:ch} also suggests that the simple and commonly used choice $c_h = a \cdot \lambda_\mathrm{max}$ may be inappropriate for \emph{any} value of $a \in (0,1]$.

Note that it is possible to select values for $c_h$ that exceed \eqref{eq:ch}. However, this would lead to the approximation being dominated by the $\psi$-wave and, hence, will shrink the time step size. Nevertheless, this will result in (even) faster correction of the magnetic field divergence, although the authors have not felt this necessary in the numerical results obtained for this work. Another possibility, although rarely seen in the literature, could be a local, instead of a global value for the hyperbolic cleaning speed, $c_h$. However, this would add $c_h$ as yet another field variable which we want to avoid in the highly parallelized targeted framework. Nonetheless, we mention this as a feasible part of future works.

\subsection{Alternative non-conservative terms}
From the derivations of the generalized ideal MHD equations for non-vanishing magnetic field divergence, we found that particular non-conservative terms are necessary to ensure the validity of the numerical scheme in situations in which \eqref{eq:divB} is not fulfilled exactly. However, it is known that schemes which do not preserve exact conservation of the physical quantities can produce erroneous shock speeds, \eg \cite{Janhunen2000}.
In cases where $\nabla\cdot\vec{B}=0$, there is no non-conservative contribution to any of the physical quantities. Note that this is not only the case for vanishing magnetic fields, but for arbitrary configurations, given that a numerical approximation is properly initialized.
Even in the presence of $\nabla\cdot\vec{B}\ne0$, the non-conservative part in \eg the total energy equation could be attributed to excess energy in the magnetic field caused by sources/sinks present in the solution. The importance of the non-conservative term in the momentum equations has already been discussed in Sec.~\ref{Sec:idealMHDdivB} and is important to ensure $\vec{F}_L \perp \vec{B}$ wherever $\nabla\cdot\vec{B}\ne0$.

Nevertheless, there are at least three possibilities to construct an entropy stable scheme for ideal MHD:
\begin{enumerate}
	\item The full non-conservative terms as derived in Sec.~\ref{Sec:idealMHDdivB}:
	\begin{equation}\label{eq:fullNonConsTerm}
	\vec{\Upsilon}^\mathrm{magnetic} = \underbrace{\left(\pderivative{B_1}{x} + \pderivative{B_2}{y} + \pderivative{B_3}{z}\right)}_{\nabla\cdot\vec{B}}  \begin{bmatrix} 0 & B_1 & B_2 & B_3 & \vec{u}\cdot\vec{B} & u & v & w & 0 \end{bmatrix}^\intercal.
	\end{equation}
	\item Only the non-conservative terms needed to ensure $\vec{F}_L \perp \vec{B}$:
	\begin{equation}
	\vec{\Upsilon}_\mathrm{BB}^\mathrm{magnetic} := \underbrace{\left(\pderivative{B_1}{x} + \pderivative{B_2}{y} + \pderivative{B_3}{z}\right)}_{\nabla\cdot\vec{B}}  \begin{bmatrix} 0 & B_1 & B_2 & B_3 & 0 & 0 & 0 & 0 & 0 \end{bmatrix}^\intercal.
	\end{equation}
	This non-conservative terms have first been suggested by Brackbill and Barnes \cite{Brackbill1980}.
	\item Furthermore, Janhunen \cite{Janhunen2000} presented non-conservative terms that add the advection of the magnetic field, see \eqref{eq:divBadvection}, but preserve the conservation in all thermodynamics quantities:
	\begin{equation}
	\vec{\Upsilon}_\mathrm{J}^\mathrm{magnetic} := \underbrace{\left(\pderivative{B_1}{x} + \pderivative{B_2}{y} + \pderivative{B_3}{z}\right)}_{\nabla\cdot\vec{B}}  \begin{bmatrix} 0 & 0 & 0 & 0 & 0 & u & v & w & 0 \end{bmatrix}^\intercal.
	\end{equation}
\end{enumerate}

The conservative formulation, \ie $\vec{\Upsilon}_0 := \vec{0}$, of the ideal MHD equations is \emph{not} entropy consistent as demonstrated by Godunov in the 1970s \cite{Godunov1972}.
Because $\vec{v}\cdot\vec{\Upsilon}^\mathrm{magnetic} = \vec{v}\cdot\vec{\Upsilon}_\mathrm{BB}^\mathrm{magnetic} = \vec{v}\cdot\vec{\Upsilon}_\mathrm{J}^\mathrm{magnetic} = 2\beta(\vec{u}\cdot\vec{B})$, all of the non-conservative terms mentioned are interchangeable in an entropic sense. That is, they all ensure entropy consistency of the scheme. We note that only the first one, $\vec{\Upsilon}^\mathrm{magnetic}$, symmetrizes the PDE system \cite{Godunov1972} and complies with our derivation of the ideal MHD equations in the general case where $\nabla\cdot\vec{B}\ne0$.

Earlier publications, \eg \cite[Sec.~6.1]{Yee2017}, investigated all four non-conservative terms choices, including the non entropy consistent fully conservative formulation, and found that the full non-conservative terms, $\vec{\Upsilon}^\mathrm{magnetic}$, has the best properties with respect to numerical stability and accuracy in the sense of convergence to analytic solutions. We found little difference between the full and the Janhunen non-conservative terms in all of our numerical tests, however, due to the reasoning given in Sec.~\ref{Sec:idealMHDdivB}, we perform all numerical tests in the following section using the full non-conservative terms \eqref{eq:fullNonConsTerm}. An exception to this is a two-dimensional shock tube test described in Sec.~\ref{Sec:2DShocktube}, as this test is specifically designed to show a breakdown of the eight-wave formulation. Here, we explicitly test the scheme also against the non-conservative terms that was suggested by Janhunen, $\vec{\Upsilon}_\mathrm{J}^\mathrm{magnetic}$, but find no notable difference to using the full non-conservative terms.

\subsection{Alternative GLM ansatz}
We also considered alternative entropy-consistent GLM-modifications. If one uses an Eulerian instead of a Lagrangian ansatz for the $\psi$ evolution equation (compare to \eqref{eq:mod_maxwell}),
\begin{equation}\label{eq:mod_maxwell2}
	\derivative{}{t}\vec{B} \vphantom{\frac{\partial\psi}{\partial t}} = \nabla\times\left(\vec{u}\times\vec{B}\right)
	\quad \Longrightarrow \quad
	\derivative{}{t}\vec{B} = \nabla\times\left(\vec{u}\times\vec{B}\right) \mathcolor{red}{-  c_h \nabla\psi}\, ,\quad \mathcolor{red}{\pderivative{}{t}\psi := -c_h(\nabla\cdot\vec{B})},
\end{equation}
we obtain the same fluxes as before \eqref{eq:3DIDEALGLMMHD}, however with a slightly simpler source term:
\begin{equation}
	\vec{\Upsilon}_\mathrm{GLM,\,alt} := (\nabla \cdot \vec{B})
	\begin{bmatrix}
	0 & \vphantom{\big(}\vec{B} & \vphantom{\big(}\vec{u}\cdot\vec{B} & \vec{u} & 0
	\end{bmatrix}^\intercal
\end{equation}
Although one can see the absence of the non-conservative $\psi \vec{u}\cdot\nabla\psi$ term on the right hand side of the total energy equation as an advantage, the resulting scheme is not Galilean invariant. This is immediately seen by computing the eigenstructure of this alternative system. While all other wave speeds stay the same, the GLM wave speeds takes the form
\begin{equation}
	\lambda_{\pm\psi,\mathrm{alt}} = \frac{1}{2} \left(u \pm \sqrt{u^2 + 4 c_h^2}\right).
\end{equation}
They are clearly not Galilean invariant. However, Galilean invariance is an important physical property for our new scheme and therefore we do not pursue this alternative ansatz any further in this work. Nevertheless, it should be noted that using \eqref{eq:mod_maxwell2}, one can build an entropy stable scheme, \eg \cite{Bohm2017}. We derived and implemented a numerical scheme for this alternative ansatz and find similar results without Galilean invariance. However, we observe slightly reduced robustness.

\section{Deriving an entropy stable numerical scheme}\label{Sec:entropy}
In this section, we briefly introduce the concept of entropy conservation and stability on the continuous level for the ideal GLM-MHD equations and derive numerical flux functions that can be used to implement an entropy stable numerical scheme for ideal MHD simulations. Furthermore, we perform an entropy analysis for several known GLM formulations. A broader introduction to the concept of entropy has been given by the authors in \cite{Derigs2016,Winters2016} as well as in the pioneering works of Tadmor \cite{Tadmor1984,Tadmor1987} and Barth \cite{barth2006role}.

We want to construct a numerical scheme that not only complies with a subset of the thermodynamical laws, but that is in agreement with \emph{all} universally valid laws of thermodynamics - including the second law of thermodynamics, \ie the entropy inequality.

We define the physical entropy density, divided by the constant $(\gamma -1)$ for convenience to be
\begin{equation}\label{eq:entropy}
	S(\vec{q}) = -\frac{\rho s}{\gamma - 1} \qquad \mbox{with} \qquad s = \ln\frac{p}{\rho^\gamma} = -(\gamma-1)\ln(\rho) - \ln(\beta)-\ln(2) \qquad\mbox{and}\qquad\beta=\frac{\rho}{2p} \propto \frac{1}{T},
\end{equation}
where $s$ is the entropy per particle, and $\beta$ is the inverse temperature. An approximation obeys the second law of thermodynamics in two regimes:
\refstepcounter{equation}\label{eq:EntropyEqualityies}
\begin{enumerate}
	\item For smooth solutions, we can design numerical methods to be \textbf{\emph{entropy conservative}} if, discretely, the local changes of entropy are the same as predicted by the continuous entropy conservation law
	\begin{equation*}\label{eq:EntropyEquality}\tag{\theequation a}
		\pderivative{}{t}S + \nabla \cdot (\vec{u}S) = 0.
	\end{equation*}

	\item For discontinuous solutions, the approximation is said to be \textbf{\emph{entropy stable}} if the entropy always possesses the correct sign (where we use the mathematical notation that entropy is a decaying function) and the numerical scheme produces more entropy than an entropy conservative scheme and satisfies the entropy inequality
	\begin{equation*}\label{eq:EntropyInequality}\tag{\theequation b}
		\pderivative{}{t}S + \nabla \cdot (\vec{u}S) \le 0,
	\end{equation*}
	that can be interpreted as the entropy conservation law \eqref{eq:EntropyEquality} augmented with a non-positive entropy source term.
\end{enumerate}

For switching from conserved to entropy space, we introduce the entropy variables
\begin{equation}\label{eq:EntropyVars}
	\vec{v} = \pderivative{S}{\vec{q}} = \left[\frac{\gamma - s}{\gamma - 1}-\beta \lVert\vec{u}\rVert^2,\, 2\beta u,\, 2\beta v,\, 2\beta w,\, -2\beta,\, 2\beta B_1,\, 2\beta B_2,\, 2\beta B_3,\, 2 \beta \psi \right]^\intercal.
\end{equation}

\subsection{Continuous entropy analysis}\label{scn:entropyanalysis}
Now that we prepared the necessary framework for an entropy analysis, we are interested in the agreement of the proposed new ideal GLM-MHD system \eqref{eq:3DIDEALGLMMHD} with \eqref{eq:EntropyEquality}. We examine how the individual components of the ideal GLM-MHD flux contract into entropy space to see if it is possible to construct schemes from \eqref{eq:3DIDEALGLMMHD} which comply with the continuous entropy conservation law. In addition, we analyze the applicability of entropy analysis to different ideal MHD + GLM systems already available in the literature.

To increase the clarity of the following derivations, we split the new flux into three pieces:
\begin{equation}\label{eq:1DidealGLMMHDsplitted}
\vec{f}^x = \vec{f}^{x,\mathrm{hydro}} + \vec{f}^{x,\mathrm{magnetic}} + \vec{f}^{x,\mathrm{\psi}} =
\begin{bmatrix}\rho \, u \\ \rho u^2 + p \\ \rho \, u \, v \\ \rho \, u \, w \\ u \big(\frac{1}{2}\rho\|\vec{u}\|^2 + \frac{\gamma p}{\gamma - 1}\big) \\ 0 \\ 0 \\ 0 \\ 0 \end{bmatrix}
+
\begin{bmatrix} 0 \\ \frac{1}{2}\|\vec{B}\|^2 -B_1 B_1 \\ - B_1  B_2 \\ - B_1  B_3 \\[.4em] u \|\vec{B}\|^2 - B_1 \big(\vec{u} \cdot \vec{B}\big) \\ 0 \\ u\, B_2 - v \, B_1 \\ u \, B_3 - w\, B_1 \\ 0 \end{bmatrix}
+ c_h
\begin{bmatrix} 0 \\ 0 \\ 0 \\ 0 \\ \psi B_1 \\ \psi \\ 0 \\ 0 \\ B_1 \end{bmatrix}.
\end{equation}

We contract the ideal GLM-MHD system \eqref{eq:3DIDEALGLMMHD} into entropy space using the entropy variables \eqref{eq:EntropyVars} and, for convenience, multiply by $\frac{1}{2\beta}$. The non-conservative term, $\vec{\Upsilon}$, is defined in \eqref{eq:GodunovSource}. Looking at the individual components of the flux one after another we obtain
\begin{align}
	\frac{1}{2\beta}\vec{v} \cdot \pderivative{}{x}\vec{f}^{x,\mathrm{hydro}} &= \left(\frac{1}{2\beta} \frac{\gamma - s}{\gamma - 1}-\frac{1}{2} \lVert\vec{u}\rVert^2 \right) (\rho u)_x + u (\rho u^2 + p) + v (\rho u v)_x + w (\rho u w)_x \ - \notag \\
	&\qquad \left( u\left( \frac{\gamma p}{\gamma - 1} + \frac{1}{2} \rho \|\vec{u}\|^2\right) \right)_x
	= \dots = -\frac{1}{2\beta} \bigg(\frac{\rho u s}{\gamma-1}\bigg)_x = \frac{1}{2 \beta} (uS)_x,
\end{align}
\begin{align}
	\frac{1}{2\beta}\vec{v} \cdot \bigg(\pderivative{}{x}\vec{f}^{x,\mathrm{magnetic}} + \vec{\Upsilon}^{x,\mathrm{magnetic}}\bigg) &= u \bigg(\frac{1}{2} \|\vec{B}\|^2 - B_1 B_1\bigg)_x - v \big(B_1 B_2\big)_x - w \big(B_1 B_3\big)_x \, - \notag\\
	&\qquad\Big(u \| \vec{B}^2 \| - B_1 (\vec{u}\cdot\vec{B})\Big)_x
	+ B_2 (u B_2 - v B_1)_x + B_3 (u B_3 - w B_1)_x \ - \notag\\
	&\qquad u (B_1)_x B_1 + v (B_1)_x B_2 + w (B_1)_x B_3 \ - \notag\\
	&\qquad ((B_1)_x u B_1 + (B_1)_x v B_2 + (B_1)_x w B_3) \ + \notag\\
	&\qquad B_1 (B_1)_x u + B_2 (B_1)_x v + B_3 (B_1)_x w = \dots = 0,
\end{align}
\begin{align}
	\frac{1}{2\beta}\vec{v} \cdot \bigg(\pderivative{}{x}\vec{f}^{x,\psi} + \vec{\Upsilon}^{x,\psi}\bigg) &= -c_h (\psi B_1)_x - u\psi(\psi)_x + c_h B_1 (\psi)_x + c_h \psi (B_1)_x + \psi u(\psi)_x \notag\\
	&= c_h \left[- (B_1 \psi)_x + B_1 (\psi)_x + \psi (B_1)_x \right] + u \psi (\psi)_x - u\psi(\psi)_x = 0
\end{align}
where we introduce the abbreviated notation $(\cdot)_x = \pderivative{(\cdot)}{x}$.
This gives an overall contribution of
\begin{equation}
	\vec{v} \cdot\bigg(\pderivative{}{x}\vec{f}^x + \vec{\Upsilon}^x\bigg) = \vec{v} \cdot \bigg(\pderivative{}{x}\vec{f}^{x,\mathrm{hydro}} + \pderivative{}{x}\vec{f}^{x,\mathrm{magnetic}} + \pderivative{}{x}\vec{f}^{x,\psi} + \vec{\Upsilon}^{x,\mathrm{magnetic}} + \vec{\Upsilon}^{x,\psi}\bigg) = (uS)_x.
\end{equation}
From the definition of the entropy variables, $\vec{v} \cdot \vec{q}_t = S_t$, we immediately obtain the entropy conservation law in one spatial dimension,
\begin{equation}
	\vec{v} \cdot\bigg(\vec{q}_t + \pderivative{}{x}\vec{f}^x + \vec{\Upsilon}^x\bigg) = S_t + (uS)_x = 0.
\end{equation}
We perform the same computations for the remaining two spatial dimensions
\begin{align}
	\vec{v} \cdot\bigg(\pderivative{}{y}\vec{f}^y + \vec{\Upsilon}^y\bigg) = (uS)_y &= 0,\\
	\vec{v} \cdot\bigg(\pderivative{}{z}\vec{f}^z + \vec{\Upsilon}^z\bigg) = (uS)_z &= 0,
\end{align}
and find the entropy balance law in three dimensions
\begin{equation}
	\vec{v} \cdot (\nabla \cdot \vec{f} + \vec{\Upsilon}) = \nabla \cdot (\vec{u}S)
	\quad\Longrightarrow\quad
	\vec{v} \cdot (\vec{q}_t + \nabla \cdot \vec{f} + \vec{\Upsilon}) = S_t + \nabla \cdot (\vec{u}S) = 0.\label{eq:conanalysis3D}
\end{equation}
This equation is identical to the continuous entropy conservation law \eqref{eq:EntropyEquality}. Therefore, the new ideal GLM-MHD system is suitable for building an entropy conserving numerical scheme. Note that when omitting the non-conservative terms, entropy consistency is lost for the ideal MHD equations \cite{Barth1999,Winters2016}.

In the remainder of this section, we compute the entropy balance equation for GLM-modified ideal MHD systems which have been presented in the literature to highlight that our system is the first presented consistent ideal MHD system with GLM divergence treatment that is fully compatible with thermodynamics.

\subsubsection{Continuous entropy analysis of Dedner et al.'s~ansatz}\label{Sec:Dedner}
Dedner et al.~\cite{Dedner2002} presented the first GLM modified ideal MHD system. Their hyperbolic and conservative modification of the ideal MHD equations (their eq.~(25)) reads
\begin{equation}\label{eq:Dedner}
	\pderivative{}{t}\, \vec{q} + \nabla\cdot\vec{f}_\mathrm{D} =
	\pderivative{}{t}\begin{bmatrix} \rho \\ \rho\vec{u} \\ E \\ \vec{B}\\ \psi \end{bmatrix}+
	\nabla\cdot\begin{bmatrix} \rho\vec{u} \\
	\rho(\vec{u}\otimes\vec{u}) + \big(p+\frac{1}{2}\|\vec{B}\|^2\big)\mat{I}-\vec{B}\otimes\vec{B} \\
	\vec{u}\big(E + p + \frac{1}{2}\|\vec{B}\|^2 \big) - \vec{B}(\vec{u}\cdot\vec{B}) \\
	\vec{u}\otimes\vec{B} - \vec{B}\otimes\vec{u} + \psi\mat{I} \\
	c_h^2 \vec{B}
	\end{bmatrix}=
	-\frac{c_h^2}{c_p^2}\begin{bmatrix} 0 \\ \vec{0} \\ 0 \\ \vec{0}\\ \psi \end{bmatrix}
	= -\vec{\Upsilon}_\mathrm{D},
\end{equation}
with
\begin{equation}
	p = (\gamma-1)\left(E - \frac{1}{2}\rho\lVert\vec{u}\rVert^2-\frac{1}{2}\lVert\vec{B}\rVert^2 \right).
\end{equation}
The entropy variables for this system are
\begin{equation}
	\vec{v}_\mathrm{D} = \pderivative{S}{\vec{q}} = \left[\frac{\gamma - s}{\gamma - 1}-\beta \lVert\vec{u}\rVert^2,\, 2\beta u,\, 2\beta v,\, 2\beta w,\, -2\beta,\, 2\beta B_1,\, 2\beta B_2,\, 2\beta B_3,\, 0 \right]^\intercal.
\end{equation}
Repeating the computations presented above for Dedner et al.'s equations, we obtain
\begin{equation}\label{eq:DednerEnt}
	\vec{v}_\mathrm{D} \cdot (\nabla \cdot \vec{f}_{\mathrm{D}} + \vec{\Upsilon}_{\mathrm{D}}) =
	\nabla\cdot (\vec{u}S) + 2\beta \big[ -(\vec{u} \cdot \vec{B}) \cdot (\nabla \cdot \vec{B}) + \vec{B} \cdot \nabla \psi \big] \ne \nabla\cdot (\vec{u}S),
\end{equation}
which is not conformable with the continuous entropy conservation law \eqref{eq:EntropyEquality} and, as such, it cannot be used to construct an entropy conserving scheme. At first glance it seems like we could still fulfill \eqref{eq:EntropyInequality} to construct an entropy stable scheme. However, this is not possible either, because we cannot guarantee the correct sign of the term $(\vec{u} \cdot \vec{B}) \cdot (\nabla \cdot \vec{B})  - \vec{B} \cdot \nabla \psi$.

It is well known that a non-conservative term that is proportional to the magnetic field divergence is necessary for entropy consistency, \eg \cite{Barth1999,Winters2016}. However, when we repeat the computation including the non-conservative terms we found in this work, we cancel only one of the extraneous terms from \eqref{eq:DednerEnt}. Even when assuming $c_p \rightarrow \infty$, one further term remains
\begin{equation}
	\vec{v}_\mathrm{D} \cdot (\nabla \cdot \vec{f}_{\mathrm{D}} + \vec{\Upsilon}_{\mathrm{D}} + \vec{\Upsilon}) =
	\nabla\cdot (\vec{u}S) + 2\beta \vec{B} \cdot \nabla \psi \ne \nabla\cdot (\vec{u}S),
\end{equation}
which is still not conformable with the continuous entropy conservation law or stability as we cannot predict the sign of the product $\vec{B} \cdot \nabla \psi$.

Note that due to the zero value in the ninth component of the entropy variables in the scheme of Dedner et al., \eqref{eq:DednerEnt}, the mapping between physical and entropy space is not bijective. As such, a one-to-one correspondence between conservative and entropy space does not exist. Hence, the equations \eqref{eq:Dedner} are not suitable for constructing an entropy stable scheme \cite[Section 2]{Tadmor1984}.

\subsubsection{Continuous entropy analysis of Dedner et al.'s~ansatz (extended version)}
In the same work, Dedner et al.~also presented an \emph{extended} GLM system, which involves additional non-conservative terms. They call this system (their eq.~(24)) the \emph{extended} GLM (EGLM) formulation of the MHD equations which has been adapted in many other works, \eg \cite{Domingues2013,Mignone2010,Jiang2012}. This extended system is given by
\begin{equation}\label{eq:DednerEGLM}
	\pderivative{}{t}\, \vec{q} + \nabla\cdot\vec{f}_\mathrm{D} =
	\pderivative{}{t}\begin{bmatrix} \rho \\ \rho\vec{u} \\ E \\ \vec{B}\\ \psi \end{bmatrix}+
	\nabla\cdot\begin{bmatrix} \rho\vec{u} \\
	\rho(\vec{u}\otimes\vec{u}) + \big(p+\frac{1}{2}\|\vec{B}\|^2\big)\mat{I}-\vec{B}\otimes\vec{B} \\
	\vec{u}\big(E + p + \frac{1}{2}\|\vec{B}\|^2 \big) - \vec{B}(\vec{u}\cdot\vec{B}) \\
	\vec{u}\otimes\vec{B} - \vec{B}\otimes\vec{u} + \psi\mat{I} \\
	c_h^2 \vec{B}
	\end{bmatrix}=
	\begin{bmatrix} 0 \\ -(\nabla\cdot \vec{B})\vec{B} \\ -\vec{B} \cdot \nabla\psi \\ \vec{0}\\ -\frac{c_h^2}{c_p^2}\psi \end{bmatrix} = - \vec{\Upsilon}_{\mathrm{EGLM}}.
\end{equation}
Doing the entropy contraction for Dedner et al.'s EGLM equations, we find
\begin{equation}
	\vec{v}_\mathrm{D} \cdot (\nabla \cdot \vec{f}_{\mathrm{D}} + \vec{\Upsilon}_{\mathrm{EGLM}}) = \nabla\cdot (\vec{u}S),
\end{equation}
which is in agreement with the continuous entropy conservation law \eqref{eq:EntropyEquality}. {However, the zero in the ninth entropy variable makes the construction of an entropy conservative scheme impossible.} Furthermore, it is in conflict with the general term we found when deriving the ideal MHD equations in Sec.~\ref{Sec:idealMHDdivB}. The non-conservative terms in the total energy contribution is not proportional to the magnetic field divergence and hence may be of significant magnitude.
Nevertheless, our finding underlines Dedner et al.'s observation that their EGLM scheme has superior robustness properties, since it is in agreement with thermodynamics.

\subsubsection{Continuous entropy analysis of Dedner et al.'s~ansatz (extended version, Galilean invariant)}
In the same work, Dedner et al.~presented a third scheme which is a variant of his \emph{extended} GLM system (their eq.~(38)) that includes the eight-wave formulation to achieve Galilean invariance. We will call this system Galilean invariant extended GLM (GI-EGLM) for convenience. It is given by
\begin{equation}\label{eq:DednerEGLMGalileo}
	\pderivative{}{t}\, \vec{q} + \nabla\cdot\vec{f}_\mathrm{D} =
	\pderivative{}{t}\begin{bmatrix} \rho \\ \rho\vec{u} \\ E \\ \vec{B}\\ \psi \end{bmatrix}+
	\nabla\cdot\begin{bmatrix} \rho\vec{u} \\
	\rho(\vec{u}\otimes\vec{u}) + \big(p+\frac{1}{2}\|\vec{B}\|^2\big)\mat{I}-\vec{B}\otimes\vec{B} \\
	\vec{u}\big(E + p + \frac{1}{2}\|\vec{B}\|^2 \big) - \vec{B}(\vec{u}\cdot\vec{B}) \\
	\vec{u}\otimes\vec{B} - \vec{B}\otimes\vec{u} + \psi\mat{I} \\
	c_h^2 \vec{B}
	\end{bmatrix}= - \vec{\Upsilon}_{\mathrm{GI\text{-}EGLM}},
\end{equation}
with
\begin{equation}
	\vec{\Upsilon}_{\mathrm{GI\text{-}EGLM}} := \begin{bmatrix} 0 \\ (\nabla\cdot \vec{B})\vec{B} \\ (\vec{u}\cdot\vec{B})(\nabla\cdot\vec{B}) + \vec{B} \cdot \nabla\psi \\ \vec{u}(\nabla \cdot \vec{B})\\ \vec{u}\cdot\nabla \psi + \frac{c_h^2}{c_p^2}\psi \end{bmatrix}.
\end{equation}
Doing the entropy contraction for Dedner et al.'s GI-EGLM equations, we find
\begin{equation}
	\vec{v}_\mathrm{D} \cdot (\nabla \cdot \vec{f}_{\mathrm{D}} + \vec{\Upsilon}_{\mathrm{GI\text{-}EGLM}}) = \nabla\cdot (\vec{u}S).
\end{equation}
We have again obtained the continuous entropy conservation law \eqref{eq:EntropyEquality}, however, the same limitations as with the EGLM terms (preceding section) apply. These equations seem to be the most preferable of the equations Dedner et al.~presented as they are both in agreement with the continuous entropy conservation law as also contain the non-conservative we found to be necessary in Section \ref{Sec:idealMHDdivB} for $\nabla\cdot\vec{B}\ne0$. Nevertheless, they are not suitable for constructing entropy stable numerical schemes as the mapping between physical and entropy space, given by \eqref{eq:DednerEnt}, is not bijective.

\subsubsection{Continuous entropy analysis of Mackey and Lim's~ansatz}
Mackey and Lim~\cite{Mackey} presented a version of Dedner et al.'s ansatz with improved performance. They modified the total energy flux to be
\begin{equation}
	\vec{f}_{\mathrm{ML}} = \vec{f}_{\mathrm{D}} +
	\begin{bmatrix}
		0 \\
		\vec{0} \\
		\begin{bmatrix}
			B_1\, f^{B_1}_{\mathrm{D}} \\
			B_2\, f^{B_2}_{\mathrm{D}} \\
			B_3\, f^{B_3}_{\mathrm{D}} \\
		\end{bmatrix} \\
		\vec{0} \\
		0 \\
	\end{bmatrix},
\end{equation}
to address gas pressure dips which appear ahead of oblique shocks in axisymmetric models of magnetized jets. The entropy variables remain unchanged, $\vec{v}_\mathrm{ML} = \vec{v}_\mathrm{D}$, and we obtain
\begin{equation}
	\vec{v}_\mathrm{ML} \cdot (\nabla \cdot \vec{f}_{\mathrm{ML}}) =
	\nabla\cdot (\vec{u}S) - 2\beta \big[ (\vec{u} \cdot \vec{B}) \cdot (\nabla \cdot \vec{B}) + \vec{B} \cdot (\nabla \psi) + \nabla \cdot (\psi \vec{B})\big] \ne \nabla\cdot (\vec{u}S),
\end{equation}
which is not conformable with the continuous entropy inequality (\ref{eq:EntropyEqualityies}a,b) either.

We repeat the computation including the non-conservative terms we found and obtain
\begin{equation}
	\vec{v}_\mathrm{ML} \cdot (\nabla \cdot \vec{f}_{\mathrm{ML}} + \vec{\Upsilon}) =
	\nabla\cdot (\vec{u}S) - 2\beta \big[ \vec{B} \cdot (\nabla \psi) + \nabla \cdot (\psi \vec{B})\big] \ne \nabla\cdot (\vec{u}S),
\end{equation}
which is still not conformable with the continuous entropy conservation law as we cannot predict the sign of neither $\vec{B} \cdot (\nabla \psi)$ nor $\nabla \cdot (\psi \vec{B})$.

\subsubsection{Continuous entropy analysis of Tricco and Price's~ansatz}
Tricco and Price~\cite{Tricco2012,Tricco2016} presented a constrained formulation of Dedner et al.'s hyperbolic divergence cleaning for SPH. The constraint they impose is that magnetic energy modified due to the cleaning process must be balanced by a new kind of correction energy which is correlated to $\psi$. We repeat their derivations with our definition of \eqref{eq:Epsi} and verified the same behavior for our scheme.
Their modified form of the ideal MHD equations for SPH reads (their eq.~(5) and (6) in \cite{Tricco2016})
\begin{equation}
	\pderivative{}{t}\, \vec{q} + \nabla\cdot\vec{f}_\mathrm{TP} =
	\pderivative{}{t}\begin{bmatrix} \rho \\ \rho\vec{u} \\ E \\ \vec{B}\\ \psi \end{bmatrix}
	+
	\nabla\cdot\begin{bmatrix} \rho\vec{u} \\
	\rho(\vec{u}\otimes\vec{u}) + \big(p+\frac{1}{2}\|\vec{B}\|^2\big)\mat{I}-\vec{B}\otimes\vec{B} \\
	\vec{u}\big(\frac{1}{2}\rho\|\vec{u}\|^2 + \frac{\gamma p}{\gamma - 1} + \|\vec{B}\|^2 \big) - \vec{B}(\vec{u}\cdot\vec{B}) \\
	\vec{u}\otimes\vec{B} - \vec{B}\otimes\vec{u} + \psi\mat{I} \\
	c_h^2 \vec{B}
	\end{bmatrix}
	=  - \vec{\Upsilon}_{\mathrm{TP}},
\end{equation}
with
\begin{equation}
	\vec{\Upsilon}_{\mathrm{TP}} := \begin{bmatrix} 0 \\ \vec{0} \\ 0 \\ \vec{u}(\nabla \cdot \vec{B})\\ \frac{1}{2}\psi(\nabla\cdot\vec{u}) + \vec{u}\cdot\nabla \psi \end{bmatrix},
\end{equation}
and
\begin{equation}
	p = (\gamma-1)\left(E - \frac{1}{2}\rho\lVert\vec{u}\rVert^2-\frac{1}{2}\lVert\vec{B}\rVert^2 - \frac{1}{2} \frac{\psi^2}{c_h^2}\right).
\end{equation}
The entropy variables for this system are
\begin{equation}
	\vec{v}_\mathrm{TP} = \pderivative{S}{\vec{q}} = \left[\frac{\gamma - s}{\gamma - 1}-\beta \lVert\vec{u}\rVert^2,\, 2\beta u,\, 2\beta v,\, 2\beta w,\, -2\beta,\, 2\beta B_1,\, 2\beta B_2,\, 2\beta B_3,\, 2 \beta c_h^{-2} \psi \right]^\intercal.
\end{equation}
We find
\begin{equation}\label{eq:TPentropy}
	\vec{v}_\mathrm{TP}\cdot\big(\nabla \cdot \vec{f}_{\mathrm{TP}} +\vec{\Upsilon}_{\mathrm{TP}}\big) =
	\nabla\cdot (\vec{u}S) + \frac{\beta}{c_h^2} \nabla \cdot (\psi^2 \vec{u}) \ne \nabla\cdot (\vec{u}S),
\end{equation}
which is inconsistent with the entropy conservation law such that - although the modified ideal MHD system of Tricco and Price are valid in the framework of SPH - it can neither be used to construct an entropy conserving nor an entropy stable scheme for a FV scheme {as we can not predict the sign of the term $\nabla\cdot (\psi^2 \vec{u})$.
Repeating the computation including the non-conservative terms we found earlier also in the momentum and total energy equations does not change the result \eqref{eq:TPentropy} as the two additional terms cancel in entropy space.}

To summarize, we demonstrated that it is important to account for energy transfers between the magnetic field and a correcting field used to control numerical errors in the divergence-free constraint. This lead to a modification of the total energy equation (as well as the induction equations) that ensures the model remains in agreement with the second law of thermodynamics. Additionally, we showed that the entropy analysis of other proposed GLM-type hyperbolic divergence methods from the literature are incompatible to build numerical approximations that discretely satisfy entropy conservation. 

For implementing our new mathematical model as an algorithm usable for computer simulations, it has to be \emph{discretized}. We will see that although the divergence diminishing property of the ideal GLM-MHD system rather trivially extends from continuous into discrete space, we have to pay special attention to the transfer of the entropy consistency property into a discrete numerical algorithm.

\subsection{Derivation of an entropy conserving numerical scheme}\label{Sec:EC}
In this section we describe the derivation of an entropy conserving approximation for the ideal GLM-MHD equations \eqref{eq:3DIDEALGLMMHD}. We drop the superscripts $(\cdot)^x$ for convenience as the following derivation is concerned with the derivation of the numerical scheme in $x-$direction only. Note that we do this without loss of generality. As shown in \cite[Sec.~3.1]{Derigs2016} and \cite[Appendix~A]{Winters2016}, the flux derivations easily extend to higher spatial dimensions. The derivations shown herein are closely related to the derivations done by Winters and Gassner \cite{Winters2016} and Chandrashekar and Klingenberg \cite{Chandrashekar2016} for entropy stable fluxes.

\subsubsection{Discrete entropy conservation}\label{Sec:discreteEC}
When we contract the ideal GLM-MHD equations with the entropy variables, we obtain the entropy conservation law with additional terms proportional to the magnetic field divergence,
\begin{equation}
	\pderivative{}{t}S + \nabla\cdot\vec{F} + 2 \beta (\vec{u}\cdot\vec{B})(\nabla\cdot\vec{B}) = 0.
\end{equation}

Hence, to ease the following derivations, we assume that we can rewrite the non-conservative terms using a homogeneous function of degree one, with respect to the entropy variables, in the form \cite{Barth1999,Chandrashekar2016}
\begin{equation}\label{eq:phi}
	\phi (\vec{v}) := \vec{v} \cdot \pderivative{\phi}{\vec{v}} = 2\beta (\vec{u}\cdot\vec{B}).
\end{equation}
A suitable candidate function is
\begin{equation}
	\phi(\vec{v}) = \frac{v_2 v_6 + v_3 v_7 + v_4 v_8}{v_5},
\end{equation}
where differentiating with respect to the entropy variables,
\begin{equation}
\pderivative{\phi}{\vec{v}} = \begin{bmatrix}0 & \vec{B} & \vec{u}\cdot\vec{B} & \vec{u} & 0\end{bmatrix}^\intercal =: \vec{\Phi}
\end{equation}
reveals the vector components of the non-conservative term \eqref{eq:GodunovSource}, which is now connected through
\begin{equation}
\vec{\Upsilon} = (\nabla \cdot \vec{B})\vec{\Phi} = (\nabla \cdot \vec{B})\pderivative{\phi}{\vec{v}}.
\end{equation}
Hence, in the following, we consider the one dimensional PDE system in the form
\begin{equation}
	\pderivative{}{t}\vec{q} + \pderivative{}{x}\vec{f} + \vec{\Phi}(\vec{v})(\nabla\cdot\vec{B})=\vec{0},
\end{equation}
that is identical to \eqref{eq:1DIDEALGLMMHD} but contains the order one homogeneity condition \eqref{eq:phi}.

We assume left and right cell-averages, denoted by L and R, with cell sizes $\Delta x_\L$ and $\Delta x_\R$ separated by a common interface. We discretize the one-dimensional ideal GLM-MHD equations \eqref{eq:1DIDEALGLMMHD} semi-discretely and derive an approximation for the fluxes at the interface in between the two adjacent cells (the $i+1/2$ interface):
\begin{equation}\label{eq:semidiscrete}
	\Delta x_\L \pderivative{}{t} \vec{q}_\L = \vec{f}_\L - \vec{f}^* - \frac{\Delta x_\L}{2} \frac{{\jump{B_1}}}{\Delta x_\mathrm{L}} \vec{\Phi}_\mathrm{L}, \qquad \mbox{and}\qquad \Delta x_\R \pderivative{}{t} \vec{q}_\R = \vec{f}^* - \vec{f}_\R - \frac{\Delta x_\R}{2} \frac{{\jump{B_1}}}{\Delta x_\mathrm{R}} \vec{\Phi}_\mathrm{R},
\end{equation}
where the adjacent states L and R are separated by a numerical interface flux $\vec{f}^*$. We define the jump in a quantity as $\jump{\cdot} := (\cdot)_\R - (\cdot)_\L$. Note that both cells are also affected by the physical fluxes $\vec{f}_{\L,\R}$.

Next, we convert \eqref{eq:semidiscrete} from physical to entropy space to get the semi-discrete entropy update in each cell
\refstepcounter{equation}
\begin{equation}\label{eq:semidiscreteentropy}
	\Delta x_\L \pderivative{}{t} S_\L = \vec{v}_\L \cdot \left(\vec{f}_\L - \vec{f}^* - \frac{\jump{B_1}}{2} \vec{\Phi}_\mathrm{L}\right) \qquad \mbox{and}\qquad \Delta x_\R \pderivative{}{t} S_\R = \vec{v}_\R \cdot \left(\vec{f}^* - \vec{f}_\R - \frac{\jump{B_1}}{2} \vec{\Phi}_\mathrm{R} \right), \tag{\theequation a,b}
\end{equation}
where we again use that $S_t = \vec{v}\cdot \vec{q}_t$.

By combining \eqref{eq:semidiscreteentropy}, setting $\Delta x_\L = \Delta x_\R =: \Delta x$ and using the homogeneity condition $\vec{v}\cdot\vec{\Phi} = \phi$ (see \eqref{eq:phi}), we obtain the total entropy update
\begin{equation}\label{eq:discreteEntropyEq}
	\Delta x \pderivative{}{t} \big(S_\L + S_\R) = \jump{\vec{v}} \cdot \vec{f}^* - \jump{\vec{v}\cdot\vec{f}} - \avg{\phi}\jump{B_1},
\end{equation}
where the average of a state is defined as $\avg{\cdot} := ((\cdot)_\L + (\cdot)_\R)/2$. When applied to vectors, the average and jump operators are evaluated separately for each vector component.

To have the finite volume update satisfy the discrete entropy conservation law, the entropy flux due to the finite volume flux must coincide with the discrete entropy flux $\vec{u}S$ from \eqref{eq:EntropyEquality}.
We define the entropy flux potential as \cite{Chandrashekar2016}
\begin{equation}\label{eq:GLMentropypotential}
	\Psi = \vec{v}\cdot\vec{f} - uS + \phi B_1 = \rho u + \beta u \lVert\vec{B}^2\rVert + 2\beta c_h B_1 \psi
\end{equation}
and rewrite \eqref{eq:discreteEntropyEq} using the linearity of the jump operator to obtain
\begin{equation} \label{eq:discreteentropy}
	\jump{\vec{v}}\cdot\vec{f}^* = \jump{\Psi} - \avg{B_1}\jump{\phi}
	= \jump{\rho u} + \jump{\beta u \lVert\vec{B}^2\rVert} + 2 c_h \jump{\beta\psi B_1} - 2\avg{B_1} \jump{\beta (\vec{u}\cdot\vec{B})},
\end{equation}
where we used that
\begin{equation}
	\avg{\phi}\jump{B_1} = \jump{\phi B_1} - \avg{B_1}\jump{\phi}.
\end{equation}

We denote \eqref{eq:discreteentropy} as the \emph{discrete entropy conservation condition} for the ideal GLM-MHD equations. Since this is a scalar equation, there are several possible solutions for the numerical flux vector $\vec{f}^*$. However, there is the additional requirement that the numerical flux must be consistent, \ie $\vec{f}^*(\vec{q},\vec{q}) = \vec{f}$ severely limiting the number of possible solutions.

With all necessary components collected, we solve \eqref{eq:discreteentropy} to obtain the new entropy conserving numerical flux. The full derivation is shown in \ref{Sec:ECderivation}. The numerical flux function reads
\begin{equation}\label{eq:ECfluxfunction}
	\vec{f}^\mathrm{KEPEC,GLM} =
\begin{bmatrix}
	\rholn \avg{u}\\
	\rholn \avg{u}^2  + \overline{p}_\mathrm{tot} - \avg{B_1}^2\\
	\rholn\avg{u} \avg{v} - \avg{B_1}\avg{B_2}\\
	\rholn\avg{u} \avg{w} - \avg{B_1}\avg{B_3}\\
	f_5^*\\
	c_h \avg{\psi}\\
	\avg{u}\avg{B_2} - \avg{v}\avg{B_1}\\
	\avg{u}\avg{B_3} - \avg{w}\avg{B_1}\\
	c_h \avg{B_1}
\end{bmatrix},
\end{equation}
with
\begin{align}
	\overline{p}_\mathrm{tot} &= \avgp + \frac{1}{2} \Big(\avg{B_1^2} + \avg{B_2^2} + \avg{B_3^2}\Big), \text{ and}\\
	f_5^* &= f_1^*\bigg[\frac{1}{2 (\gamma-1) \betaln} - \frac{1}{2} \left(\avg{u^2} + \avg{v^2} + \avg{w^2}\right) \bigg] + f_2^* \avg{u} + f_3^* \avg{v} + f_4^* \avg{w} + \notag\\
	&\quad + f_6^* \avg{B_1} + f_7^* \avg{B_2} + f_8^* \avg{B_3} + f_9^* \avg{\psi} - \frac{1}{2} \Big(\avg{u B_1^2}+\avg{u B_2^2}+\avg{u B_3^2}\Big) +\notag\\
	&\quad + \avg{B_1}\Big(\avg{u B_1}+\avg{v B_2}+\avg{w B_3}\Big) - c_h \avg{B_1 \psi},
\end{align}
the logarithmic mean $\avgln{(\cdot)} = \frac{\jump{\cdot}}{\jump{\ln(\cdot)}}$, and the specifically averaged pressure, $\avgp = \frac{\avg{\rho}}{2\avg{\beta}}$. A numerically stable procedure to compute the logarithmic mean is described by Ismail and Roe \cite[App. B]{IsmailRoe2009}.

We compute the magnetic field divergence in the discretized non-conservative term using central differencing:
\begin{equation}
	\vec{\Upsilon}_i^{x,\mathrm{magnetic}} :=
	\underbrace{\left(\frac{(B_1)_{i+1} - (B_1)_{i-1}}{2\Delta x}\right)}_\text{central derivative}
	\begin{bmatrix}
	0\\\vec{B}_i\\(\vec{u}\cdot\vec{B})_i\\
	\vec{u}_i\\
	0
	\end{bmatrix}=
	\frac{1}{2\Delta x}
	\left\{\!\!
	\left(\jump{B_1}
	\begin{bmatrix}
	0\\
	\vec{B}_\mathrm{L}\\
	(\vec{u}\cdot\vec{B})_\mathrm{L}\\
	\vec{u}_\mathrm{L}\\
	0
	\end{bmatrix}\right)_{\!\!i-1/2}\hspace*{-4mm}
	\!\!+
	\left(\jump{B_1}
	\begin{bmatrix}
	0\\
	\vec{B}_\mathrm{R}\\
	(\vec{u}\cdot\vec{B})_\mathrm{R}\\
	\vec{u}_\mathrm{R}\\
	0
	\end{bmatrix}\right)_{\!\!i+1/2}
	\right\}\label{eq:discreteSourceterm_mag}
\end{equation}
Similarly, we also find the $\psi$ correlated non-conservative term
\begin{equation}
	\vec{\Upsilon}_i^{x,\psi} :=
	\left(\frac{(\psi)_{i+1} - (\psi)_{i-1}}{2\Delta x}\right)
	\begin{bmatrix}
	0\\\vec{0}\\(u\psi)_i\\\vec{0}\\u_i
	\end{bmatrix}=
	\frac{1}{2\Delta x}
	\left\{\!\!
	\left(\jump{\psi}
	\begin{bmatrix}
	0\\
	0\\
	(u\psi)_\mathrm{L}\\
	0\\
	u_\mathrm{L}
	\end{bmatrix}\right)_{\!\!i-1/2}\hspace*{-4mm}
	\!\!+
	\left(\jump{\psi}
	\begin{bmatrix}
	0\\
	0\\
	(u\psi)_\mathrm{R}\\
	0\\
	u_\mathrm{R}
	\end{bmatrix}\right)_{\!\!i+1/2}
	\right\}\label{eq:discreteSourceterm_psi}
\end{equation}

The full discrete non-conservative term is simply given by the sum of the two terms presented above:
\begin{equation}
	\vec{\Upsilon}_i^x = \vec{\Upsilon}_i^{x,\mathrm{magnetic}} + \vec{\Upsilon}_i^{x,\psi}
	\label{eq:discreteSourceterm}
\end{equation}

We highlight that the newly derived numerical flux function \eqref{eq:ECfluxfunction} conserves the discrete entropy by construction. Furthermore, in the case of vanishing magnetic fields, the scheme is not only \emph{entropy conserving} (EC), but also \emph{kinetic energy preserving} (KEP) \cite{Chandrashekar2012}. As has been shown by the authors, the kinetic energy preserving property is favorable in terms of robustness of the scheme particularly at high Mach numbers \cite{Derigs2016_2}.

When investigating the consistency of the obtained numerical flux function \eqref{eq:ECfluxfunction} we assume that the left/right states are identical and find
\begin{align}
	\vec{f}^\mathrm{KEPEC,GLM} =
	\begin{bmatrix}
	\rho u\\
	\rho u^2 + p + \frac{1}{2}\|\vec{B}\|^2 - B_1^2 \\
	\rho u v - B_1 B_2\\
	\rho u w - B_1 B_3\\
	u \big(\frac{1}{2}\rho\|\vec{u}\|^2 + \frac{\gamma p}{\gamma - 1} + \|\vec{B}\|^2\big) - B_1 \big(\vec{u} \cdot \vec{B}\big) + c_h \psi B_1\\
	c_h \psi\\
	u B_2 - v B_1\\
	u B_3 - w B_1\\
	c_h B_1
	\end{bmatrix}
	= \vec{f}^x,
\end{align}
where we used that
\begin{equation}
	u (B_2^2 + B_3^2) - B_1 (v B_2 + w B_3) = u \lVert\vec{B}^2\rVert - B_1 (\vec{u}\cdot\vec{B}).
\end{equation}
Thus, we have shown that the newly derived entropy conservative numerical flux for the ideal GLM-MHD equations given by \eqref{eq:ECfluxfunction} is consistent with the physical flux, and, together with the discretization of the non-conservative terms \eqref{eq:discreteSourceterm}, conserves the discrete entropy by construction. We note that the non-conservative terms \eqref{eq:discreteSourceterm} vanish when the left/right states are identical, reflecting convergence to the continuous case where the divergence of the magnetic field should vanish.

We stress that the presented way of deriving the entropy conservative scheme is not unique. We prefer the numerical flux presented herein as it avoids problematic non-conservative term discretizations which is what we found in previous works \cite{Winters2017}.

\subsection{Derivation of an entropy stable numerical scheme}\label{Sec:ES}
The entropy of a closed system is only conserved if the solution remains smooth.
If additional dissipation is not included in an entropy conservative method, spurious oscillations will develop near discontinuities as energy is re-distributed at the smallest resolvable scale \citep{Mishra2011}.
Hence, entropy conserving schemes may suffer breakdown in the presence of discontinuities.
From the second law of thermodynamics we know that kinetic and/or magnetic energy can be transformed irreversibly into heat, which we denote as \emph{dissipation} (also known as ``thermalization'').
Accordingly, we require dissipation is added to the approximation such that discrete satisfaction of the entropy inequality \eqref{eq:EntropyInequality} is guaranteed.
A numerical scheme requires a diffusion operator to match such a physical process.

In order to create an entropy stable numerical flux function, we use the KEPEC flux \eqref{eq:ECfluxfunction} as a baseline flux and add a general form of numerical dissipation to compute a \emph{kinetic energy preserving} and \emph{entropy stable} (KEPES) numerical flux that is applicable to arbitrary flows
\begin{equation}\label{eq:KEPES_Dmatrix}
	\vec{f}^\mathrm{KEPES} = \vec{f}^\mathrm{KEPEC} - \frac{1}{2} \mat{D} \jump{\vec{q}},
\end{equation}
where $\mat{D}$ is a suitable dissipation matrix that is guaranteed to cause a negative contribution in \eqref{eq:EntropyInequality}.

\subsubsection{Scalar dissipation term (Lax-Friedrichs and Rusanov schemes)}
\renewcommand{\He}{\ensuremath{\bm{\mathcal{H}}}}
If we make the simple choice of $\mat{D}$ to be
\begin{equation}
	\mat{D}_\mathrm{LF} = |\lambda_{\rm max}^\mathrm{global}| \mat{I},
\end{equation}
where $\lambda_{\rm max}^\mathrm{global} = \max(\lambda_{\mathrm{max},i=1,\dots,N})$ is the largest eigenvalue of the ideal GLM-MHD system in the whole computational domain, we can rewrite the dissipation term
\begin{align}
	\frac{1}{2} \mat{D}_\mathrm{LF} \jump{\vec{q}} = \frac{1}{2}  |\lambda_{\rm max}^\mathrm{global}| \mat{I} \jump{\vec{q}} = \frac{1}{2} |\lambda_{\rm max}^\mathrm{global}| \H{} \jump{\vec{v}} \simeq \frac{1}{2} |\lambda_{\rm max}^\mathrm{global}| \He{} \jump{\vec{v}}, \label{eq:dissipation_term3}
\end{align}
where $\H{} = \pderivative{\vec{q}}{\vec{v}}$ is a matrix that relates the variables in conserved and entropy space. This choice for the dissipation term leads to a \emph{scalar dissipation term}, also called \emph{Lax-Friedrichs} (LF) type dissipation.
While the entropy Jacobian, \H{}, is easily found in continuous space, it was shown in \cite{Derigs2016_2} that it is highly non-trivial to discretize this matrix for use in a numerical scheme. The requirement is to average the quantities in such a way that the relation $\jump{\vec{q}} = \He{} \jump{\vec{v}}$ holds whenever possible.
The reformulation of the dissipation term to incorporate the jump in entropy variables (rather than the jump in conservative variables) is done to ensure entropy stability by guaranteeing a negative contribution in the entropy inequality \cite{Winters2016}.

The entries of the matrix $\He{}$ are derived step-by-step through the solution of 81 equations similar to what was done in \cite[Section 4]{Derigs2016_2} for the unmodified ideal MHD equations. Details are given in~\ref{app:entropyJacobian}. We summarize the symmetric $\He{}$ matrix:
\begin{equation}\label{eq:H}
\He = \resizebox{0.95\hsize}{!}{$ %
	\begin{bmatrix}
	\rholn & \rholn\avg{u} & \rholn\avg{v} & \rholn\avg{w} & \Eline & 0 & 0 & 0 & 0 \\
	\rholn\avg{u} & \rholn\avg{u}^2 + \avgp & \rholn\avg{u}\avg{v} & \rholn\avg{u}\avg{w} & \left(\Eline + \avgp \right) \avg{u} & 0 & 0 & 0 & 0 \\
	\rholn\avg{v} & \rholn\avg{v}\avg{u} & \rholn\avg{v}^2 + \avgp & \rholn\avg{v}\avg{w} & \left(\Eline + \avgp \right) \avg{v} & 0 & 0 & 0 & 0 \\
	\rholn\avg{w} & \rholn\avg{w}\avg{u} & \rholn\avg{w}\avg{v} & \rholn\avg{w}^2 + \avgp & \left(\Eline + \avgp \right) \avg{w} & 0 & 0 & 0 & 0 \\
	\Eline & \left(\Eline + \avgp \right) \avg{u} & \left(\Eline + \avgp \right) \avg{v} & \left(\Eline + \avgp \right) \avg{w} & \He_{5,5} & \tau \avg{B_1} & \tau \avg{B_2} & \tau \avg{B_3} & \tau\avg{\psi} \\
	0 & 0 & 0 & 0 & \tau \avg{B_1} & \tau & 0 & 0 & 0 \\
	0 & 0 & 0 & 0 & \tau \avg{B_2} & 0 & \tau & 0 & 0 \\
	0 & 0 & 0 & 0 & \tau \avg{B_3} & 0 & 0 & \tau & 0 \\
	0 & 0 & 0 & 0 & \tau\avg{\psi} & 0 & 0 & 0 & \tau \\
	\end{bmatrix},
	$}
\end{equation}\vspace*{-1\baselineskip}
\begin{equation}
	\overline{E} = \frac{1}{2}\rholn\uavg + \frac{\pln}{\gamma-1}, \quad \tau = \frac{\avg{p}}{\avg{\rho}} = \frac{1}{2 \avg{\beta}}, \quad\mbox{and}
\end{equation}
\begin{equation}
	\He_{5,5} = \frac{1}{\rholn}\bigg(\frac{(\pln)^2}{\gamma-1} + {\Eline^2}\bigg) + \avgp \big(\avg{u}^2 + \avg{v}^2 + \avg{w}^2\big) + \tau \sum_{i=1}^{3} \avg{B_i}^2 + \tau\avg{\psi}^2.
\end{equation}
With the particular averaging of the matrix $\He$ it can be shown that
\begin{equation}\label{eq:almostEquality}
	(\jump{\vec{q}})_i = (\He{}\jump{\vec{v}})_i,\quad i = 1,2,3,4,6,7,8,9\quad\textrm{and}\quad (\jump{\vec{q}})_5 \simeq (\He{}\jump{\vec{v}})_5.
\end{equation}
So, the equality holds for each term except for the jump in total energy. The relation that the jump in total energy only holds asymptotically was necessary to create a discrete dissipation operator that is still symmetric \cite[Sec.~4]{Derigs2016_2}. 
It is straightforward, using Sylvester's criterion, to verify that the discrete matrix \eqref{eq:H} is symmetric positive definite (SPD) \cite[Appendix A]{Derigs2016_2}.

Due to the structure of the dissipation term \eqref{eq:dissipation_term3}, the SPD property of the discrete matrix guarantees that the numerical flux
\begin{equation}\label{eq:KEPES_LF}
	\vec{f}^\mathrm{KEPES,LF} = \vec{f}^\mathrm{KEPEC} - \frac{1}{2} |\lambda_{\rm max}^\mathrm{global}| \He{} \jump{\vec{v}},
\end{equation}
complies with the entropy inequality \eqref{eq:EntropyInequality} on the discrete level.
Although the LF flux is quite dissipative (especially for slow waves), it has the advantage that it very numerically stable, non-oscillatory \cite{Winters2016}, and easy to implement.

A natural reason for the diffusivity of LF lies in its global nature. The wave speeds involved are the maximum allowed speeds in the computational domain and do not take into account the local wave speeds of the solution. Indeed, Rusanov \cite{Rusanov} showed that a less diffusive, yet stable scheme can be built using a local, instead of a global, wave speed measure. The resulting dissipation term is the \emph{Rusanov} or \emph{local Lax-Friedrichs} (LLF) stabilization term:
\begin{equation}\label{eq:KEPES_LLF}
	\vec{f}^\mathrm{KEPES,LLF} = \vec{f}^\mathrm{KEPEC} - \frac{1}{2} |\lambda_{\rm max}^\mathrm{local}| \He{} \jump{\vec{v}} \qquad\mbox{with}\qquad \lambda_{\rm max}^\mathrm{local} = \max(\lambda_{\mathrm{max},\R},\lambda_{\mathrm{max},\L}).
\end{equation}

\subsubsection{Matrix dissipation term}

We create a less diffusive operator than LF or LLF if we select the dissipation matrix in \eqref{eq:KEPES_Dmatrix} to be
\begin{equation}\label{eq:Roe}
	\mat{D}_{\rm{MD}} = \Ree{}|\boldsymbol{\Lambda}|\Ree{}^{-1},
\end{equation}
where $\Ree{}$ is the matrix of right eigenvectors and $\boldsymbol\Lambda$ is the diagonal matrix of the eigenvalues of the flux Jacobian for the ideal MHD system. Here we focus on the particular mean states at which the matrix $\Ree$ is evaluated. For entropy stable schemes there exists a particular scaling of the eigenvectors that relates the matrix $\Ree{}$ to the entropy Jacobian $\He$ \cite{Barth1999} such that
\begin{equation}\label{eq:scaling}
	\He = \Ree\Tee\Ree^\intercal.
\end{equation}
The derivation and entropy scaling of the eigenvectors is provided in \ref{sec:eigenstructure}. 
From this scaling we rewrite the dissipation term
\begin{equation}\label{eq:matDiss}
	\begin{aligned}
	\frac{1}{2}\mat{D}_{\rm{MD}}\jump{\vec{q}} &= \frac{1}{2}\Ree|\bm\Lambda|\Ree^{-1}\jump{\vec{q}} \simeq \frac{1}{2}\Ree|\bm\Lambda|\Ree^{-1}\He\jump{\vec{v}}\textrm{\quad (in the sense of \eqref{eq:almostEquality})}\\
	&= \frac{1}{2}\Ree|\bm\Lambda|\Ree^{-1}\Ree\Tee\Ree^\intercal\jump{\vec{v}} = \frac{1}{2}\Ree|\bm\Lambda|\Tee\Ree^\intercal\jump{\vec{v}}.
	\end{aligned}
\end{equation}

We already know the particular averaging needed for the matrix $\He$ from \eqref{eq:H}. Next, we use these known averages and the condition \eqref{eq:scaling} to determine the mean state evaluations for $\Ree$ and $\Tee$. This creates a unique averaging procedure for the matrix dissipation term \eqref{eq:matDiss} while retaining the almost equal property \eqref{eq:almostEquality}.

It is straightforward, albeit laborious, to relate the entries of the matrix $\He$ and determine the 81 individual components of the matrices $\Ree$ and $\Tee$. An outline of the general technique and justification of the somewhat unconventional averaging strategies that result in the final form is provided in \cite{Winters2017}. We forgo the algebraic details and after many manipulations, present the unique averaging procedure for the discrete eigenvector and scaling matrices.

Due to the complicated structure of the eigenvectors the presentation of the final form is divided into three parts. First, we give the specific averages for several convenience variables
\begin{equation}\label{eq:alotofequations_discrete}
	\begin{aligned}
	\hat{\Psi}_{\pm\mathrm{s}} &= \frac{\hat{\alpha}_\mathrm{s} \rho^{\ln} \overline{\lVert\vec{u}\rVert^2}}{2} - a^{\beta} \hat{\alpha}_\mathrm{f} \rho^{\ln} \bar{b}_\perp + \frac{\hat{\alpha}_\mathrm{s} \rho^{\ln} (a^{\ln})^2}{\gamma-1} \pm \hat{\alpha}_\mathrm{s} \hat{c}_\mathrm{s} \rho^{\ln} \avg{u} \pm \hat{\alpha}_\mathrm{f} \hat{c}_\mathrm{f} \rho^{\ln} \sigma(\bar{b}_1) (\avg{v} \bar{\chi}_2 + \avg{w} \bar{\chi}_3), \\
	\hat{\Psi}_{\pm\mathrm{f}} &= \frac{\hat{\alpha}_\mathrm{f} \rho^{\ln} \overline{\lVert\vec{u}\rVert^2}}{2} + a^{\beta} \hat{\alpha}_\mathrm{s} \rho^{\ln} \bar{b}_\perp + \frac{\hat{\alpha}_\mathrm{f} \rho^{\ln} (a^{\ln})^2}{\gamma-1} \pm \hat{\alpha}_\mathrm{f} \hat{c}_\mathrm{f} \rho^{\ln} \avg{u} \mp \hat{\alpha}_\mathrm{s} \hat{c}_\mathrm{s} \rho^{\ln} \sigma(\bar{b}_1) (\avg{v} \bar{\chi}_2 + \avg{w} \bar{\chi}_3),  \\
	\hat{c}_\mathrm{a}^2& = \bar{b}_1^2, \quad \hat{c}_{\f,\s}^2 = \frac{1}{2}\left((\bar{a}^2+\bar{b}^2) \pm \sqrt{(\bar{a}^2+\bar{b}^2)^2 - 4\bar{a}^2 \bar{b}_1^2}\right),\quad
	\avgp = \frac{\avg{\rho}}{2\avg{\beta}}, \quad \bar{a}^2 = \gamma \frac{\avgp}{\rho^{\ln}}, \\ (a^{\ln})^2 &= \gamma\frac{p^{\ln}}{\rho^{\ln}},\quad (a^{\beta})^2 = \gamma\frac{1}{2\avg{\beta}},\quad
	\bar{b}^2 = \bar{b}_1^2 + \bar{b}_2^2 + \bar{b}_3^2, \quad \bar{b}_\perp^2 = \bar{b}_2^2 + \bar{b}_3^2, \quad \bar{\chi}_{2,3} = \frac{\bar{b}_{2,3}}{\bar{b}_\perp}, \\ \bar{b}^2_{1,2,3} &= \frac{\avg{B_{1,2,3}}^2}{\rho^{\ln}}, \quad \hat{\alpha}_\mathrm{f}^2 = \frac{\bar{a}^2 - \hat{c}_\mathrm{s}^2}{\hat{c}_\mathrm{f}^2 - \hat{c}_\mathrm{s}^2}, \quad \hat{\alpha}_\mathrm{s}^2 = \frac{\hat{c}_\mathrm{f}^2 -\bar{a}^2}{\hat{c}_\mathrm{f}^2 - \hat{c}_\mathrm{s}^2},\quad
	\sigma(\omega) = \begin{cases}
	+1 &\mbox{if } \omega \ge 0, \\
	-1 &\text{otherwise}
	\end{cases}.
	\end{aligned}
\end{equation}
Next, we give the average of the right eigenvector matrix
\begin{equation}\label{eq:rightEigen}
	\Ree = \left[\,\ree{}_{\rm +{f}} \,|\, \ree{}_{\rm +{a}} \,|\, \ree{}_{\rm +{s}} \,|\, \ree{}_{\rm +\psi} \,|\, \ree{}_{\rm E} \,|\, \ree{}_{\rm -\psi} \,|\, \ree{}_{\rm -{s}} \,|\, \ree{}_{\rm -{a}} \,|\, \ree{}_{\rm -{f}} \, \right].
\end{equation}
Where each of the discrete eigenvectors are
\begin{itemize}
	\item[] \underline{GLM Waves}: ${\lambda}_{  \pm \psi}$\hspace{20mm} \underline{Entropy Wave} : ${\lambda}_{\rm E}$\hspace{20mm} \underline{Alfv\'{e}n Waves}: ${\lambda}_{\pm \rm a}$
	\begin{equation}
	\ree_{\pm\psi} = \begin{bmatrix} 0 \\
	 0 \\
	 0 \\
	 0 \\
	\avg{B_1}\pm\avg{\psi} \\[0.05cm]
	 1 \\
	 0 \\
	 0 \\
	\pm 1\\
	 \end{bmatrix},\quad
	\ree_{\rm E}  = \begin{bmatrix} 1 \\ \avg{u} \\ \avg{v} \\ \avg{w} \\[0.1cm] \frac{1}{2}\uavg\\[0.05cm]0 \\0 \\0 \\0\end{bmatrix},\quad
	\ree_{\rm \pm a} = \begin{bmatrix}
	0 \\
	0 \\
	\pm \rho^{\ln}\sqrt{\avg{\rho}}\,\bar{\chi}_3 \\[0.1cm]
	\mp \rho^{\ln}\sqrt{\avg{\rho}}\,\bar{\chi}_2 \\[0.1cm]
	\mp \rho^{\ln}\sqrt{\avg{\rho}}(\bar{\chi}_2 \avg{w} - \bar{\chi}_3 \avg{v})\\[0.1cm]
	0 \\
	-\rho^{\ln} \bar{\chi}_3 \\[0.1cm]
	\rho^{\ln} \bar{\chi}_2  \\[0.1cm]
	0
	\end{bmatrix},\tag{\refstepcounter{equation}\theequation~/~\refstepcounter{equation}\theequation~/~\refstepcounter{equation}\theequation}
	\end{equation}
	\item[] \underline{Fast and Slow Magnetoacoustic Waves}:  ${\lambda}_{\rm \pm f,\pm s}$
	\begin{equation}
	\ree_{\rm \pm f} = \begin{bmatrix}
	\hat{\alpha}_{\rm f}\rho^{\ln} \\[0.1cm]
	\hat{\alpha}_{\rm f}\rho^{\ln}(\avg{u} \pm \hat{c}_{f}) \\[0.1cm]
	\rho^{\ln}\left(\hat{\alpha}_{\rm f} \avg{v} \mp \hat{\alpha}_{\rm s} \hat{c}_{\rm s} \bar{\chi}_2 \sigma(\bar{b}_1) \right) \\[0.1cm]
	\rho^{\ln}\left(\hat{\alpha}_{\rm f} \avg{w} \mp \hat{\alpha}_{\rm s} \hat{c}_{\rm s} \bar{\chi}_3 \sigma(\bar{b}_1) \right) \\[0.1cm]
	\hat{\Psi}_{\pm\rm f} \\[0.1cm]
	0 \\[0.1cm]
	\hat{\alpha}_{\rm s} a^{\beta} \bar{\chi}_2 \sqrt{\rho^{\ln}} \\[0.1cm]
	\hat{\alpha}_{\rm s} a^{\beta} \bar{\chi}_3 \sqrt{\rho^{\ln}} \\[0.1cm]
	0
	\end{bmatrix},
	\qquad
	\ree_{\rm \pm s} = \begin{bmatrix}
	\hat{\alpha}_{\rm s}\rho^{\ln} \\[0.1cm]
	\hat{\alpha}_{\rm s}\rho^{\ln}\left(\avg{u} \pm \hat{c}_{\rm s}\right) \\[0.1cm]
	\rho^{\ln}\left(\hat{\alpha}_{\rm s} \avg{v} \pm \hat{\alpha}_{\rm f} \hat{c}_{\rm f} \bar{\chi}_2 \sigma(\bar{b}_1)\right) \\[0.1cm]
	\rho^{\ln}\left(\hat{\alpha}_{\rm s} \avg{w} \pm \hat{\alpha}_{\rm f} \hat{c}_{\rm f} \bar{\chi}_3 \sigma(\bar{b}_1)\right) \\[0.1cm]
	\hat{\Psi}_{\pm\rm s} \\[0.1cm]
	0 \\[0.1cm]
	-\hat{\alpha}_{\rm f} a^{\beta} \bar{\chi}_2 \sqrt{\rho^{\ln}} \\[0.1cm]
	-\hat{\alpha}_{\rm f} a^{\beta} \bar{\chi}_3 \sqrt{\rho^{\ln}} \\[0.1cm]
	0
	\end{bmatrix}.\tag{\refstepcounter{equation}\theequation~/~\refstepcounter{equation}\theequation}
	\end{equation}
\end{itemize}
The mean state for the diagonal scaling matrix is
\begin{equation}
	\Tee = \diag\left(\frac{1}{2\gamma\rho^{\ln}},\frac{1}{4\avg{\beta}(\rho^{\ln})^2},\frac{1}{2\gamma\rho^{\ln}},\frac{1}{4\avg{\beta}},\frac{\rho^{\ln}(\gamma-1)}{\gamma},\frac{1}{4\avg{\beta}},\frac{1}{2\gamma\rho^{\ln}},\frac{1}{4\avg{\beta}(\rho^{\ln})^2},\frac{1}{2\gamma\rho^{\ln}}\right),
\end{equation}
and the diagonal matrix of eigenvalues for the ideal GLM-MHD system is
\begin{align}
	\mat{\Lambda} &= \diag\left(\ \hat{\lambda}_{+{\f}} \ |\  \hat{\lambda}_{+{\a}} \ |\  \hat{\lambda}_{+{\s}} \ |\ \hat{\lambda}_{+\psi} \ |\ \hat{\lambda}_{\E} \ |\  \hat{\lambda}_{-\psi} \ |\  \hat{\lambda}_{-{\s}} \ |\  \hat{\lambda}_{-{\a}} \ |\  \hat{\lambda}_{-{\f}} \ \right),
\end{align}
where we describe the discrete evaluation of the wave speeds in the following subsection.

The final form of the entropy stable numerical flux with the matrix dissipation term takes the form \cite{Winters2016,Barth1999}
\begin{equation}\label{eq:entropystableflux2}
	\vec{f}^{*,\rm KEPES, MD} = \vec{f}^{*,\rm KEPEC} -\half\Ree|\boldsymbol{\Lambda}|\Tee\Ree^\intercal\jump{\vec{v}}.
\end{equation}

\subsection{Discrete eigenvalues of the ideal GLM-MHD system}
An important aspect of utmost concern for robustness and stability of the numerical scheme we construct is how to define the discrete wave speeds, $\lambda$, at the interfaces. From \eqref{eq:eigenvalues} we know that the wave speeds in continuous space are given by
\begin{equation}
	\lambda_{\pm \f} = u \pm c_\f, \qquad \lambda_{\pm \s} = u \pm c_\s, \qquad \lambda_{\pm \a} = u \pm c_\a, \qquad \lambda_\E = u,\quad \mbox{and} \quad \lambda_{\pm \psi} = u \pm c_h.
\end{equation}

However, in discretized space, we have to compute the eigenvalues at each interface from the discrete left and right states, $\vec{q}_{\L,\R}$. First, we compute a discrete flux Jacobian as was described, in a different context, in \cite{Derigs2016_2}. From this matrix we compute the eigenvalues of the discrete ideal GLM-MHD system at the interface between the left and right cells and obtain:
\begin{equation}
	\hat{\vec{\lambda}} =
	\begin{bmatrix}
	\hat{\lambda}_{+\f}\vphantom{\hat{\lambda}_{+\f}}\\
	\hat{\lambda}_{+\a}\vphantom{\avg{u}+\hat{c}_\a}\\
	\hat{\lambda}_{+\s}\vphantom{\avg{u}+\hat{c}_\s}\\
	\hat{\lambda}_{+\psi}\vphantom{\avg{u}+\hat{c}_h}\\
	\hat{\lambda}_\E\vphantom{\avg{u}}\\
	\hat{\lambda}_{-\psi}\vphantom{\avg{u}+\hat{c}_h}\\
	\hat{\lambda}_{-\s}\vphantom{\avg{u}-\hat{c}_\s}\\
	\hat{\lambda}_{-\a}\vphantom{\avg{u}-\hat{c}_\a}\\
	\hat{\lambda}_{-\f}\vphantom{\avg{u}-\hat{c}_\f}
	\end{bmatrix} =
	\begin{bmatrix}
	\avg{u}+\hat{c}_\f\vphantom{\hat{\lambda}_{-\f}}\\
	\avg{u}+\hat{c}_\a\vphantom{\hat{\lambda}_{-\f}}\\
	\avg{u}+\hat{c}_\s\vphantom{\hat{\lambda}_{-\f}}\\
	\avg{u}+\avg{c_h}\\
	\avg{u}\\
	\avg{u}-\avg{c_h}\\
	\avg{u}-\hat{c}_\s\vphantom{\hat{\lambda}_{-\f}}\\
	\avg{u}-\hat{c}_\a\vphantom{\hat{\lambda}_{-\f}}\\
	\avg{u}-\hat{c}_\f\vphantom{\hat{\lambda}_{-\f}}
	\end{bmatrix}\quad
	\begin{matrix*}[l]
	\mbox{right going fast magnetoacoustic wave}\vphantom{\hat{\lambda}_{-\f}}\\
	\mbox{right going Alfv\'en wave}\vphantom{\hat{\lambda}_{-\f}}\\
	\mbox{right going slow magnetoacoustic wave}\vphantom{\hat{\lambda}_{-\f}}\\
	\mbox{right going GLM wave}\vphantom{\vphantom{\avg{u}+\hat{c}_h}}\\
	\mbox{entropy wave}\vphantom{\avg{u}}\\
	\mbox{left going GLM wave}\vphantom{\vphantom{\avg{u}+\hat{c}_h}}\\
	\mbox{left going slow magnetoacoustic wave}\vphantom{\hat{\lambda}_{-\f}}\\
	\mbox{left going Alfv\'en wave}\vphantom{\hat{\lambda}_{-\f}}\\
	\mbox{left going fast magnetoacoustic wave}\vphantom{\hat{\lambda}_{-\f}}
	\end{matrix*}
\end{equation}
The precise form of the discrete speeds ($\hat{c}_{\f,\s,\a}$) as well as technical details are summarized in \ref{App:discreteEigenvalues}.

\subsection{Mixed hyperbolic/parabolic GLM ansatz}\label{Sec:mixedansatz}
Dedner et al.~\cite{Dedner2002}, Wesenberg~\cite{Wesenberg2003} and Tricco \& Price \cite{Tricco2016} found that the best approximation of $\psi$ may be obtained by a mixed hyperbolic/parabolic ansatz. Hence, we supplement (\ref{eq:3DIDEALGLMMHD}e) with an additional source term
\begin{equation}\label{eq:psisrcterm}
	\vec{\Upsilon}^\alpha =
	\begin{bmatrix}
	0 & \vec{0} & 0 & \vec{0} & \alpha \psi
	\end{bmatrix}^\intercal,
\end{equation}
with the parabolic diffusion rate $\alpha \in [0,\infty)$ that controls the damping of the $\psi$ field. Through the addition of this source term, the $\psi$ field is no longer a conserved quantity, but is actively dissipated. Clearly, including such a dissipative term makes the derivation of an entropy \emph{conserving} scheme impossible. However, if we carefully compute the entropy budget of the source term \eqref{eq:psisrcterm}, we find that the contribution to the entropy is guaranteed to have the correct sign, \ie it fulfills the entropy inequality and is still suitable for creating an entropy \emph{stable} scheme,
\begin{equation}
	\vec{v} \cdot (\vec{q}_t + \frac{\partial}{\partial x} \vec{f}^x + \vec{\Upsilon}^x + \vec{\Upsilon}^\alpha) \quad\Rightarrow\quad \pderivative{}{t}S + \nabla \cdot (\vec{u}S) = -2\beta \alpha\psi^2 \le 0 \quad \mbox{with} \quad \alpha,\beta \ge 0.
\end{equation}

The effect of this additional source term on the evolution of the total energy is found by looking at the temporal evolution of the $\psi$ field,
\begin{equation}
	\pderivative{}{t}\left(\frac{1}{2}\psi^2\right) = \psi \cdot \pderivative{\psi}{t} = - c_h \psi (\nabla \cdot  \vec{B})+ \vec{u}\psi(\nabla\psi) \mathcolor{red}{+ \alpha\psi^2}.
\end{equation}
The resulting system is given by \eqref{eq:3DIDEALGLMMHD} with an additional source term $\alpha\psi^2$ on the right hand side of the total energy.
We chose to ignore this additional source term on the total energy equation, thus any energy dissipated from the $\psi$ field directly enters the thermal pressure.

The source term \eqref{eq:psisrcterm} introduces a new free parameter $\alpha$ which requires further analysis.
We see that for the purely hyperbolic case, \ie $\alpha = 0$, we can derive an entropy conserving scheme. Furthermore, for any $\alpha > 0$, entropy is guaranteed to be dissipated but never destroyed. There are several choices one can make in selecting the damping parameter $\alpha$. As discussed by Dedner et al.~\cite[Section 4]{Dedner2002}, a favorable choice for the damping parameter is a fixed proportion of decay (parabolic) to transport (hyperbolic) with a ratio $c_r := c_p^2/c_h$ at all times. In their observation this choice of $\alpha$ resulted in satisfactory numerical results that are independent of the grid resolution or the scheme used. Dedner et al.~\cite[p.~661]{Dedner2002} define the optimal ratio as $c_r = 0.18$,
\begin{equation}\label{eq:alpha_Dedner}
	\alpha = \frac{c_h^2}{c_p^2} = \frac{c_h}{c_r} = \frac{c_h}{0.18}.
\end{equation}
As we show in the numerical results section, the mixed GLM ansatz gives very good results and is in fact even necessary for periodic boundary conditions.

\subsection{High-order accurate entropy stable scheme}\label{sec:high-order-extension}
The scheme we discussed so far is first order in space and still continuous in time. In the following subsections, we discuss its extension to the fully discrete case with high-order accuracy. 

\subsubsection{Temporal accuracy}
A very simple time integrator is the Euler scheme, where the solution is advanced in small time steps in which the flux is assumed to be constant. The Euler scheme is only first-order accurate in \emph{time}, \ie the solution is accurate to $\mathcal{O}(\Delta t)$. Fortunately, the temporal order can be increased by replacing the time integrator by a suitable higher-order scheme, \eg strong stability preserving (SSP) Runge-Kutta (RK) schemes \cite{Gottlieb2001}. However, high-order accurate time integrators come at significant additional computational costs so one has to always find a compromise between (temporal) accuracy and computational resources. For the numerical tests we present in this work, we use a third order SSPRK time integrator.

\subsubsection{Spatial accuracy}
Unfortunately, higher spatial accuracy is harder to obtain for finite volume schemes because we only have cell-averaged quantities available, although we need interface values for computing the numerical fluxes. There are, essentially, two ways to achieve high-order accuracy in space: The first, and most often used one, is the technique of ``spatial reconstruction'' where an algorithm is used to deduce an interface value based on a certain stencil on the cell-averaged quantities. The simplest approximation is to assume that the values at the interfaces are identical to the cell-averages. Unsurprisingly, the solution obtained using such an approximation is only first-order accurate in \emph{space}, \ie the solution is accurate to $\mathcal{O}(\Delta x)$. There exists a vast amount of literature on the technique of spatial reconstruction. The authors give an extensive introduction in \cite[Section 4.1]{DMV}.

The second approach, uses the fact that the entropy conserving flux describes the rate of change for the quantities over an interface and, as such, is a first order derivative with respect to time over a fixed volume.
Using a suitable extrapolation, we can construct arbitrarily accurate entropy conservative interface fluxes through linear combinations of our computationally inexpensive entropy conservative flux derived in Section \ref{Sec:EC} as shown below.

Given suitable coefficients $\{\xi_{p,r}\}_{r=1}^p$ (see Table~\ref{tab:alphas}), the entropy conserving flux
\begin{equation}
	{}^{2p}\!\vec{f}^\mathrm{EC}_{i-1/2} := \sum_{r=1}^{p} \xi_{p,r} \sum_{s=1}^{r} \vec{f}^\mathrm{EC}(\vec{u}_{i-s},\vec{u}_{i-s+r})
\end{equation}
is $2p$th-order accurate in space, \ie
\begin{equation}
	\frac{1}{\Delta x}\left({}^{2p}\!\vec{f}^\mathrm{EC}_{i+1/2} - {}^{2p}\!\vec{f}^\mathrm{EC}_{i-1/2}\right) = \pderivative{}{x}\vec{f}(\vec{u})\Big|_{x=x_i} + \mathcal{O}(\Delta x^{2p})
\end{equation}
\cite[Theorem 4.4]{LeFloch2002}.
The coefficients $\{\xi_{p,r}\}_{r=1}^p$ are obtained by solving the $p$ linear equations given by
\begin{equation}\label{eq:alphas}
	\sum_{r=1}^{p} i \xi_{p,r} = 1, \quad \sum_{r=1}^{p} r^{2s-1}\xi_{p,r} = 0 \qquad (s=2,\dots,p).
\end{equation}

\begin{table}[h]
	\centering
	\begin{tabular}{ccccccccc}
		\toprule
		Accuracy $2p$ & $\xi_{p,1}$&$\xi_{p,2}$&$\xi_{p,3}$&$\xi_{p,4}$&$\xi_{p,5}$&$\xi_{p,6}$&$\xi_{p,7}$&$\xi_{p,8}$\\
		\midrule
		2 & 1&&&&&&&\\
		4 & $\frac{4}{3}$&$ -\frac{1}{6}$&&&&&&\\[.3em]
		6 & $\frac{3}{2}$&$ -\frac{3}{10}$&$ \frac{1}{30}$&&&&&\\[.3em]
		8 & $\frac{8}{5}$&$ -\frac{2}{5}$&$ \frac{8}{105}$&$ -\frac{1}{140}$&&&&\\[.3em]
		10 & $\frac{5}{3}$&$ -\frac{10}{21}$&$ \frac{5}{42}$&$ -\frac{5}{252}$&$ \frac{1}{630}$&&&\\[.3em]
		12 & $\frac{12}{7}$&$ -\frac{15}{28}$&$ \frac{10}{63}$&$ -\frac{1}{28}$&$ \frac{2}{385}$&$ -\frac{1}{2772}$&&\\[.3em]
		14 & $\frac{7}{4}$&$ -\frac{7}{12}$&$ \frac{7}{36}$&$ -\frac{7}{132}$&$ \frac{7}{660}$&$ -\frac{7}{5148}$&$\frac{1}{12012}$&\\[.3em]
		16 & $\frac{16}{9}$&$ -\frac{28}{45}$&$ \frac{112}{495}$&$ -\frac{7}{99}$&$ \frac{112}{6435}$&$ -\frac{4}{1287}$&$ \frac{16}{45045}$&$ -\frac{1}{51480}$\\
		\bottomrule
	\end{tabular}
	\caption{High-order coefficients $\{\xi_{p,r}\}_{r=1}^p$ for up to 16${}^\mathrm{th}$ order accuracy. Coefficients for even higher order can be computed from \eqref{eq:alphas}.}
	\label{tab:alphas}
\end{table}

As an example, we summarize the second- to sixth-order accurate entropy conserving fluxes below:
\begin{itemize}
	\item Second-order accurate EC interface flux ($p=1$)
	\begin{equation}
	\includegraphics[scale=1]{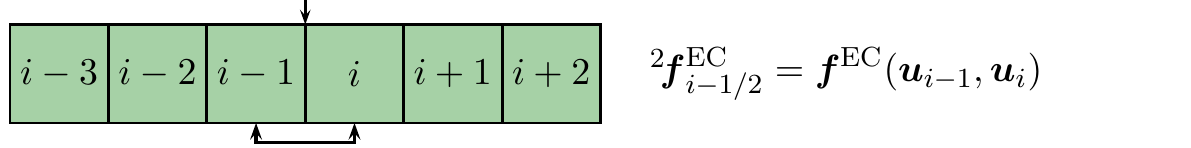}
	\end{equation}\\[-12mm]
	\item Fourth-order accurate EC interface flux ($p=2$)
	\begin{equation}
	\includegraphics[scale=1]{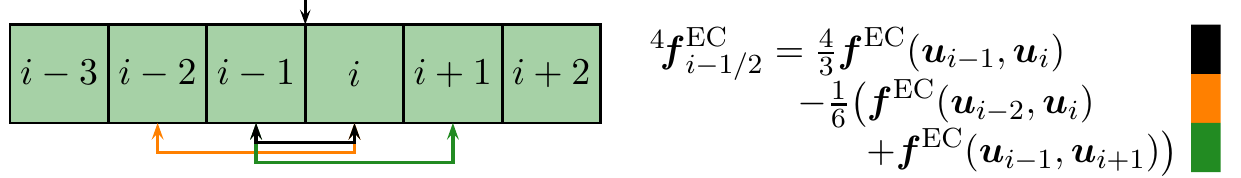}
	\end{equation}\\[-14mm]
	\item Sixth-order accurate EC interface flux ($p=3$)
	\begin{align}
	{}^{6}\!\vec{f}^\mathrm{EC}_{i-1/2} = &\frac{3}{2}\vec{f}(\vec{u}_{i-1}, \vec{u}_i) -\frac{3}{10}\big(\vec{f}(\vec{u}_{i-2}, \vec{u}_i) + \vec{f}(\vec{u}_{i-1}, \vec{u}_{i+1}) \big) \notag\\
	&+\frac{1}{30}\big(\vec{f}(\vec{u}_{i-3}, \vec{u}_i) + \vec{f}(\vec{u}_{i-2}, \vec{u}_{i+1}) + \vec{f}(\vec{u}_{i-1}, \vec{u}_{i+2}) \big)
	\end{align}
\end{itemize}
Up to now it is unknown how to discretize the non-conservative terms to obtain a high-order accurate approximation. The source term added to the numerical fluxes is given by
\begin{equation}
\vec{\Upsilon}^x = (\nabla\cdot\vec{B})^x
\begin{bmatrix}0 & \vec{B} & \vec{u}\cdot\vec{B} & \vec{u} & 0\end{bmatrix}^\intercal
+ (\vec{u}\cdot(\nabla\psi))^x
\begin{bmatrix}0 & \vec{0} & \psi & \vec{0} & 1\end{bmatrix}^\intercal,
\end{equation}
where, in its discretized version, it is a high-order representation of the divergence and gradient operators in combination with the cell-centered quantities in cell $i$,
\begin{equation}
{}^{2p}\vec{\Upsilon}_{i}^x = {}^{2p}(\nabla\cdot\vec{B})_i^x \begin{bmatrix}0 & \vec{B} & \vec{u}\cdot\vec{B} & \vec{u} & 0\end{bmatrix}_i^\intercal + {}^{2p}(\vec{u}\cdot(\nabla\psi))_i^x \begin{bmatrix}\vec{0} & 0 & \psi & 0 & 1\end{bmatrix}_i^\intercal
\end{equation}
with 
\begin{equation}
{}^{2p}(\nabla\cdot\vec{B})_i^x := \frac{1}{2\Delta x}\sum_{r=1}^{p}\xi_{p,r}(B_{1,i+r}-B_{1,i-r})
\qquad\text{and}\qquad
{}^{2p}(\vec{u}\cdot(\nabla\psi))_i^x := \frac{u_i}{2\Delta x}\sum_{r=1}^{p}\xi_{p,r}( \psi_{i+r}-\psi_{i-r})
\end{equation}
Again, we summarize the resulting non-conservative terms for second- to sixth-order accuracy below:
\begin{itemize}
	\item Second-order accurate magnetic field divergence ($p=1$)
	\begin{align}
	{}^{2}(\nabla\cdot\vec{B})_i^x &= \frac{B_{1,i+1}-B_{1,i-1}}{2\Delta x}\\
	{}^{2}(\vec{u}\cdot\nabla\psi)_i &= u_i \frac{\psi_{i+1}-\psi_{i-1}}{2\Delta x}\\
	{}^{2}\vec{\Upsilon}_{i} &= {}^{2}(\nabla\cdot\vec{B})_i^x \begin{bmatrix}0 & \vec{B} & \vec{u}\cdot\vec{B} & \vec{u} & 0\end{bmatrix}_i^\intercal
	+ {}^{2}(\vec{u}\cdot(\nabla\psi))_i^x
	\begin{bmatrix}0 & \vec{0} & \psi & \vec{0} & 1\end{bmatrix}_i^\intercal
	\end{align}
	\item Fourth-order accurate magnetic field divergence ($p=2$)
	\begin{align}
	{}^{4}(\nabla\cdot\vec{B})_i^x &= \frac{\frac{4}{3}(B_{1,i+1}-B_{1,i-1}) - \frac{1}{6}(B_{i,i+2}-B_{i,i-2})}{2\Delta x}\\
	{}^{4}(\vec{u}\cdot\nabla\psi)_i &= u_i\frac{\frac{4}{3}(\psi_{i+1}-\psi_{i-1}) - \frac{1}{6}(\psi_{i+2}-\psi_{i-2})}{2\Delta x}\\
	{}^{4}\vec{\Upsilon}_{i} &= {}^{4}(\nabla\cdot\vec{B})_i^x \begin{bmatrix}0 & \vec{B} & \vec{u}\cdot\vec{B} & \vec{u} & 0\end{bmatrix}_i^\intercal
	+ {}^{4}(\vec{u}\cdot(\nabla\psi))_i^x
	\begin{bmatrix}0 & \vec{0} & \psi & \vec{0} & 1\end{bmatrix}_i^\intercal
	\end{align}
	\item Sixth-order accurate magnetic field divergence ($p=3$)
	\begin{align}
	{}^{6}(\nabla\cdot\vec{B})_i^x &= \frac{\frac{3}{2}(B_{1,i+1}-B_{1,i-1}) - \frac{3}{10}(B_{1,i+2}-B_{1,i-2}) + \frac{1}{60}(B_{1,i+3}-B_{1,i-3})}{2\Delta x}\\
	{}^{6}(\vec{u}\cdot\nabla\psi)_i^x &= u_i\frac{\frac{3}{2}(\psi_{i+1}-\psi_{i-1}) - \frac{3}{10}(\psi_{i+2}-\psi_{i-2}) + \frac{1}{60}(\psi_{i+3}-\psi_{i-3})}{2\Delta x}\\
	{}^{6}\vec{\Upsilon}_{i} &= {}^{6}(\nabla\cdot\vec{B})_i^x \begin{bmatrix}0 & \vec{B} & \vec{u}\cdot\vec{B} & \vec{u} & 0\end{bmatrix}_i^\intercal
	+ {}^{6}(\vec{u}\cdot(\nabla\psi))_i^x
	\begin{bmatrix}0 & \vec{0} & \psi & \vec{0} & 1\end{bmatrix}_i^\intercal
	\end{align}
\end{itemize}

Unfortunately, one cannot apply the same technique for the entropy stable part of the numerical fluxes.
As detailed by Fjordholm \cite[Sec.~3.2]{Fjordholm2012}, a specific reconstruction procedure (preferably done in entropy rather than in conservative variables) can be used to ensure high-order entropy stability. To do so, we ensure that the sign of the reconstructed jump, ${}^{k}\!\jump{\vec{v}}_{i-1/2}$, is the same sign as the naive jump, $\jump{\vec{v}}_{i-1/2}$.

\section{Numerical tests}\label{Sec:numtests}
We demonstrate the numerical magnetic divergence evolution of the new entropy stable numerical scheme for ideal GLM-MHD derived in this work by computing several ideal MHD test problems. We use the finite volume code \texttt{FLASH} in version 4.5. The technical aspects of our testbed are described in great detail in \cite{Derigs2016}.
We use fourth-order accurate entropy-conservative as well as third-order accurate entropy-stable fluxes (limited reconstruction \cite{Schmidtmann2015}) in space as well as a third-order accurate SSP RK time integration scheme \cite{Gottlieb2001} and a CFL coefficient of 0.8 for all tests.
With our numerical tests we focus on the magnetic field divergence cleaning effectiveness of the new entropy stable ideal GLM-MHD system.
A numerical validation of the entropy conservation properties of the new numerical flux is given as a supplementary test case.

Note that, given its nature, a suitable choice for the initial value for $\psi$ is $\psi_0 = \psi(t=0)=0$. This has two reasons:
\begin{enumerate}
	\item Only gradients of $\psi$ appear in the numerical fluxes. Hence, given divergence-free initial conditions, we should initialize the $\psi$ field with a constant value everywhere such that $\nabla\psi_0 = \vec{0}$, initially.
	\item We define the energy in the $\psi$ field as $E_\psi = \frac{1}{2}\psi^2$. In a divergence-free state, it makes sense to have this ``correction field energy'' equal to zero, suggesting $\psi_0 = 0$ everywhere.
\end{enumerate}

\subsection{Artificial non-zero magnetic field divergence test (1D)}\label{sec:divClean1D}
\begin{figure}[H]
	\centering
	\begin{minipage}{0.47\textwidth}
		\includegraphics[scale=1]{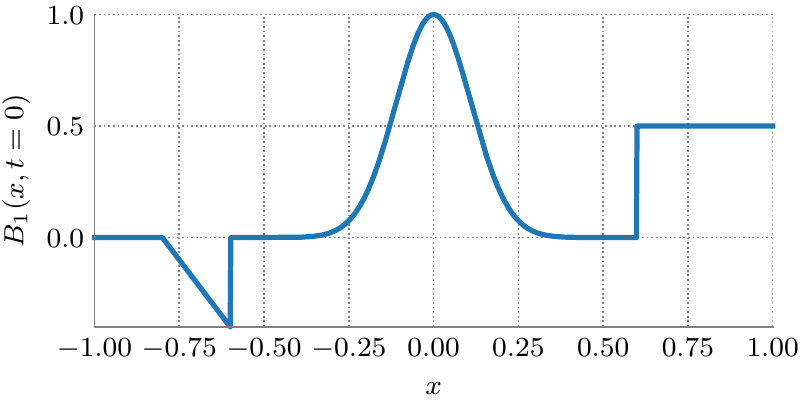}
	\end{minipage}
	\begin{minipage}{0.47\textwidth}\vspace*{-1.5em}
		\begin{equation}\label{eq:divBB1}
		B_1(x) = \begin{cases}
		0.0 & \hphantom{-0.8 < \; } x \le -0.8\\
		-2(x+0.8) & -0.8 < x \le -0.6\\
		\exp\left({-\frac{(x/0.11)^2}{2}}\right) & -0.6 < x \le 0.6\\
		0.5 & \hphantom{-0.8 < \; } x > 0.6
		\end{cases}
		\end{equation}
	\end{minipage}
	\caption{Initial $x$ component of the magnetic field of the artificial non-zero magnetic field divergence test (1D).}
	\label{fig:divBinitial1D}
\end{figure}

The behavior of any numerical scheme, given non-zero initial divergence is of high interest as non-zero divergence may also be caused by poorly chosen initial conditions. The scheme must deal with the divergence errors properly in order to produce a credible solution.
This artificial numerical test starts from a magnetic field with non-zero divergence involving not only smooth gradients but also steps making it a more challenging test for the divergence cleaning method. The remaining quantities are flat. We select a fixed resolution of 256 uniformly distributed cells.
\begin{table}[h]
	\setlength\extrarowheight{3pt}
	\centering
	\begin{minipage}[t]{0.315\textwidth}
		\begin{tabular}[t]{|l|l|}\hline
			Density	$\rho$		& 1.0\\\hline
			Pressure $p$		& 1.0\\\hline
			Velocity $\vec{u}$	& $\vec{0}$\\\hline
			Mag.~field $\vec{B}$& $[B_1(x) \quad 0 \quad 0]^\intercal$\\ \hline
		\end{tabular}
	\end{minipage}
	\begin{minipage}[t]{0.47\textwidth}
		\begin{tabular}[t]{|l|l|}\hline
			Domain size &$\{x_\mathrm{min},x_\mathrm{max}\} = \{-1,1\}$\\\hline
			Boundary conditions & outflow or periodic\\\hline
			Simulation end time & $t_\mathrm{max} = 5.0$ \\\hline
			Adiabatic index & $\gamma = 1.4$ \\\hline
		\end{tabular}
	\end{minipage}
	\vspace*{-1em}
	\caption{Initial conditions and runtime parameters: Artificial non-zero magnetic field divergence test (1D).}
	\label{tab:divB}
\end{table}
We present the initial conditions for this test in Table~\ref{tab:divB}. The magnetic field in the $x-$direction is given by \eqref{eq:divBB1} in Fig.~\ref{fig:divBinitial1D}. Note that these initial conditions intentionally violate the constraint $\nabla\cdot\vec{B} = 0$ by a significant amount.
\begin{figure}[H]
	\centering
	\vspace*{-1\baselineskip}
	\includegraphics[scale=.9]{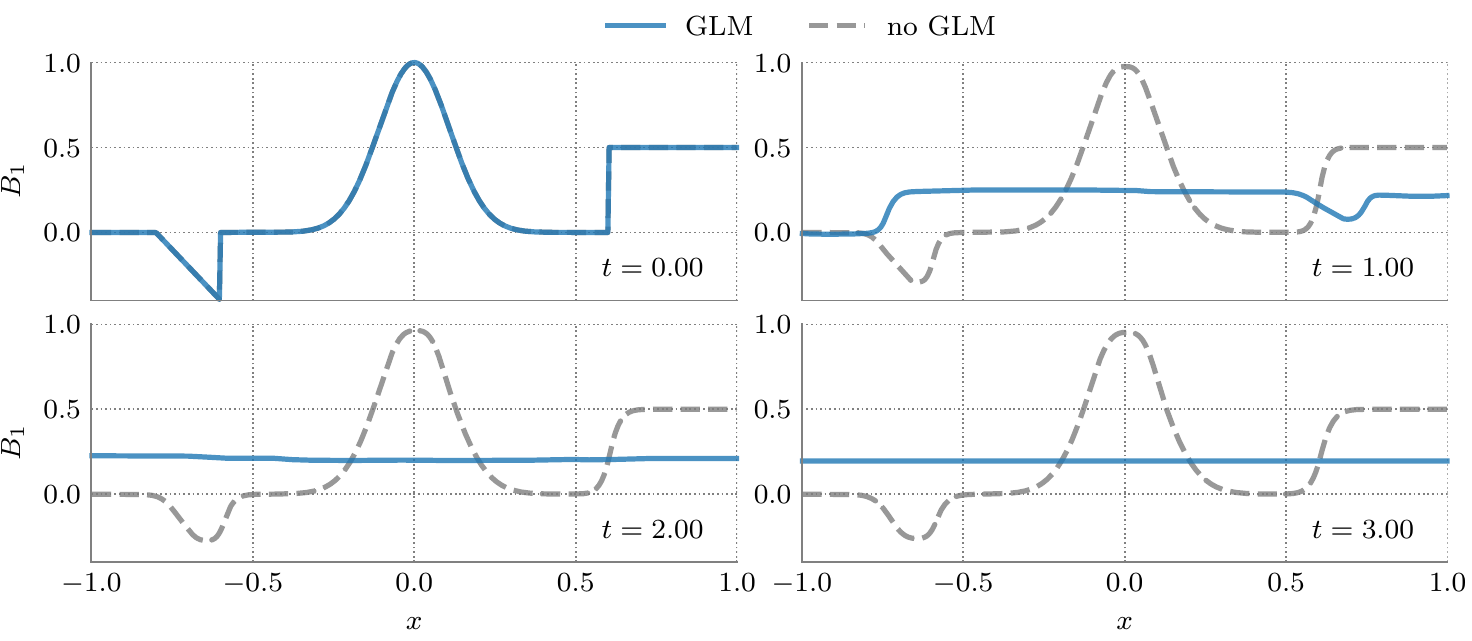}
	\caption{Magnetic field $B_1$ of the artificial non-zero magnetic field divergence test at four different times. We show the magnetic field evolution computed using the new GLM-KEPES flux (blue lines), together with the solution obtained using the KEPES flux that is not using the GLM technique for divergence cleaning (dashed, orange lines). Note that $t={1,2,3}$ corresponds to one, two, and three Alfv\'en crossing times.}
	\label{fig:divBfinal1D}
\end{figure}
In Fig.~\ref{fig:divBfinal1D}, we compare the result of this test against the one obtained using the kinetic energy preserving entropy stable (KEPES) solver for ideal MHD \cite{Derigs2016_2} at different times. The new GLM-KEPES solver treats the substantial initial divergence error correctly and removes any divergence in the magnetic field quickly. The KEPES solver is, however, only capable of dissipating the magnetic field slightly due to numerical dissipation caused by the spatial and temporal discretizations. Note that we can obtain similar results with our scheme if we enforce $c_h=0$ throughout the simulation.
It is obvious that a scheme that is not capable of removing significant divergence errors in the solution generates unsatisfactory numerical results. We set $\alpha=0$ in this test case (hyperbolic cleaning only) and use outflow boundaries.
In the following, we use periodic boundary conditions. Note that for outflow boundary conditions, the errors can quickly leave the computational domain.
\begin{figure}[h]
	\centering
	\includegraphics[scale=1]{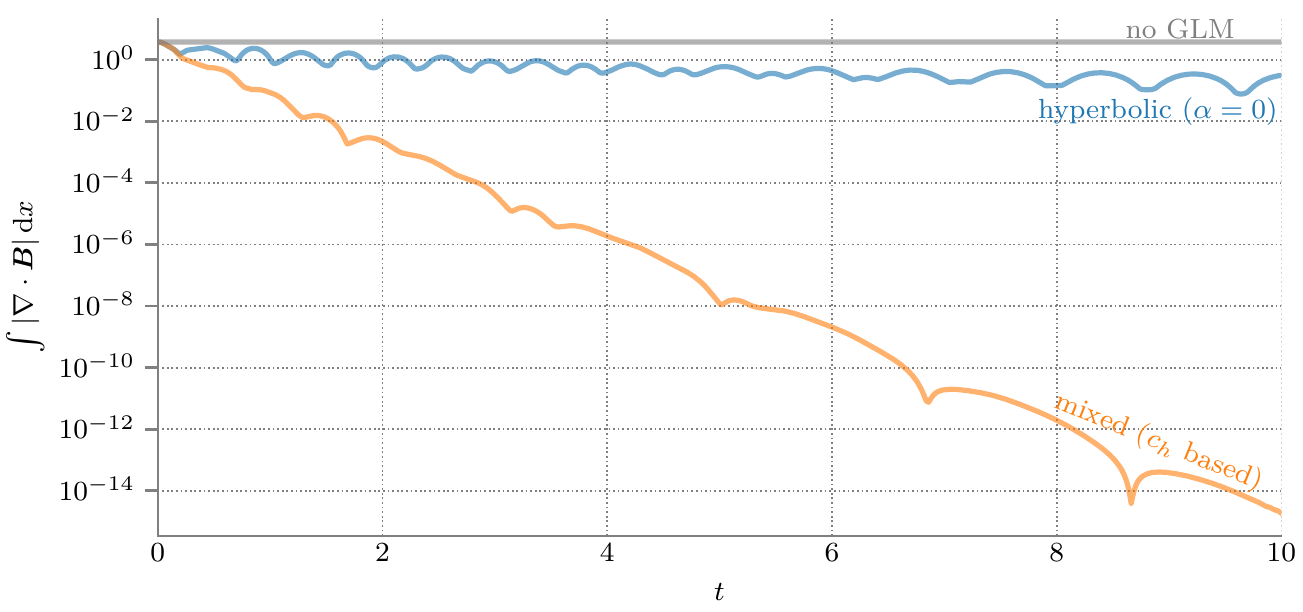}
	\caption{Artificial non-zero magnetic field divergence test: Evolution of the integrated absolute magnetic field divergence in the artificial non-zero magnetic field divergence test in one dimension using periodic boundary conditions. It is clear the the magnetic field divergence evolution with mixed cleaning is favorable.}
	\label{fig:divB_1D_periodic}
\end{figure}
However, as can be seen in Fig.~\ref{fig:divB_1D_periodic}, a solely hyperbolic cleaning ($\alpha = 0$) is not sufficient to reduce the divergence for periodic boundary conditions. This is easily understood as the divergence error cannot be advected ``out'' of the computational domain anymore. Hence, additional damping is essential in order to reduce the divergence error over time, as suggested by Dedner et al.~\cite[Section 4]{Dedner2002}, denoted by ``mixed ($c_h$ based)'' in the figure.

\subsection{Divergence advection test (2D)}\label{sec:divClean2D}
Another simple test case for the cleaning efficiency is the two-dimensional divergence advection test presented by Tricco \& Price \cite[Sec.~5.2]{Tricco2012}. It consists of divergence in the magnetic field artificially induced in the initial conditions that is advected by a uniform flow and is a variant of the ``peak in $B_1$'' test presented by Dedner et al.~\cite[Sec.~5]{Dedner2002}. This test includes a density step that is used to examine the reflection and refraction of the divergence waves as they transit between media of different densities. The initial conditions are listed in Table \ref{tab:divBinitial2D}.
\begin{table}[h]
	\centering
	\begin{minipage}[t]{0.5\textwidth}
		\begin{align*}
		\rho &= \begin{cases}
		1.0 & \text{if } x \le 0.5\\
		2.0 & \text{else}
		\end{cases}\\
		\vec{B} &= \begin{bmatrix} B_x & 0 & \frac{1}{\sqrt{4\pi}} \end{bmatrix}^\intercal, \quad \vec{u} = \vec{0}, \quad p = 6.0 \\
		B_x &= \begin{cases}
		\frac{1}{\sqrt{4\pi}}\left(\left(\frac{r}{r_0}\right)^8\!\!\! - 2 \left(\frac{r}{r_0}\right)^4\!\!\! + 1\right) & \text{if } r \le r_0 , \\
		0 & \text{else.}
		\end{cases}
		\end{align*}
	\end{minipage}
	\hspace*{5mm}
	\begin{minipage}[t]{0.4\textwidth}
		\begin{tabular}[t]{|l|l|}
			\hline
			Domain size &$x,y \in [-0.5,1.5]$ \\
			\hline
			Radial extent	&  $r_0 = {1}/{\sqrt{8}}$ \\
			\hline
			Boundary conditions & all: periodic \\
			\hline
			Simulation end time & $t_\mathrm{max} = 1.0$ \\
			\hline
			Adiabatic index & $\gamma = 5/3$ \\
			\hline
		\end{tabular}
		with $r=\sqrt{x^2+y^2}$
	\end{minipage}
	\caption{Initial conditions and runtime parameters: Divergence advection test (2D) \cite{Tricco2012}.}
	\label{tab:divBinitial2D}
\end{table}

\begin{figure}[h]
	\centering
	\includegraphics[scale=1]{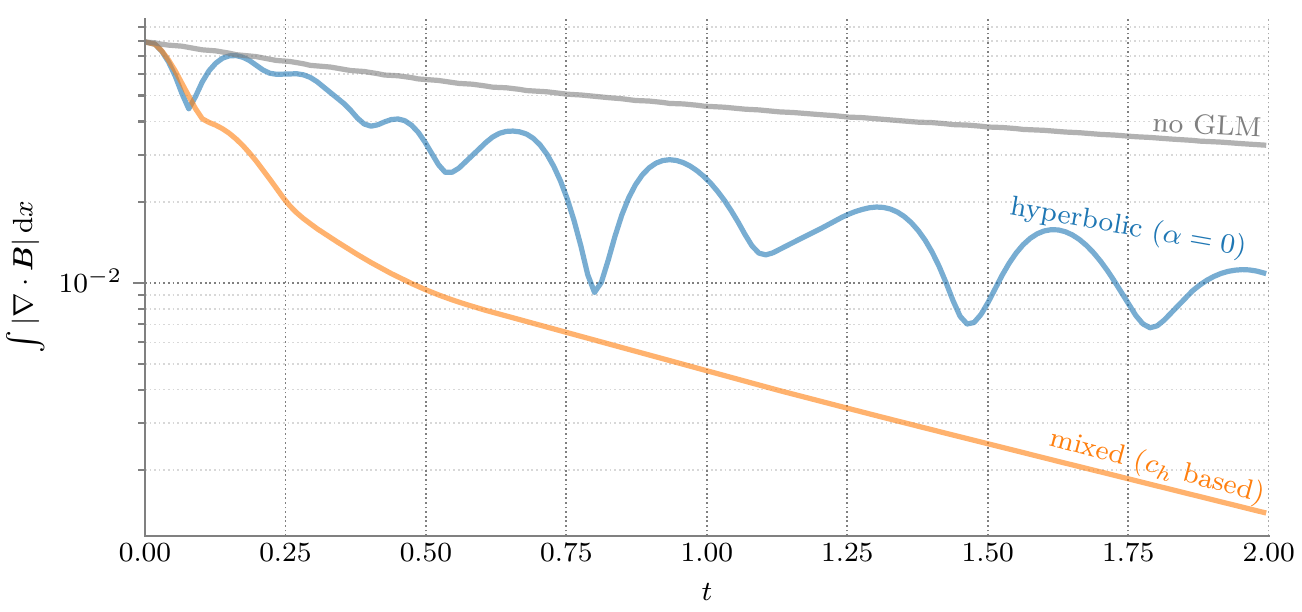}
	\caption{Divergence advection test: Evolution of the integrated absolute magnetic field divergence in the advection test in two dimensions. We test various choices of the damping parameter.}
	\label{fig:divBevolution2Dlogdissipation}
\end{figure}

Our findings are plotted in Fig.~\ref{fig:divBevolution2Dlogdissipation}. We find similar results compared to the one-dimensional test with periodic boundary conditions as the semi-adaptive choice suggested by Dedner et al.~gives the best cleaning behavior.
\FloatBarrier

\subsection{MHD Rotor test}
The MHD rotor problem \cite{Balsara1999} describes a rapidly spinning dense cylinder embedded in a magnetized, homogeneous medium at rest. Due to centrifugal forces, the dense cylinder is not in equilibrium. As the rotor spins with the given initial rotating velocity, the initially uniform magnetic field will wind up. The wrapping of the rotor by the magnetic field leads to strong toroidal Alfv\'en waves launched into the ambient fluid. The initial conditions are listed in Table~\ref{tab:Rotor}.
\begin{table}[h]
	\centering
	\begin{minipage}[t]{0.41\textwidth}
		\begin{tabular}[t]{l|ccc}
			&	{$r \le r_0$}				& {$r \in (r_0,r_1)$}		& {$r \ge r_1$}\\
			\midrule
			$\rho$		&	$10.0$ 						& $1.0 + 9.0 f(r)$			& $1.0$	\\
			$p$		&	$1.0$						& $1.0$  					& $1.0$\\
			$B_1$		&	$5/\sqrt{4\pi}$				& $5/\sqrt{4\pi}$			& $5/\sqrt{4\pi}$\\
			$B_2$		&	$0.0$						& $0.0$						& $0.0$\\
			$B_3$		&	$0.0$						& $0.0$						& $0.0$\\
			$u$		&	$-20.0 \Delta y$		& $-20.0 f(r) \Delta y$		& $0.0$\\
			$v$		&	$20.0 \Delta x$		& $20.0 f(r) \Delta x$		& $0.0$\\
			$w$		&	$0.0$		& $0.0$		& $0.0$\\
		\end{tabular}\\[.6em]
		with $f(r) = \frac{r_1-r}{r_1-r_0}$, \par $r=\sqrt{(x-x_\mathrm{center})^2+(y-y_\mathrm{center})^2}$, \par
		$\Delta x = (x-x_\mathrm{center})$, $\Delta y = (y-y_\mathrm{center})$
	\end{minipage}
	\hspace*{5mm}
	\begin{minipage}[t]{0.54\textwidth}
		\begin{tabular}[t]{|l|l|}
			\hline
			Domain size & $x,y \in [0,1]$ \\
			\hline
			Inner radius	&  $r_0 = 0.1$ \\
			\hline
			Outer radius	&  $r_1 = 0.115$ \\
			\hline
			$x$-center		& $x_\mathrm{center} = 0.5$ \\
			\hline
			$y$-center		& $y_\mathrm{center} = 0.5$ \\
			\hline
			Boundary conditions & all: outflow \\
			\hline
			Adaptive refinement on	& density, magnetic field \\
			\hline
			Simulation end time & $t_\mathrm{max} = 0.15$ \\
			\hline
			Adiabatic index & $\gamma = 1.4$ \\
			\hline
		\end{tabular}
	\end{minipage}
	\caption{Initial conditions and runtime parameters: 2D MHD rotor test \cite{Derigs2016}.}
	\label{tab:Rotor}
\end{table}

\begin{figure}[!ht]
	\centering
	\includegraphics[scale=1]{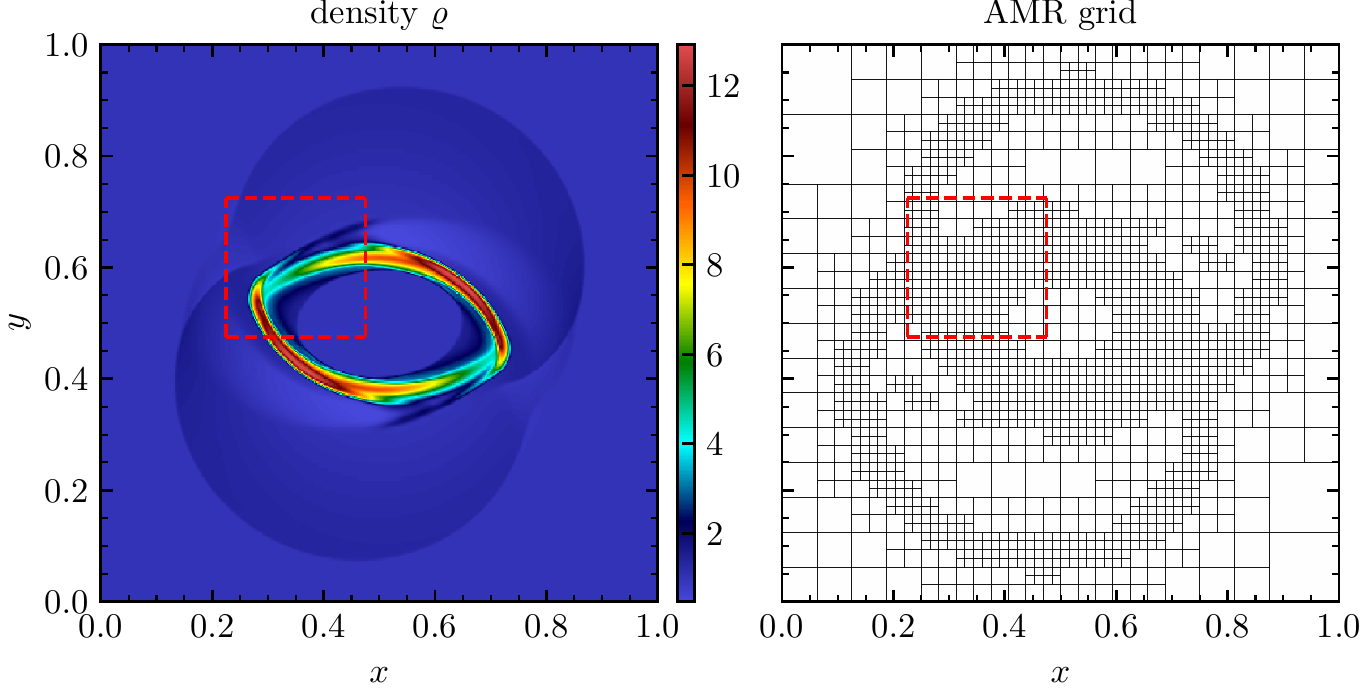}
	\caption{MHD rotor test ($t=0.15$): Adaptive grid resolution up to $512\times512$ (each shown rectangle encloses $8\times8$ cells). The marked area is shown in Fig.~\ref{fig:RotorZoom}.}
	\label{fig:Rotor}
\end{figure}

\begin{figure}[!hp]
	\centering
	\includegraphics[scale=1]{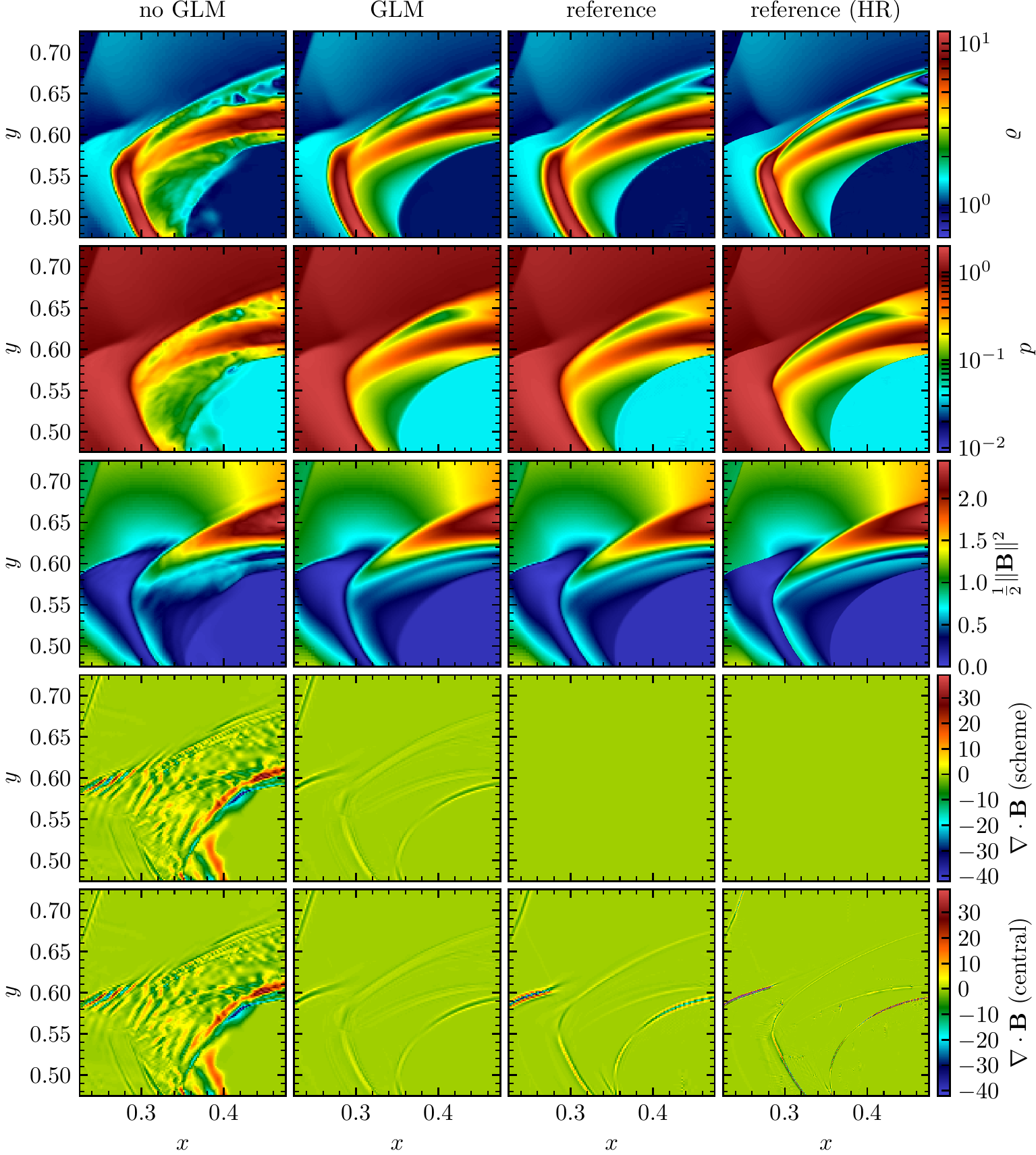}
	\caption{MHD rotor test: Zoom-in plot of Fig.~\ref{fig:Rotor}. 
	Left to right: Disabled GLM, enabled GLM, and {two} reference solution obtained using the unsplit staggered mesh solver implemented in \texttt{FLASH}. {The left reference solution is obtained on a similar (adaptive) grid, while the right reference solution (labeled ``HR'') is obtained on a grid that was $4\times$ finer resolution in each spatial direction.}
	Top to bottom: Gas density $\rho$, gas pressure $p$, magnetic pressure, $\frac{1}{2}\|\vec{B}\|^2$, and magnetic field divergence (two plots). The upper magnetic field divergence plot shows the magnetic field divergence computed using the method that is used in the corresponding numerical scheme. The lower plot shows the magnetic field divergence computes using central-differencing over the cell-center variables \eqref{eq:divBcell}. We see that in this context our GLM scheme shows a comparable result to the constrained transport scheme.}
	\label{fig:RotorZoom}
\end{figure}

The reference solution has been obtained using the unsplit staggered mesh (USM) solver implemented in \texttt{FLASH} \cite{Lee2009}. The USM solver uses constrained transport to ensure the solenoidal constraint of the magnetic field on a staggered mesh geometry. An advantage of our numerical scheme, compared to the staggered mesh USM solver, is that it requires appreciably less memory (see \eg \cite[Sec.~4.5]{Derigs2016}). This makes our scheme computationally attractive on many modern supercomputing systems where simulations are commonly memory-limited, \eg \cite{Walch2014}.
As can be seen in Figs.~\ref{fig:Rotor} and \ref{fig:RotorZoom}, the results obtained using the KEPES-GLM scheme show much smaller divergence artifacts and hence are much closer to the reference solution than the solution computed with the same numerical solver but without GLM correction (where we artificially set $c_h = 0$).

{In the zoom-in figure (Fig.~\ref{fig:RotorZoom}) we see density (top row) and pressure (2\textsuperscript{nd} row) minima visible in the GLM solution at $(x,y) \approx (0.4,0.64)$ which are absent in the reference solution. To check for a possible wrong behavior of our scheme we re-ran the reference solution, obtained with the unsplit staggered mesh solver of \texttt{FLASH}, with an adaptive resolution of up to $2048\times2048$ (four-fold). With the higher resolution run we confirm the extrema seen with our scheme and conclude that, in this test case, our scheme is able to capture finer details than the reference simulation on the same resolution.}

The USM solver uses a formulation that ensures that the numerical divergence is zero at the cell face-centered magnetic fields,
\begin{equation}\label{eq:divBface}
	(\nabla\cdot\vec{B})_{i,j}^\text{face} = \frac{b_{x,i+1/2,j} - b_{x,i-1/2,j}}{\Delta x} + \frac{b_{y,i,j+1/2} - b_{y,i,j-1/2}}{\Delta y},
\end{equation}
where $b_{x,y,z}$ describes the face-centered magnetic field components \cite[eq.~(20)]{Lee2009}.
Note that this way of computing the magnetic field divergence is different from our treatment of the magnetic fields at the cell-centers as defined in \eqref{eq:discreteSourceterm}:
\begin{equation}\label{eq:divBcell}
	(\nabla\cdot\vec{B})_{i,j}^\text{cell} = \frac{B_{x,i+1,j} - B_{x,i-1,j}}{2\Delta x} + \frac{B_{y,i,j+1} - B_{y,i,j-1}}{2\Delta y}
\end{equation}
We see that for the USM solver, the face-centered magnetic field divergence \eqref{eq:divBface} is indeed on the order of machine precision at any time. Hence, the \texttt{USM} solver itself solves the ideal MHD equations always in regions with vanishing magnetic field divergence. However, the cell-centered magnetic field divergence \eqref{eq:divBcell} is not guaranteed to vanish. While this is not an issue for the scheme itself, it may be relevant for schemes that use the cell-centered values for post-processing the numerical results. In fact, we find that the cell-centered magnetic field divergence is comparable between our scheme with GLM correction and the USM result which highlights the effectiveness of our scheme, cf.~the bottom panels in Fig.~\ref{fig:RotorZoom}.

\subsection{Orszag-Tang MHD vortex}\label{Sec:OT}
The Orszag-Tang vortex problem \citep{Orszag1979,Derigs2016} is a two-dimensional, spatially periodic problem well suited for studies of MHD turbulence. Thus, it has become a classical test for numerical MHD schemes. It includes dissipation of kinetic and magnetic energy, magnetic reconnection, the formation of high-density jets, dynamic alignment and the emergence and manifestation of small-scale structures.
The Orszag-Tang MHD vortex problem starts from non-random, smooth initial data. As the flow evolves it gradually becomes increasingly complex, forming intermediate shocks. Thus, this problem demonstrates the transition from initially smooth data to compressible, supersonic MHD turbulence.
\begin{table}[h]
	\centering
	\begin{minipage}[t]{0.43\textwidth}
		\begin{tabular}[t]{|l|l|}
			\hline
			Density $\rho$ & $1.0$ \\
			\hline
			Pressure $p$ & $1.0/\gamma$ \\
			\hline
			Velocity $\vec{u}$ & $\begin{bmatrix}-\sin(2\pi y)&\sin(2\pi x)&0\end{bmatrix}^\intercal$ \\
			\hline
			Mag.~field $\vec{B}$ & $\gamma^{-1}\begin{bmatrix}-\sin(2\pi y)&\sin(4\pi x)&0\end{bmatrix}^\intercal$\\
			\hline
		\end{tabular}
	\end{minipage}
	\hspace*{10mm}
	\begin{minipage}[t]{0.48\textwidth}
		\begin{tabular}[t]{|l|l|}
			\hline
			Domain size & $x,y \in [0,1]$ \\
			\hline
			Boundary conditions & all: periodic\\
			\hline
			Adaptive refinement on	& density, magnetic field \\
			\hline
			Simulation end time & $t_\mathrm{max} = 0.5$ \\
			\hline
			Adiabatic index & $\gamma = 5/3$\\
			\hline
		\end{tabular}
	\end{minipage}
	\caption{Initial conditions and runtime parameters: Orszag-Tang MHD vortex \cite{Derigs2016}.}
	\label{tab:OrszagTang}
\end{table}

\begin{figure}[!ht]
	\centering
	\includegraphics[scale=1]{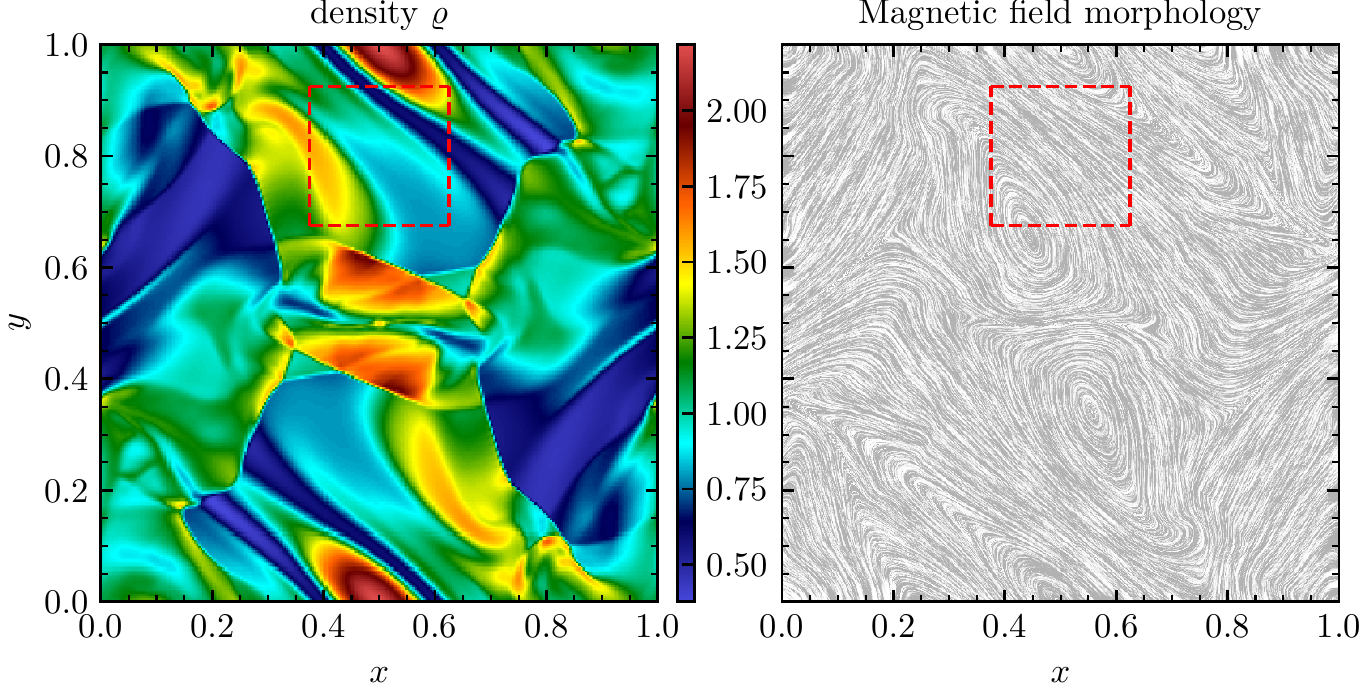}
	\caption{Orszag-Tang MHD vortex: Adaptive grid resolution up to $256\times256$. The simulation domain is fully refined at the shown time, $t=0.5$, for the given refinement criteria. The marked area is shown in Fig.~\ref{fig:OrszagTangZoom}.}
	\label{fig:OrszagTang}
\end{figure}

\begin{figure}[!ht]
	\centering
	\includegraphics[scale=1]{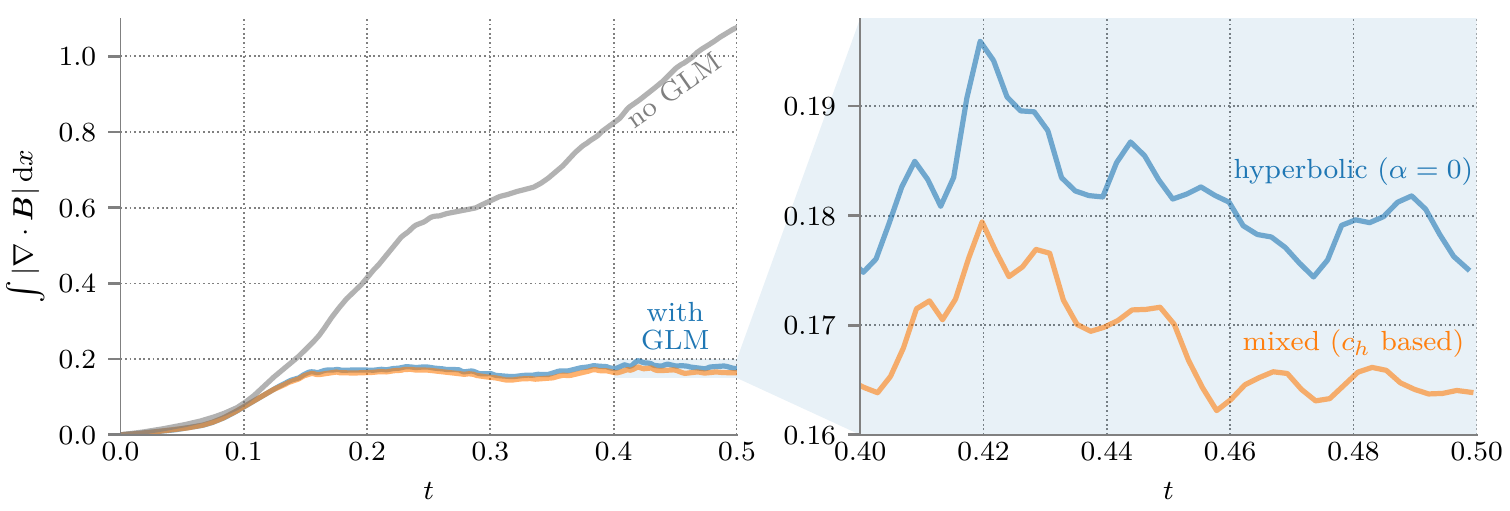}
	\caption{Orszag-Tang MHD vortex: Evolution of the integrated absolute magnetic field divergence. We test various choices of the damping parameter. The numerical results shown in this section use the $c_h$ based damping.}
	\label{fig:OrszagTangdivB}
\end{figure}

\begin{figure}[!hp]
	\centering
	\includegraphics[scale=1]{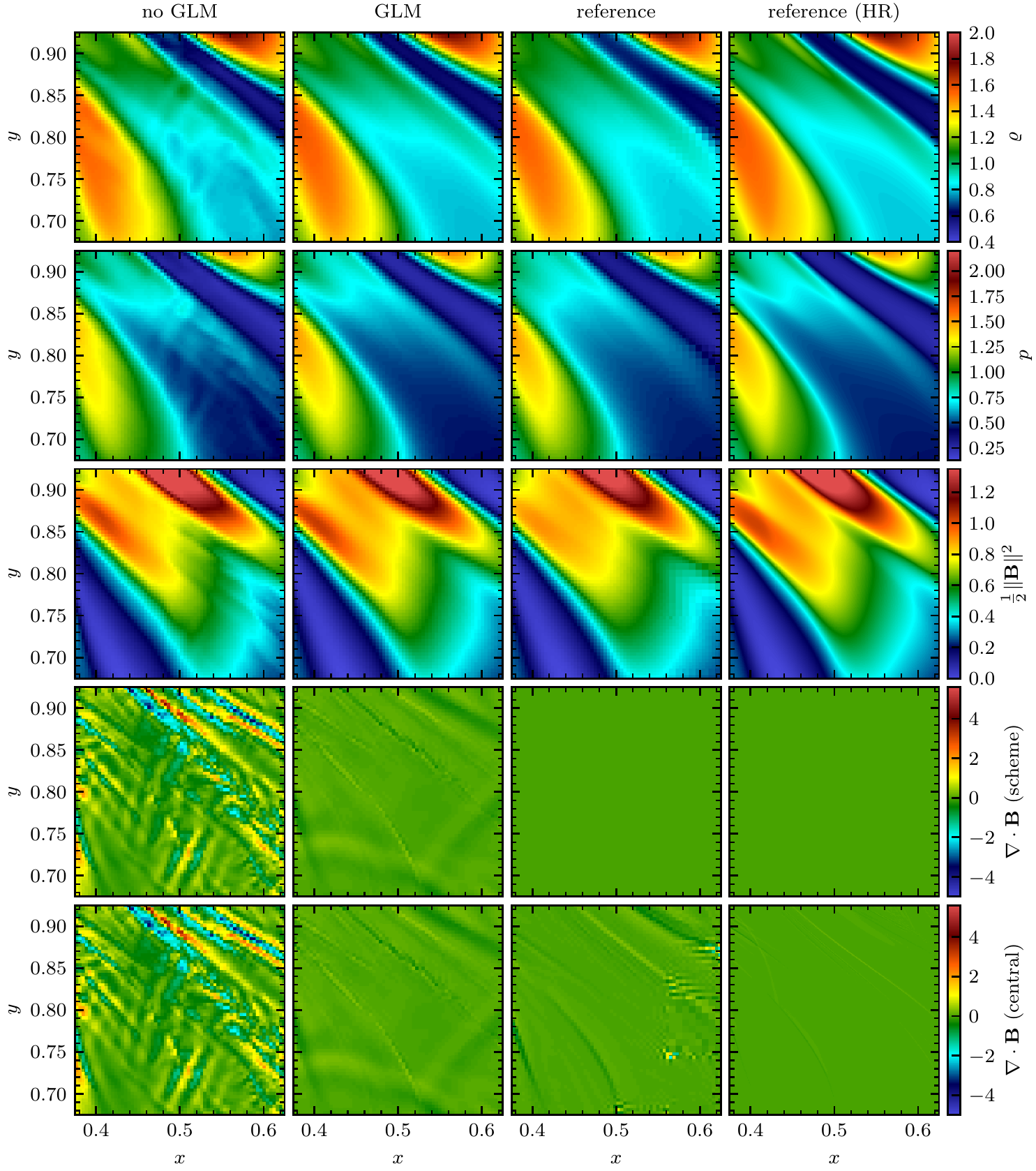}
	\caption{Orszag-Tang MHD vortex: Zoom-in plot of Fig.~\ref{fig:OrszagTang}.
	Left to right: Disabled GLM, enabled GLM, and two reference solutions obtained using the unsplit staggered mesh solver implemented in \texttt{FLASH}. {The left reference solution is obtained on a similar (adaptive) grid, while the right reference solution (labeled ``HR'') is obtained on a uniform grid that was $4\times$ finer in each spatial direction.}
	Top to bottom: Gas density $\rho$, gas pressure $p$, magnetic pressure, $\frac{1}{2}\|\vec{B}\|^2$, and magnetic field divergence (two plots). The upper magnetic field divergence plot shows the magnetic field divergence computed using the method that is used in the corresponding numerical scheme. The lower plot shows the magnetic field divergence computes using central-differencing over the cell-center variables \eqref{eq:divBcell}. We see that our GLM scheme, again, shows a similar result to the constrained transport scheme (on the same grid) when looking at the cell-centered magnetic field divergence. Visible artifacts are caused by recent adaptive mesh refinements.}
	\label{fig:OrszagTangZoom}
\end{figure}

Fig.~\ref{fig:OrszagTang} shows the density of the plasma at $t=0.5$ given the initial conditions listed in Table~\ref{tab:OrszagTang}. As the solution evolves in time, the initial vortex splits into two vortices. Sharp gradients accumulate and the vortex pattern becomes increasingly complex due to highly non-linear interactions between multiple intermediate shock waves traveling at different speeds. We compute the solution using a comparably low resolution of up to $256\times256$ in order to demonstrate that our numerical scheme is able to resolve sharp features on low to intermediate resolutions. The result compares very well with results given in the literature, \eg \cite{Balbas2005,Dai1998,Londrillo2000}.

{In Fig.~\ref{fig:OrszagTangdivB}, we plot the temporal evolution of the integrated absolute magnetic field divergence in this test. In contrast to the first two test cases presented in this work, this test is different in that there is no initial magnetic field divergence but the magnetic field divergence is naturally generated by the numerical scheme as the simulation evolves. It is not surprising that the eight-wave (``no GLM'') solution shows the largest divergence error. Again, the $c_h$ based damping leads to an efficient divergence treatment.}

As before, we see very good agreement between the GLM-KEPES and the USM reference solution (cf.~Fig.~\ref{fig:OrszagTangZoom}). Again, the divergence errors contaminate the solution notably in the uncorrected case (``no GLM''), leading to unphysical oscillations in both, density and pressure.

\subsection{2D shock tube test for the effect of the non-conservative terms}\label{Sec:2DShocktube}
This oblique magnetized shock tube was first proposed by Ryu et al. \cite{Ryu1995}. T\'oth used it later to show the failure of the conventional eight-wave scheme to obtain the correct values of the magnetic fields. The formulation he tested used the same non-conservative source term \eqref{eq:3DgeneralizedidealMHD} in the evolution of the ideal MHD equations \cite{Toth2000}. This shock tube is bounded by a left- and a right-going fast shock wave as well as a left- and right-facing slow rarefaction, a right-going slow shock wave, and a contact discontinuity. Thus, it is suitable for testing the correct behavior of a numerical code that is facing a variety of different MHD waves within the same solution.

\begin{figure}[!ht]
	\centering
	\includegraphics[scale=1]{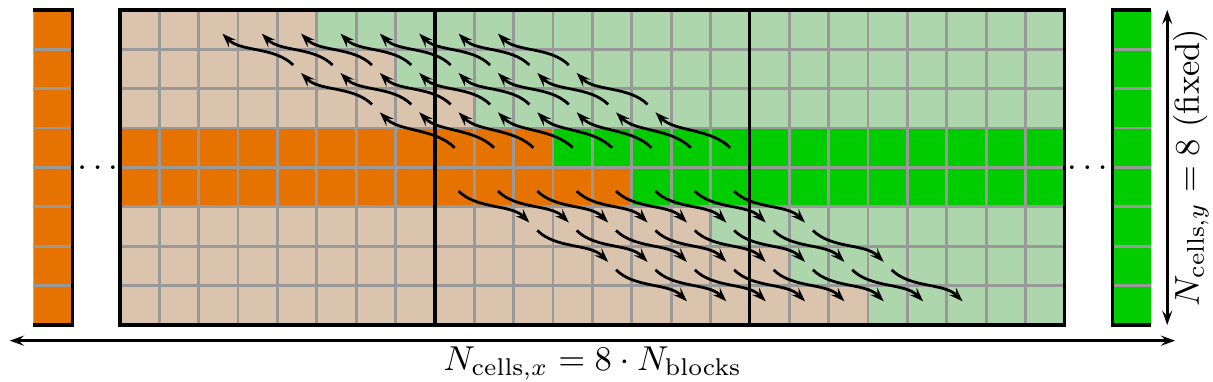}
	\caption{Special boundary conditions for the 2D shock tube test. The outermost blocks are fixed to the initial conditions indicated by orange (left) and green (right) (color online). The inner blocks use special shifted periodic boundary conditions where the values of the inner computation domain (center unshaded cells) are copied to the surrounding cells (shaded cells) as indicated by the arrows. This can be done without changing the results of the computation due to $(-2,1)$ translational symmetry \cite[Sec.~6.3.2.]{Toth2000}.}
	\label{fig:TothBC}
\end{figure}

The shock tube is rotated by an angle of $\theta = \tan^{-1}(2)\approx\SI{63}{\degree}$. Since the magnetic field is initially uniform, the initial conditions trivially fulfill \eqref{eq:divB} for any discretization of the divergence-free condition at any rotation angle. The computational domain is a narrow strip where the top and bottom boundaries are given by ``shifted'' periodic boundary conditions. We depict our realization of these boundary conditions in Fig.~\ref{fig:TothBC}. Note that the realization of the shifted boundary conditions is done by copying the cell values after each solver step according to the specific $(-2,1)$ translational symmetry resulting from the chosen rotation angle. We ensure that the outermost cells still contain the initial conditions by ensuring that the simulation is ended before the shock reaches the boundaries of the computational domain. These specific boundary conditions ensure that effects coming from the boundaries of the computational domains do not influence the flow in our region of interest. Similar to T\'oth, we use a uniform grid with $N_{\mathrm{cells},x} = 256$, which translates into 32 blocks in $x$ and 1 block in $y$ direction in \texttt{FLASH}'s grid configuration. We compare our numerical results to the analytic solution of this test.

\begin{table}[h]
	\centering
	\begin{minipage}[t]{0.49\textwidth}
		\begin{tabular}[t]{|l|c|c|}
			\hline
			 & $x<x_\mathrm{shock}$ & $x>x_\mathrm{shock}$\\
			\hline
			Density $\rho$ & \multicolumn{2}{c|}{$1.0$} \\
			\hline
			Pressure $p$ & $20.0$ & $1.0$ \\
			\hline
			Velocity ${u}_\parallel$ & $10$ & $-10$ \\
			\hline
			\phantom{Velocity} ${u}_\perp$,${w}$ & \multicolumn{2}{c|}{$0$} \\
			\hline
			Mag.~field $[B_\parallel,B_\perp,B_z]$ & \multicolumn{2}{c|}{$(4\pi)^{-0.5}\begin{bmatrix}5 & 5 & 0\end{bmatrix}^\intercal$}\\
			\hline
		\end{tabular}
	\end{minipage}
	\hspace*{3mm}
	\begin{minipage}[t]{0.42\textwidth}
		\begin{tabular}[t]{|l|l|}
			\hline
			Domain size & $x \in [0,1]$ \\
			& $y \in [0,8/N]$ \\
			\hline
			Boundary conditions & see text\\
			\hline
			Shock position & $x_\mathrm{shock}=0.5$\\
			\hline
			Simulation end time & $t_\mathrm{max} = 0.08/\sqrt{5}$\\
			\hline
			Adiabatic index & $\gamma = 5/3$\\
			\hline
		\end{tabular}
	\end{minipage}
	\caption{Initial conditions and runtime parameters: }
	\label{tab:TothTube}
\end{table}

\begin{figure}[!ht]
	\centering
	\includegraphics[scale=1]{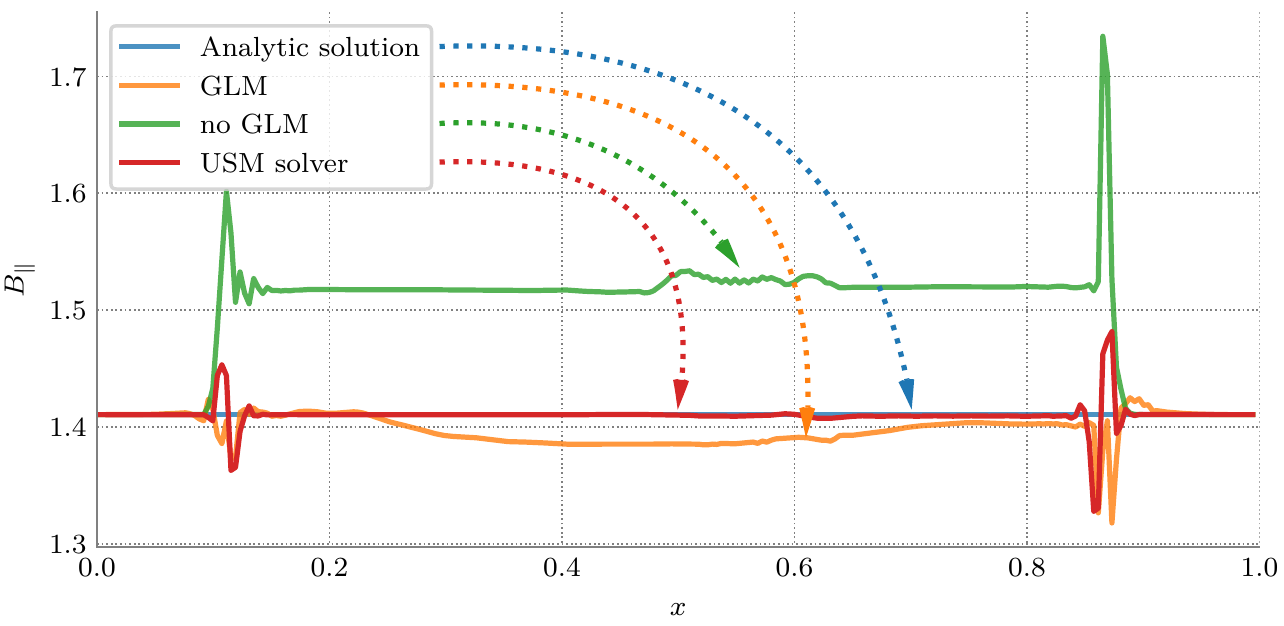}
	\caption{2D shock tube test: Plot of the parallel component of the magnetic field at $t=t_\mathrm{max}$. As can be seen, the non-conservative terms have a notable influence on the perpendicular magnetic field, $B_\perp$. We see that the GLM correction improves the accuracy of the solution significantly. Furthermore, we note that the reduced form of the non-conservative terms (the ``Janhunen'' non-conservative terms) gives similar results as the results obtained using the full non-conservative terms as derived in Sec.~\ref{Sec:idealMHDdivB}.}
	\label{fig:TothPlot}
\end{figure}

The parallel component of the magnetic field, $B_\parallel = B_1 \cos \theta + B_2 \sin \theta$ should be a constant for all time. Note that, we obtain similar errors to those found by T\'oth for {an entropy stable eight-wave scheme without GLM correction} \cite[Fig.~12]{Toth2000}. Also, our baseline scheme yields similar results to what he found earlier for his scheme \cite[Fig.~11]{Toth2000}. If we, however, use our entropy stable GLM treatment at the same time, we see that the errors introduced by the non-conservative terms are reduced significantly (cfm.~Fig.~\ref{fig:TothPlot}). Note that the solutions obtained when using the full vs.~the reduced non-conservative terms are basically indistinguishable.

We chose this test as T\'oth used it specifically to point to a potential weakness of the non-conservative formulation. However, he also points out that this scheme performs well in many other test cases. We see that our entropy-stable scheme which contains both, the eight-wave as well as GLM methods, greatly enhances the solution quality of this specific test case, making the scheme preferable in comparison to a standard eight-wave formulation.

\subsection{Entropy conservation test for the new ideal GLM-MHD system}
The mathematical entropy conservation obtained in Sec. \ref{Sec:EC} is in the semi-discrete sense. That is, the discrete entropy is conserved up to the errors introduced by the temporal approximation. Hence, we can use the error in the conservation of the total entropy with respect to the chosen time step size as a measurement for the temporal discretization error. We use a 3\textsuperscript{rd} order accurate SSP RK time integration scheme as mentioned in Sec.~\ref{sec:high-order-extension}. Hence, we expect the entropy conservation error,
\begin{equation*}
	\Delta S := \left|S(t=0) - S(t=t_\mathrm{end})\right|,
\end{equation*}
to behave like
\begin{equation}
	\Delta S \propto (\Delta t)^3.
\end{equation}

As our test of choice, we run the two-dimensional version of the Brio and Wu magnetohydrodynamical shock tube problem \cite{Brio1988} with a number of different fixed time step lengths $\Delta t$.
This test includes discontinuities, a magnetic field and is performed in multiple dimensions and starts from discontinuous initial conditions. Hence, it utilizes the full set of features of the entropy aware scheme we derived in this work.
We keep the previously used periodic boundary conditions to eliminate any possible influence from the boundaries of the domain and to ensure that we observe a closed system.
We construct the two-dimensional initial conditions by rotating the one-dimensional conditions (see Table~\ref{tab:BrioWu}) at a $45{}^\circ$ angle. The fluid is initially at rest on either side of the interface.

Note that the entropy conserving scheme cannot describe systems with discontinuities as it cannot add the physically needed dissipation.
We limit the end time step to $t_\mathrm{end} = 10^{-3}$ when the oscillations have not grown too large as to cause numerical instabilities. Our sole intention is to show that even under high stresses and with active divergence cleaning our scheme is still capable of conserving the thermodynamic entropy correctly.

\begin{table}[h]
	\centering
	\begin{minipage}[t]{0.32\textwidth}
		\begin{tabular}[t]{l|cc}
			&	{$x < x_\mathrm{shock}$}	& {$x \ge x_\mathrm{shock}$}\\
			\midrule
			$\rho$		& 1	& 0.125	\\
			$p$			& 1	& 0.1 \\
			$\vec{u}$	& $\vec{0}$	& $\vec{0}$ \\
			$B_1$		& 0.75	& 0.75 \\
			$B_2$		& 1	& -1 \\
			$B_3$		& 0	& 0 \\
		\end{tabular}\\[.4em]
	\end{minipage}
	\hspace*{5mm}
	\begin{minipage}[t]{0.47\textwidth}
		\setlength\extrarowheight{3pt}
		\begin{tabular}[t]{|l|l|}
			\hline
			Domain size &$x_\mathrm{min} = 0, x_\mathrm{max} = 1$ \\
			\hline
			Initial shock position &$x_\mathrm{shock} = 0.5$ \\
			\hline
			Boundary conditions & all: periodic\\
			\hline
			Simulation end time & $t_\mathrm{max} = 0.1$ \\
			\hline
			Adiabatic index & $\gamma = 2.0$ \\
			\hline
		\end{tabular}
	\end{minipage}
	\caption{Initial conditions: Brio and Wu MHD shock tube \cite{Brio1988}}
	\label{tab:BrioWu}
\end{table}

\begin{figure}[!h]
	\centering
	\includegraphics[scale=1]{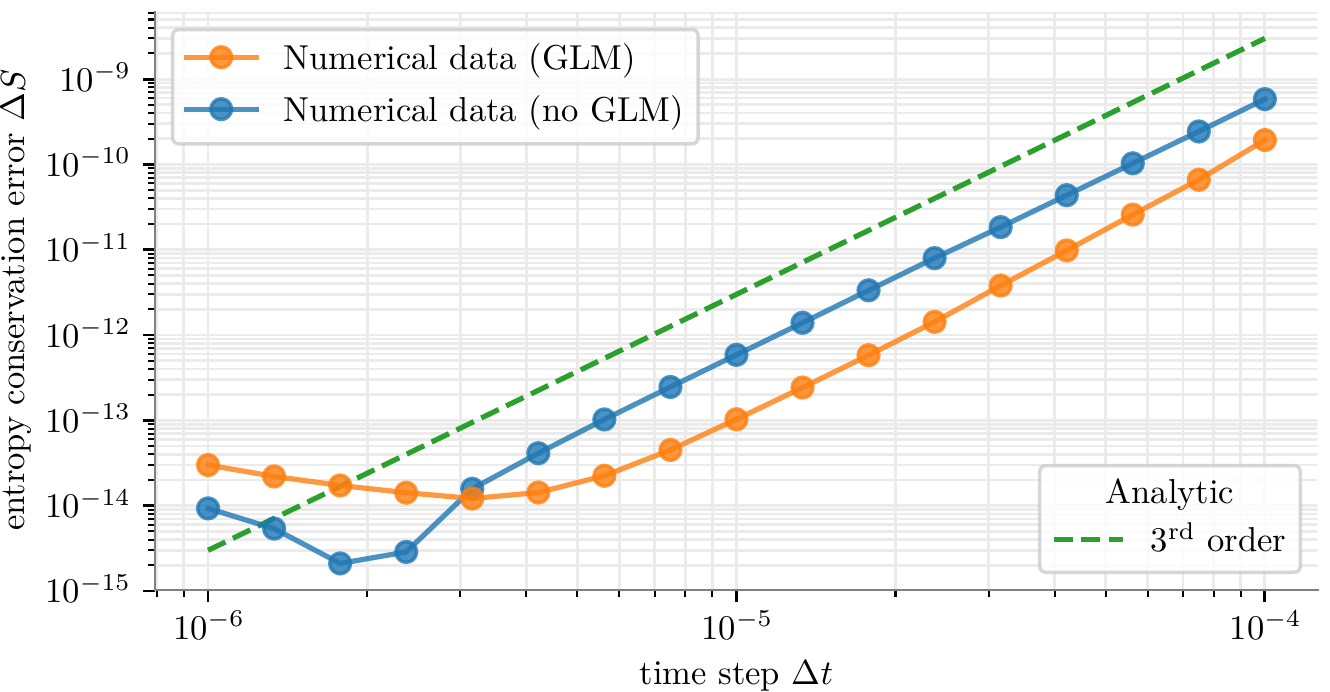}
	\caption{Entropy conservation test, SSP RK3 time integration.}
	\label{fig:ECtest3rd}
\end{figure}

We ran a number of simulations using logarithmically equally spaced time steps and plot the measured entropy conservation error. In Fig.~\ref{fig:ECtest3rd}, we see that for very fine temporal resolution (\ie very small time steps) the approximation tapers around $10^{-14}$. This is caused by the finite precision of the numerical approximation and expected due to floating point arithmetic issues \cite{Lyness1967}. We conclude there exists a natural limit for the accuracy of the entropy conservation in our scheme. This test demonstrates that we are able to successfully construct a numerical scheme that is capable of cleaning the divergence in agreement with the second law of thermodynamics.
\FloatBarrier

\subsection{Entropy consistency test}
In this section, we elaborate on the importance of entropy consistency for the proposed ideal GLM-MHD scheme. Numerical tests suitable for assessing the entropic properties of numerical models have been summarized \eg by Toro \cite[Section 11.5]{Toro2009}. We first look at the ``entropy glitch'' test and later amend this test with magnetic fields.

\begin{figure}[!h]
	\centering
	\includegraphics[scale=1]{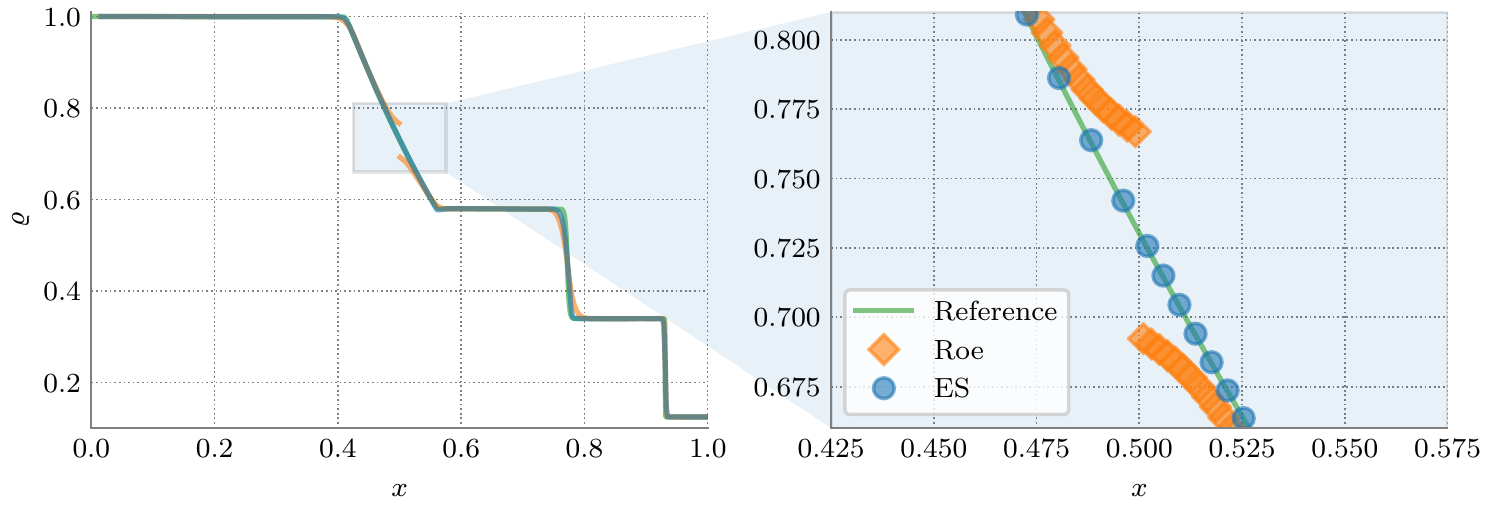}
	\caption{Entropy consistency test (HD variant): Plot of the fluid density, $\varrho$, at $t=t_\mathrm{max}$. On the right panel we show a zoom-in of the plot on the left. The entropy glitch of the original Roe scheme is apparent whereas there is no inconsistency visible in our numerical scheme.}
	\label{fig:Sod}
\end{figure}

Toro describes a modified variant of Sod's well-known shock tube test \cite[Section 3]{Sod1978}, where he added a constant velocity on the left side of the initial shock. We summarize the initial conditions in Table~\ref{tab:Sod}.
The given initial conditions result in a solution that consists of a left sonic rarefaction wave, a contact discontinuity, and a right shock wave.
This test nicely shows the necessity of entropy consistency, as some widely used schemes -- such as the original Roe scheme -- fail to numerically solve this test correctly when compared to the exact solution. The numerics can produce an unphysical jump feature in the rarefaction wave, also known as ``entropy glitch''.
It arises in the presence of sonic rarefaction waves if schemes are not constructed with entropy consistency in mind \cite{Osher1984}. In Fig.~\ref{fig:Sod}, we plot the numerical solution obtained on an adaptive grid of up to $\num{512}$ cells using an implementation of the original Roe scheme in \texttt{FLASH}'s \texttt{USM} solver (parameter \texttt{RiemannSolver = "Roe"}), our numerical entropy stable scheme (denoted by ``ES'') and a highly-resolved ($\num{2048}$ cells, uniform grid) reference solution obtained using an LLF scheme. We immediately see that the numerical solution of the original Roe scheme exhibits a discontinuity within the wave. This discontinuity is not only unphysical, it also violates the entropy condition \cite[Section 11.4.1]{Toro2009}. Our numerical scheme is not affected by this issue and behaves just as expected. Note that we verified the absence of any unphysical discontinuity also for higher and lower resolution. The size of the jump of the entropy glitch reduces for increased resolution but never disappears completely. It is apparent that, using our numerical scheme, the reference solution is matched with a comparably lower adaptive resolution as in Fig.~\ref{fig:Sod}. Although the solution obtained using the Roe scheme fully resolves the region around the ``entropy glitch'', it still fails to obtain the correct physical result.
\begin{table}[!h]
	\centering
	\begin{minipage}[t]{0.32\textwidth}
		\begin{tabular}[t]{l|cc}
			&	{$x < x_\mathrm{shock}$}	& {$x \ge x_\mathrm{shock}$}\\
			\midrule
			$\rho$		& 1	& 0.125	\\
			$p$			& 1	& 0.1 \\
			$u$			& 0.75	& 0 \\
			$v$			& 0	& 0 \\
			$w$			& 0	& 0 \\
			$\vec{B}$	& $\vec{0}$ & $\vec{0}$
		\end{tabular}\\[.4em]
	\end{minipage}
	\hspace*{5mm}
	\begin{minipage}[t]{0.47\textwidth}
		\setlength\extrarowheight{3pt}
		\begin{tabular}[t]{|l|l|}
			\hline
			Domain size &$x_\mathrm{min} = 0, x_\mathrm{max} = 1$ \\
			\hline
			Initial shock position &$x_\mathrm{shock} = 0.5$ \\
			\hline
			Boundary conditions & all: zero-gradient\\
			\hline
			Simulation end time & $t_\mathrm{max} = 0.2$ \\
			\hline
			Adiabatic index & $\gamma = 1.4$ \\
			\hline
		\end{tabular}
	\end{minipage}
	\caption{Initial conditions: Modified Sod shock tube test \cite[Section 11.5.1]{Toro2009}}
	\label{tab:Sod}
\end{table}

\FloatBarrier
For assessing entropy consistency in the presence of magnetic fields, we design a new test. It seems an obvious choice to start the design of such a test from a well known MHD shock tube just as the one described by Brio \& Wu \cite{Brio1988}. We similarly add a constant velocity on the left-hand side where we precompute the velocity of the rarefaction wave in the solution to create a rarefaction wave with a sonic point in this new test case. We summarize the initial conditions proposed for this test in Table~\ref{tab:BrioWu-mod}.

\begin{table}[!h]
	\centering
	\begin{minipage}[t]{0.32\textwidth}
		\begin{tabular}[t]{l|cc}
			&	{$x < x_\mathrm{shock}$}	& {$x \ge x_\mathrm{shock}$}\\
			\midrule
			$\rho$		& 1	& 0.125	\\
			$p$			& 1	& 0.1 \\
			$u$			& 1.75	& 0 \\
			$v$,$w$		& 0	& 0 \\
			$B_1$		& 0.75 & 0.75 \\
			$B_2$		& 1 & -1\\
			$B_3$		& 0 & 0
		\end{tabular}\\[.4em]
	\end{minipage}
	\hspace*{5mm}
	\begin{minipage}[t]{0.47\textwidth}
		\setlength\extrarowheight{3pt}
		\begin{tabular}[t]{|l|l|}
			\hline
			Domain size &$x_\mathrm{min} = 0, x_\mathrm{max} = 1$ \\
			\hline
			Initial shock position &$x_\mathrm{shock} = 0.5$ \\
			\hline
			Boundary conditions & all: zero-gradient\\
			\hline
			Simulation end time & $t_\mathrm{max} = 0.1$ \\
			\hline
			Adiabatic index & $\gamma = 2.0$ \\
			\hline
		\end{tabular}
	\end{minipage}
	\caption{Initial conditions: Modified Brio \& Wu shock tube test}
	\label{tab:BrioWu-mod}
\end{table}

\begin{figure}[!h]
	\centering
	\includegraphics[scale=1]{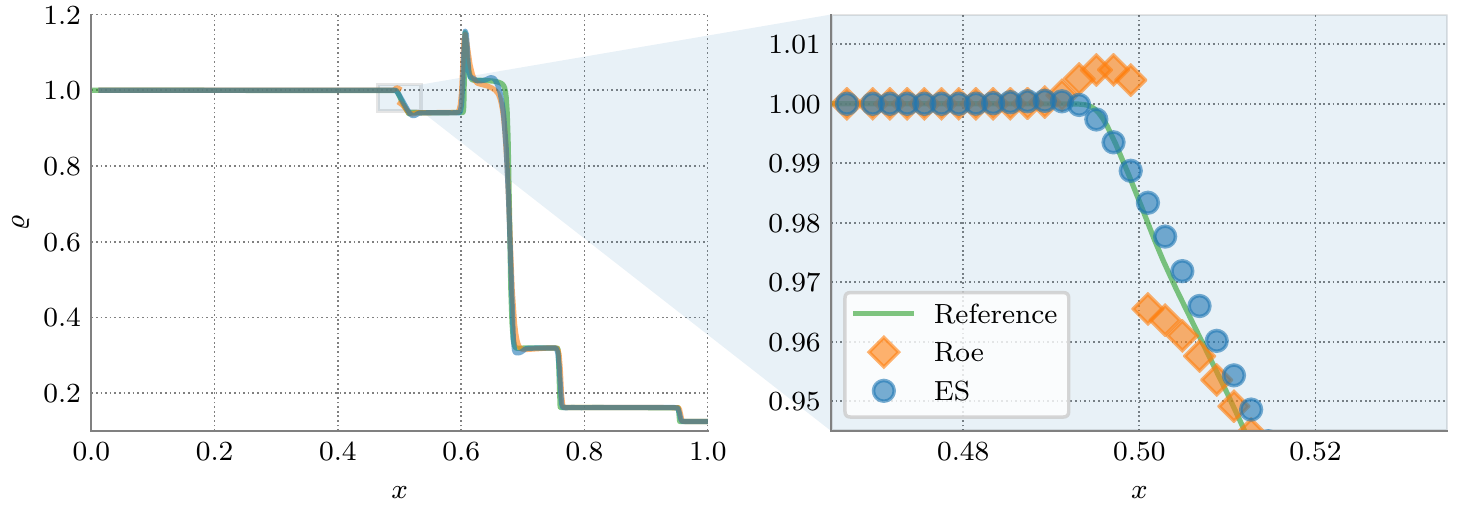}
	\caption{Entropy consistency test (MHD variant): Plot of the fluid density, $\varrho$, at $t=t_\mathrm{max}$. On the right panel we show a zoom-in of the plot on the left. The entropy glitch of the original Roe scheme is apparent whereas there is no inconsistency visible in our numerical scheme.}
	\label{fig:BrioWu-mod}
\end{figure}
In Fig.~\ref{fig:BrioWu-mod}, we plot the numerical solution. We see that, due to the compound wave in this MHD test, the left rarefaction wave is much smaller in the MHD case. Although clearly present, the entropy glitch could be overlooked with too low resolution and misinterpreted as a dissipative/dispersive effect.
As before, we see an unphysical discontinuity in the left rarefaction wave that violates the entropy condition. In contrast, the result obtained using our numerical scheme is free of any entropy violating parts in the solution.
\FloatBarrier

\section{Conclusion}\label{Sec:conclusion}
In this work, we describe a {physically motivated} mathematical model that is suitable for building numerical schemes for ideal MHD flows. We call our system the ``ideal GLM-MHD equations'' in agreement with earlier publications. The significance of our modifications is, as shown in this work, that they lead to the first entropy consistent hyperbolic formulation of the ideal MHD equations with effective inbuilt divergence cleaning. One major benefit of this approach is that divergence cleaning is done alongside the hydrodynamical flux computations so no additional communication or globally coupled computations are introduced when implementing our scheme. This underlines its usability for highly-parallelized numerical simulation codes as it does not only ease the process of parallelization but also allows unaffected scaling of the scheme on over dozens of thousands of computing cores.

We carefully investigate the properties of the proposed mathematical model and discuss the implications of, e.g., the new eigenvalues that come from the GLM waves while we explain how our new model converges to already known models in various limits such as vanishing cleaning speed, vanishing magnetic field divergence and altogether vanishing magnetic fields.

For demonstrating the numerical feasibility of our model, we derive a set of magnetic field divergence diminishing entropy conservative and entropy stable fluxes. Note that these can be used to build entropy stable numerical approximations that respect the fundamental laws of thermodynamics by construction. Our decision to build entropy stable numerical fluxes - in contrast to, \eg HLL fluxes, was due to the fact that entropy stability plays a crucial dual role in ensuring both the physical relevance of the simulation results as well as the numerical robustness of the scheme.

We conclude our analysis and derivations with a section presenting numerical results obtained using the new entropy stable solver implementation for MHD flows in multiple spatial dimensions with adaptive mesh refinement. These numerical tests serve to demonstrate the flexibility of the new solver, the utility as well as the rigor of the implemented divergence diminishing formulation.
We found that additional damping is not strictly needed when dealing with boundary conditions that allow the divergence errors to be advected out of the computational domain, but found the additionally possible $\psi$ wave damping useful when this is not possible, e.g., due to periodic boundary conditions that do not allow the divergence errors to leave the simulation domain.

\section*{Acknowledgements}
We thank the anonymous reviewers for their thorough review, insightful comments and suggestions on this work. In particular we appreciate the detailed comments and discussion about Galilean invariance.

Dominik Derigs and Stefanie Walch acknowledge the support of the Bonn-Cologne Graduate School for Physics and Astronomy (BCGS), which is funded through the Excellence Initiative, as well as the Sonderforschungsbereich (SFB) 956 on the ``Conditions and impact of star formation''.
Stefanie Walch thanks the Deutsche Forschungsgemeinschaft (DFG) for funding through the SPP 1573 ``The physics of the interstellar medium'' and the funding from the European Research Council via the ERC Starting Grant ``The radiative interstellar medium'' (\texttt{RADFEEDBACK}, project no.~679852).
Gregor Gassner thanks the European Research Council for funding through the ERC Starting Grant ``An Exascale aware and Un-crashable Space-Time-Adaptive Discontinuous Spectral Element Solver for Non-Linear Conservation Laws'' (\texttt{Extreme}, project no.~71448).
The software used in this work was in part developed by the DOE NNSA-ASC OASCR Flash Center at the University of Chicago.
This work was partially performed on the Cologne High Efficiency Operating Platform for Sciences (CHEOPS) at the Regionales Rechenzentrum K\"oln (RRZK). The authors gratefully acknowledge the Gauss Centre for Supercomputing e.V.~(www.gauss-centre.eu) for funding this project by providing computing time on the GCS supercomputer SuperMUC at Leibniz Supercomputing Centre (www.lrz.de).

\bibliography{mybibfile}

\begin{thebibliography}{10}
\expandafter\ifx\csname url\endcsname\relax
  \def\url#1{\texttt{#1}}\fi
\expandafter\ifx\csname urlprefix\endcsname\relax\def\urlprefix{URL }\fi
\expandafter\ifx\csname href\endcsname\relax
  \def\href#1#2{#2} \def\path#1{#1}\fi

\bibitem{Brackbill1980}
J.~U. Brackbill, D.~C. Barnes, The effect of nonzero $\nabla \cdot {B}$ on the
  numerical solution of the magnetohydrodynamic equations, Journal of
  Computational Physics 35~(3) (1980) 426 -- 430.
\newblock \href {http://dx.doi.org/10.1016/0021-9991(80)90079-0}
  {\path{doi:10.1016/0021-9991(80)90079-0}}.

\bibitem{Derigs2016}
D.~Derigs, A.~R. Winters, G.~J. Gassner, S.~Walch, {A Novel High-Order, Entropy
  Stable, 3D AMR MHD Solver with Guaranteed Positive Pressure}, Journal of
  Computational Physics 317 (2016) 223--256.
\newblock \href {http://dx.doi.org/10.1016/j.jcp.2016.04.048}
  {\path{doi:10.1016/j.jcp.2016.04.048}}.

\bibitem{Winters2016}
A.~R. Winters, G.~J. Gassner, {Affordable, Entropy Conserving and Entropy
  Stable Flux Functions for the Ideal {MHD} Equations}, Journal of
  Computational Physics 304 (2016) 72--108.
\newblock \href {http://dx.doi.org/10.1016/j.jcp.2015.09.055}
  {\path{doi:10.1016/j.jcp.2015.09.055}}.

\bibitem{Fjordholm2011}
U.~S. Fjordholm, S.~Mishra, E.~Tadmor, Well-blanaced and energy stable schemes
  for the shallow water equations with discontiuous topography, Journal of
  Computational Physics 230~(14) (2011) 5587--5609.
\newblock \href {http://dx.doi.org/10.1016/j.jcp.2011.03.042}
  {\path{doi:10.1016/j.jcp.2011.03.042}}.

\bibitem{Tadmor1987}
E.~Tadmor, The numerical viscosity of entropy stable schemes for systems of
  conservation laws, Mathematics of Computation 49~(179) (1987) 91--103.
\newblock \href {http://dx.doi.org/10.2307/2008251}
  {\path{doi:10.2307/2008251}}.

\bibitem{Godunov1972}
S.~Godunov, Symmetric form of the equations of magnetohydrodynamics, Numerical
  Methods for Mechanics of Continuum Medium 1 (1972) 26--34.

\bibitem{Powell1999}
K.~G. Powell, P.~L. Roe, T.~J. Linde, T.~I. Gombosi, D.~L. De~Zeeuw, A
  solution-adaptive upwind scheme for ideal magnetohydrodynamics, Journal of
  Computational Physics 154~(2) (1999) 284--309.
\newblock \href {http://dx.doi.org/10.1006/jcph.1999.6299}
  {\path{doi:10.1006/jcph.1999.6299}}.

\bibitem{Dirac31}
P.~A.~M. Dirac, {Quantised Singularities in the Electromagnetic Field},
  {Proceedings of the Royal Society of London A: Mathematical, Physical and
  Engineering Sciences} 133~(821) (1931) 60--72.
\newblock \href {http://dx.doi.org/10.1098/rspa.1931.0130}
  {\path{doi:10.1098/rspa.1931.0130}}.

\bibitem{Jackson06}
J.~D. Jackson, C.~Witte, K.~Muller, {Klassische Elektrodynamik (\"Uberarbeitete
  Auflage) (German Edition)}, 4th Edition, Walter {de Gruyter}, 2006.

\bibitem{Moulin01}
F.~{Moulin}, {{M}agnetic monopoles and {L}orentz force}, {Nuovo Cimento B
  Serie} 116 (2001) 869.
\newblock \href {http://arxiv.org/abs/math-ph/0203043}
  {\path{arXiv:math-ph/0203043}}.

\bibitem{Ogilvie2016}
G.~I. {Ogilvie}, {Lecture notes: Astrophysical fluid dynamics}, ArXiv
  e-prints\href {http://arxiv.org/abs/1604.03835} {\path{arXiv:1604.03835}}.

\bibitem{Barth1999}
T.~J. Barth, Numerical methods for gasdynamic systems on unstructured meshes,
  in: D.~Kr\"{o}ner, M.~Ohlberger, C.~Rohde (Eds.), An Introduction to Recent
  Developments in Theory and Numerics for Conservation Laws, Vol.~5 of Lecture
  Notes in Computational Science and Engineering, Springer Berlin Heidelberg,
  1999, pp. 195--285.
\newblock \href {http://dx.doi.org/10.1007/978-3-642-58535-7_5}
  {\path{doi:10.1007/978-3-642-58535-7_5}}.

\bibitem{Janhunen2000}
P.~Janhunen, A positive conservative method for magnetohydrodynamics based on
  {HLL} and {R}oe methods, Journal of Computational Physics 160~(2) (2000)
  649--661.
\newblock \href {http://dx.doi.org/10.1006/jcph.2000.6479}
  {\path{doi:10.1006/jcph.2000.6479}}.

\bibitem{Yee2017}
B.~{Sj{\"o}green}, H.~C. {Yee}, D.~{Kotov}, {Skew-Symmetric Splitting and
  Stability of High Order Central Schemes}, in: Journal of Physics Conference
  Series, Vol. 837, 2017, p. 012019.
\newblock \href {http://dx.doi.org/10.1088/1742-6596/837/1/012019}
  {\path{doi:10.1088/1742-6596/837/1/012019}}.

\bibitem{Marder1987}
B.~Marder, {A method for incorporating Gauss' law into electromagnetic {PIC}
  codes}, Journal of Computational Physics 68~(1) (1987) 48 -- 55.
\newblock \href {http://dx.doi.org/10.1016/0021-9991(87)90043-X}
  {\path{doi:10.1016/0021-9991(87)90043-X}}.

\bibitem{Zachary1994}
A.~L. Zachary, A.~Malagoli, P.~Colella, A higher-order {G}odunov method for
  multidimensional ideal magnetohydrodynamics, SIAM Journal on Scientific
  Computing 15~(2) (1994) 263--284.
\newblock \href {http://dx.doi.org/10.1137/0915019}
  {\path{doi:10.1137/0915019}}.

\bibitem{Balsara1998}
D.~S. Balsara, Total variation diminishing scheme for adiabatic and isothermal
  magnetohydrodynamics, The Astrophysical Journal Supplement Series 116~(1)
  (1998) 133--153.
\newblock \href {http://dx.doi.org/10.1086/313093} {\path{doi:10.1086/313093}}.

\bibitem{Crockett2005}
R.~K. {Crockett}, P.~{Colella}, R.~T. {Fisher}, R.~I. {Klein}, C.~F. {McKee},
  {An unsplit, cell-centered Godunov method for ideal MHD}, Journal of
  Computational Physics 203 (2005) 422--448.
\newblock \href {http://dx.doi.org/10.1016/j.jcp.2004.08.021}
  {\path{doi:10.1016/j.jcp.2004.08.021}}.

\bibitem{Evans1988}
C.~R. {Evans}, J.~F. {Hawley}, {Simulation of magnetohydrodynamic flows - A
  constrained transport method}, Astrophysical Journal 332 (1988) 659--677.
\newblock \href {http://dx.doi.org/10.1086/166684} {\path{doi:10.1086/166684}}.

\bibitem{Balsara1999}
D.~S. Balsara, D.~Spicer, A staggered mesh algorithm using high order {G}odunov
  fluxes to ensure solenoidal magnetic fields in magnetohydrodynamic
  simulations, Journal of Computational Physics 149~(2) (1999) 270--292.
\newblock \href {http://dx.doi.org/10.1006/jcph.1998.6153}
  {\path{doi:10.1006/jcph.1998.6153}}.

\bibitem{Toth2000}
G.~T\'oth, {The $\nabla\cdot {B}=0$ Constraint in Shock-Capturing
  Magnetohydrodynamics Codes}, Journal of Computational Physics 161~(2) (2000)
  605 -- 652.
\newblock \href {http://dx.doi.org/10.1006/jcph.2000.6519}
  {\path{doi:10.1006/jcph.2000.6519}}.

\bibitem{Yee66}
K.~S. Yee, Numerical solution of initial boundary value problems involving
  {M}axwell's equations in isotropic media, IEEE Trans.~Antennas and
  Propagation 14~(3) (1966) 302--307.
\newblock \href {http://dx.doi.org/10.1109/TAP.1966.1138693}
  {\path{doi:10.1109/TAP.1966.1138693}}.

\bibitem{Balsara2004b}
D.~S. {Balsara}, J.~{Kim}, {A Comparison between Divergence-Cleaning and
  Staggered-Mesh Formulations for Numerical Magnetohydrodynamics}, The
  Astrophysical Journal 602 (2004) 1079--1090.
\newblock \href {http://dx.doi.org/10.1086/381051} {\path{doi:10.1086/381051}}.

\bibitem{Waagan2009}
K.~Waagan, A positive {MUSCL}-{H}ancock scheme for ideal magnetohydrodynamics,
  Journal of Computational Physics 228~(23) (2009) 8609--8626.
\newblock \href {http://dx.doi.org/10.1016/j.jcp.2009.08.020}
  {\path{doi:10.1016/j.jcp.2009.08.020}}.

\bibitem{Munz2000}
C.-D. Munz, P.~Omnes, R.~Schneider, E.~Sonnendr{\"u}cker, U.~Vo{\ss},
  {Divergence Correction Techniques for Maxwell Solvers Based on a Hyperbolic
  Model}, Journal of Computational Physics 161~(2) (2000) 484 -- 511.
\newblock \href {http://dx.doi.org/10.1006/jcph.2000.6507}
  {\path{doi:10.1006/jcph.2000.6507}}.

\bibitem{Dedner2002}
A.~Dedner, F.~Kemm, D.~Kr{\"o}ner, C.-D. Munz, T.~Schnitzer, M.~Wesenberg,
  Hyperbolic divergence cleaning for the {MHD} equations, Journal of
  Computational Physics 175~(2) (2002) 645--673.
\newblock \href {http://dx.doi.org/10.1006/jcph.2001.6961}
  {\path{doi:10.1006/jcph.2001.6961}}.

\bibitem{Tricco2012}
T.~S. Tricco, D.~J. Price, Constrained hyperbolic divergence cleaning for
  smoothed particle magnetohydrodynamics, Journal of Computational Physics
  231~(21) (2012) 7214 -- 7236.
\newblock \href {http://dx.doi.org/j.jcp.2012.06.039}
  {\path{doi:j.jcp.2012.06.039}}.

\bibitem{Brio1988}
M.~Brio, C.~C. Wu, An upwind differencing scheme for the equations of ideal
  magnetohydrodynamics, Journal of Computational Physics 75~(2) (1988)
  400--422.
\newblock \href {http://dx.doi.org/10.1016/0021-9991(88)90120-9}
  {\path{doi:10.1016/0021-9991(88)90120-9}}.

\bibitem{Cargo}
P.~Cargo, G.~Gallice, {Roe Matrices for Ideal MHD and Systematic Construction
  of Roe Matrices for Systems of Conservation Laws}, Journal of Computational
  Physics 136~(2) (1997) 446 -- 466.
\newblock \href {http://dx.doi.org/10.1006/jcph.1997.5773}
  {\path{doi:10.1006/jcph.1997.5773}}.

\bibitem{Feng2016}
X.~{Feng}, M.~{Zhang}, {A comparative study of divergence cleaning methods of
  magnetic field in the solar coronal numerical simulation}, Frontiers in
  Astronomy and Space Sciences 3 (2016) 6.
\newblock \href {http://dx.doi.org/10.3389/fspas.2016.00006}
  {\path{doi:10.3389/fspas.2016.00006}}.

\bibitem{Walch2014}
S.~Walch, P.~Girichidis, T.~Naab, A.~Gatto, S.~C.~O. Glover, R.~W{\"u}nsch,
  R.~S. Klessen, P.~C. Clark, T.~Peters, D.~Derigs, C.~Baczynski, {The SILCC
  (SImulating the LifeCycle of molecular Clouds) project -- I. Chemical
  evolution of the supernova-driven ISM}, MNRAS 454~(1) (2015) 238--268.
\newblock \href {http://dx.doi.org/10.1093/mnras/stv1975}
  {\path{doi:10.1093/mnras/stv1975}}.

\bibitem{Bohm2017}
M.~Bohm, R.~W. Winters, D.~Derigs, G.~J. Gassner, S.~Walch, J.~Saur, {An
  entropy stable nodal discontinuous Galerkin method for the resistive MHD
  equations: Continuous analysis and GLM divergence cleaning}, submitted to
  Computers \& Fluids\href {http://arxiv.org/abs/1711.05576}
  {\path{arXiv:1711.05576}}.

\bibitem{Tadmor1984}
E.~Tadmor, Numerical viscosity and the entropy condition for conservative
  difference schemes, Mathematics of Computation 43~(168) (1984) 369--381.
\newblock \href {http://dx.doi.org/10.2307/2008282}
  {\path{doi:10.2307/2008282}}.

\bibitem{barth2006role}
T.~Barth, On the role of involutions in the discontinuous galerkin
  discretization of maxwell and magnetohydrodynamic systems, in: Compatible
  spatial discretizations, Springer, 2006, pp. 69--88.

\bibitem{Domingues2013}
{Domingues, M. O.}, {Gomes, A. K. F.}, {Gomes, S. M.}, {Mendes, O.}, {Di
  Pierro, B.}, {Schneider, K.}, Extended generalized {Lagrangian} multipliers
  for magnetohydrodynamics using adaptive multiresolution methods, ESAIM: Proc.
  43 (2013) 95--107.
\newblock \href {http://dx.doi.org/10.1051/proc/201343006}
  {\path{doi:10.1051/proc/201343006}}.

\bibitem{Mignone2010}
A.~{Mignone}, P.~{Tzeferacos}, G.~{Bodo}, {High-order conservative finite
  difference GLM-MHD schemes for cell-centered MHD}, Journal of Computational
  Physics 229 (2010) 5896--5920.
\newblock \href {http://arxiv.org/abs/1001.2832} {\path{arXiv:1001.2832}},
  \href {http://dx.doi.org/10.1016/j.jcp.2010.04.013}
  {\path{doi:10.1016/j.jcp.2010.04.013}}.

\bibitem{Jiang2012}
R.-L. Jiang, C.~Fang, P.-F. Chen, A new {MHD} code with adaptive mesh
  refinement and parallelization for astrophysics, {Computer Physics
  Communications} 183~(8) (2012) 1617 -- 1633.
\newblock \href {http://arxiv.org/abs/1204.5849} {\path{arXiv:1204.5849}},
  \href {http://dx.doi.org/https://doi.org/10.1016/j.cpc.2012.02.030}
  {\path{doi:https://doi.org/10.1016/j.cpc.2012.02.030}}.

\bibitem{Mackey}
J.~Mackey, A.~J. Lim, Effects of magnetic fields on photoionized pillars and
  globules, Monthly Notices of the Royal Astronomical Society 412~(3) (2011)
  2079--2094.
\newblock \href {http://dx.doi.org/10.1111/j.1365-2966.2010.18043.x}
  {\path{doi:10.1111/j.1365-2966.2010.18043.x}}.

\bibitem{Tricco2016}
T.~S. Tricco, D.~J. Price, M.~R. Bate, Constrained hyperbolic divergence
  cleaning in smoothed particle magnetohydrodynamics with variable cleaning
  speeds, Journal of Computational Physics 322 (2016) 326 -- 344.
\newblock \href {http://dx.doi.org/10.1016/j.jcp.2016.06.053}
  {\path{doi:10.1016/j.jcp.2016.06.053}}.

\bibitem{Chandrashekar2016}
P.~Chandrashekar, C.~Klingenberg, Entropy stable finite volume scheme for ideal
  compressible mhd on 2-d cartesian meshes, SIAM Journal on Numerical Analysis
  54~(2) (2016) 1313--1340.
\newblock \href {http://dx.doi.org/10.1137/15M1013626}
  {\path{doi:10.1137/15M1013626}}.

\bibitem{IsmailRoe2009}
F.~Ismail, P.~L. Roe, Affordable, entropy-consistent {E}uler flux functions
  {II}: Entropy production at shocks, Journal of Computational Physics 228~(15)
  (2009) 5410--5436.
\newblock \href {http://dx.doi.org/10.1016/j.jcp.2009.04.021}
  {\path{doi:10.1016/j.jcp.2009.04.021}}.

\bibitem{Chandrashekar2012}
P.~Chandrashekar, {Kinetic Energy Preserving and Entropy Stable Finite Volume
  Schemes for Compressible Euler and Navier-Stokes Equations}, Communications
  in Computational Physics 14 (2013) 1252--1286.
\newblock \href {http://dx.doi.org/10.4208/cicp.170712.010313a}
  {\path{doi:10.4208/cicp.170712.010313a}}.

\bibitem{Derigs2016_2}
D.~Derigs, A.~R. Winters, G.~J. Gassner, S.~Walch, A novel averaging technique
  for discrete entropy stable dissipation operators for ideal {MHD}, Journal of
  Computational Physics 330 (2016) 624--632.
\newblock \href {http://dx.doi.org/10.1016/j.jcp.2016.10.055}
  {\path{doi:10.1016/j.jcp.2016.10.055}}.

\bibitem{Winters2017}
A.~R. Winters, D.~Derigs, G.~J. Gassner, S.~Walch, A uniquely defined entropy
  stable matrix dissipation operator for high {M}ach number ideal {MHD} and
  compressible {E}uler simulations, {Journal of Computational Physics} 332
  (2017) 274--289.
\newblock \href {http://dx.doi.org/10.1016/j.jcp.2016.12.006}
  {\path{doi:10.1016/j.jcp.2016.12.006}}.

\bibitem{Mishra2011}
S.~Mishra, Entropy stable high-order schemes for systems of conservation laws,
  Modern Techniques in the Numerical Solution of Partial Differential
  Equations.

\bibitem{Rusanov}
V.~V. Rusanov, The calculation of the interaction of non-stationary shock waves
  with barriers, \v Z. Vy\v cisl. Mat. i Mat. Fiz. 1 (1961) 267--279.
\newblock \href {http://dx.doi.org/10.1016/0041-5553(62)90062-9}
  {\path{doi:10.1016/0041-5553(62)90062-9}}.

\bibitem{Wesenberg2003}
M.~Wesenberg, \href{https://www.freidok.uni-freiburg.de/data/792}{Efficient
  finite-volume schemes for magnetohydrodynamic simulations in solar physics},
  dissertation, Universit\"at Freiburg (2003).
\newline\urlprefix\url{https://www.freidok.uni-freiburg.de/data/792}

\bibitem{Gottlieb2001}
S.~Gottlieb, C.-W. Shu, E.~Tadmor, Strong stability-preserving high-order time
  discretization methods, SIAM Review 43~(1) (2001) 89--112.
\newblock \href {http://dx.doi.org/10.1137/S003614450036757X}
  {\path{doi:10.1137/S003614450036757X}}.

\bibitem{DMV}
D.~Derigs, G.~J. Gassner, S.~Walch, R.~W. Winters, {Entropy Stable Finite
  Volume Approximations for Ideal Magnetohydrodynamics}, submitted to
  Jahresberichte der Deutschen Mathematiker-Vereinigung\href
  {http://arxiv.org/abs/1708.03537} {\path{arXiv:1708.03537}}.

\bibitem{LeFloch2002}
P.~G. LeFloch, J.~M. Mercier, C.~Rohde, Fully discrete, entropy conservative
  schemes of arbitraryorder, SIAM Journal on Numerical Analysis 40~(5) (2002)
  1968--1992.
\newblock \href {http://dx.doi.org/10.1137/S003614290240069X}
  {\path{doi:10.1137/S003614290240069X}}.

\bibitem{Fjordholm2012}
U.~S. Fjordholm, S.~Mishra, E.~Tadmor, Arbitrarily high-order accurate entropy
  stable essentially nonoscillatory schemes for systems of conservation laws,
  SIAM Journal on Numerical Analysis 50~(2) (2012) 544--573.
\newblock \href {http://dx.doi.org/10.1137/110836961}
  {\path{doi:10.1137/110836961}}.

\bibitem{Schmidtmann2015}
B.~Schmidtmann, B.~Seibold, M.~Torrilhon, {Relations between {WENO}3 and
  Third-Order Limiting in Finite Volume Methods}, Journal of Scientific
  Computing,{ }\href {http://dx.doi.org/10.1007/s10915-015-0151-z}
  {\path{doi:10.1007/s10915-015-0151-z}}.

\bibitem{Lee2009}
D.~Lee, A.~E. Deane, An unsplit staggered mesh scheme for multidimensional
  magnetohydrodynamics, J. Comput. Phys. 228~(4) (2009) 952--975.
\newblock \href {http://dx.doi.org/10.1016/j.jcp.2008.08.026}
  {\path{doi:10.1016/j.jcp.2008.08.026}}.

\bibitem{Orszag1979}
S.~A. Orszag, C.-M. Tang, Small-scale structure of two-dimensional
  magnetohydrodynamic turbulence, Journal of Fluid Mechanics 90~(01) (1979)
  129--143.
\newblock \href {http://dx.doi.org/10.1017/s002211207900210x}
  {\path{doi:10.1017/s002211207900210x}}.

\bibitem{Balbas2005}
J.~{Balb\'as}, E.~{Tadmor}, {A Central Differencing Simulation of the {O}rszag
  {T}ang Vortex System}, IEEE Transactions on Plasma Science 33 (2005)
  470--471.
\newblock \href {http://dx.doi.org/10.1109/TPS.2005.845282}
  {\path{doi:10.1109/TPS.2005.845282}}.

\bibitem{Dai1998}
W.~Dai, P.~R. Woodward, A simple finite difference scheme for multidimensional
  magnetohydrodynamical equations, Journal of Computational Physics 142~(2)
  (1998) 331 -- 369.
\newblock \href {http://dx.doi.org/10.1006/jcph.1998.5944}
  {\path{doi:10.1006/jcph.1998.5944}}.

\bibitem{Londrillo2000}
P.~{Londrillo}, L.~{Del Zanna}, {High-Order Upwind Schemes for Multidimensional
  Magnetohydrodynamics}, The Astrophysical Journal 530 (2000) 508--524.
\newblock \href {http://arxiv.org/abs/astro-ph/9910086}
  {\path{arXiv:astro-ph/9910086}}, \href {http://dx.doi.org/10.1086/308344}
  {\path{doi:10.1086/308344}}.

\bibitem{Ryu1995}
D.~{Ryu}, T.~W. {Jones}, A.~{Frank}, {{Numerical Magnetohydrodynamics in
  Astrophysics: Algorithm and Tests for Multidimensional Flow}},
  {A}strophysical {J}ournal 452 (1995) 785.
\newblock \href {http://arxiv.org/abs/astro-ph/9505073}
  {\path{arXiv:astro-ph/9505073}}, \href {http://dx.doi.org/10.1086/176347}
  {\path{doi:10.1086/176347}}.

\bibitem{Lyness1967}
J.~N. Lyness, C.~B. Moler, Numerical differentiation of analytic functions,
  SIAM Journal on Numerical Analysis 4~(2) (1967) 202--210.
\newblock \href {http://dx.doi.org/10.1137/0704019}
  {\path{doi:10.1137/0704019}}.

\bibitem{Toro2009}
E.~F. Toro, Riemann Solvers and Numerical Methods for Fluid Dynamics: A
  Practical Introduction, Springer, 2009.

\bibitem{Sod1978}
G.~A. Sod, A survey of several finite difference methods for systems of
  nonlinear hyperbolic conservation laws, {Journal of Computational Physics}
  27~(1) (1978) 1--31.

\bibitem{Osher1984}
S.~Osher, Riemann solvers, the entropy condition, and difference, SIAM Journal
  on Numerical Analysis 21~(2) (1984) 217--235.

\bibitem{roe1996}
P.~L. Roe, D.~S. Balsara, Notes on the eigensystem of magnetohydrodynamics,
  {SIAM} Journal on Applied Mathematics 56~(1) (1996) 57--67.
\newblock \href {http://dx.doi.org/10.1137/S003613999427084X}
  {\path{doi:10.1137/S003613999427084X}}.

\end{thebibliography}

\appendix
\let\oldsection\section
\renewcommand{\section}[1]{\oldsection{#1}\setcounter{figure}{0}\setcounter{equation}{0}\setcounter{table}{0}}

\section{Derivation of the entropy conserving numerical ideal GLM-MHD flux}\label{Sec:ECderivation}
First, we use the properties of the linear jump operator
\begin{equation*}
\jump{ab} = \avg{b}\jump{a} + \avg{a}\jump{b}, \quad \jump{a^2} = 2\avg{a}\jump{a},
\end{equation*}
to expand the jump in entropy variables
\begin{equation}
\jump{\vec{v}} = \resizebox{.885\hsize}{!}{$ %
	\jump{\begin{bmatrix}\frac{\gamma - s}{\gamma - 1}-\beta \lVert\vec{u}\rVert^2 \vphantom{\Big(}
		\\2\beta  u\\2\beta  v\\2\beta  w\\-2\beta \\2\beta B_1\\2\beta B_2\\2\beta B_3\\ 2\beta \psi\end{bmatrix}}
	=
	\begin{bmatrix}
	\frac{\jump{\rho}}{\rholn}+\frac{\jump{\beta}}{\betaln(\gamma-1)}-\Big(\avg{u^2}+\avg{v^2}+\avg{w^2}\Big)\jump{\beta}-2\avg{\beta}\Big( \avg{u}\jump{u} + \avg{v}\jump{v} + \avg{w}\jump{w} \Big) \\
	2 \avg{\beta}\jump{u} + 2 \avg{u}\jump{\beta} \\
	2 \avg{\beta}\jump{v} + 2 \avg{v}\jump{\beta} \\
	2 \avg{\beta}\jump{w} + 2 \avg{w}\jump{\beta} \\
	-2 \jump{\beta} \\
	2 \avg{\beta}\jump{B_1} + 2 \avg{B_1}\jump{\beta} \\
	2 \avg{\beta}\jump{B_2} + 2 \avg{B_2}\jump{\beta} \\
	2 \avg{\beta}\jump{B_3} + 2 \avg{B_3}\jump{\beta} \\
	2 \avg{\beta}\jump{\psi} + 2 \avg{\psi}\jump{\beta} \\
	\end{bmatrix}
	$}.
\end{equation}
With the known jump in entropy variables, we expand the LHS of \eqref{eq:discreteentropy} componentwise to find
{\small
\begin{align}
	\jump{\vec{v}}\cdot\vec{f}^* &= f^*_{1}\left(\frac{\jump{\rho}}{\rholn} + \frac{\jump{\beta}}{\betaln (\gamma-1)} - \left(\avg{u^2} + \avg{v^2} + \avg{w^2}\right) \jump{\beta} - 2\avg{\beta}\Big(\avg{u}\jump{u} + \avg{v}\jump{v} + \avg{w}\jump{w}\Big)\right) \notag \\
	&+\, {f}^*_2 \left( 2 \avg{\beta}\jump{u} + 2 \avg{u}\jump{\beta} \right) + {f}^*_3 \left( 2 \avg{\beta}\jump{v} + 2 \avg{v}\jump{\beta} \right) + {f}^*_4 \left( 2 \avg{\beta}\jump{w} + 2 \avg{w}\jump{\beta} \right) + {f}^*_5 \left( - 2 \jump{\beta}\right) \notag \\
	&+\, {f}^*_6 \left(2 \avg{\beta}\jump{B_1} + 2 \avg{B_1}\jump{\beta} \right) +  {f}^*_7 \left( 2 \avg{\beta}\jump{B_2} + 2 \avg{B_2}\jump{\beta} \right) + {f}^*_8 \left( 2 \avg{\beta}\jump{B_3} + 2 \avg{B_3}\jump{\beta} \right) \notag \\
	&+\, {f}^*_9 \left(2\avg{\beta}\jump{\psi} + 2\avg{\psi}\jump{\beta} \right), \label{eq:LHS}
\end{align}}%
where we introduce the logarithmic mean $\avgln{(\cdot)} = \frac{\jump{\cdot}}{\jump{\ln(\cdot)}}$. A numerically stable procedure to compute the logarithmic mean is described by Ismail and Roe \cite[Appendix B]{IsmailRoe2009}. In the algorithm to compute $\avgln{(\cdot)}$ we chose $\epsilon = \num{1e-3}$ to increase the accuracy of the entropy conservative approximation to close to machine precision.

Next, we expand the individual components on the RHS of \eqref{eq:discreteentropy} into combinations of linear jumps
\begin{align}
	\jump{\rho u} &= \avg{u}\jump{\rho} + \avg{\rho}\jump{u}, \label{eq:RHS1}\\
	\jump{\beta u \lVert\vec{B}^2\rVert} &= \jump{\beta u B_1^2} + \jump{\beta u B_2^2} + \jump{\beta u B_3^2} \notag \\
	&= \avg{u B_1^2} \jump{\beta} + \avg{\beta}\big(\avg{B_1^2} \jump{u}+ 2\avg{u} \avg{B_1} \jump{B_1} \big)  \notag\\
	\quad &\qquad+ \avg{u B_2^2} \jump{\beta} + \avg{\beta}\big(\avg{B_2^2} \jump{u} + 2\avg{u} \avg{B_2} \jump{B_2} \big)  \notag\\
	\quad &\qquad+ \avg{u B_3^2} \jump{\beta} + \avg{\beta}\big(\avg{B_3^2} \jump{u} + 2\avg{u} \avg{B_3} \jump{B_3} \big) \notag\\
	&= \jump{\beta} \big( \avg{u B_1^2} + \avg{u B_2^2} + \avg{u B_3^2} \big) \notag\\
	\quad&\qquad+ \jump{u} \big( \avg{\beta} \avg{B_1^2} + \avg{\beta} \avg{B_2^2} + \avg{\beta} \avg{B_3^2} \big) \notag\\
	\quad&\qquad+ \jump{B_1} \big( 2\avg{\beta} \avg{u} \avg{B_1} \big) + \jump{B_2} \big( 2\avg{\beta} \avg{u} \avg{B_2} \big) + \jump{B_3} \big( 2\avg{\beta} \avg{u} \avg{B_3} \big),
	\label{eq:RHS2}\\
	\jump{\beta (\vec{u}\cdot\vec{B})} &= \jump{\beta u B_1} + \jump{\beta v B_2} + \jump{\beta w B_3} \notag \\
	&=\jump{\beta}\left( \avg{u B_1} + \avg{v B_2} + \avg{w B_3} \right) + \jump{u} \avg{\beta} \avg{B_1} + \jump{v} \avg{\beta} \avg{B_2} + \jump{w} \avg{\beta} \avg{B_3} \notag\\
	\quad&\qquad+ \jump{B_1} \avg{\beta} \avg{u} + \jump{B_2} \avg{\beta} \avg{v} + \jump{B_3} \avg{\beta} \avg{w},
	\label{eq:RHS3}\\
	\shortintertext{and}
	\jump{\beta B_1 \psi} &= \avg{\beta}\jump{B_1 \psi} + \avg{B_1 \psi}\jump{\beta} = \avg{\beta} \big( \avg{B_1} \jump{\psi} + \avg{\psi} \jump{B_1} \big) + \avg{B_1 \psi}\jump{\beta} \notag\\
	&= \jump{\psi} \avg{\beta} \avg{B_1} + \jump{B_1} \avg{\beta} \avg{\psi} + \jump{\beta} \avg{B_1 \psi}. \label{eq:RHS4}
\end{align}

After rewriting every term in the discrete entropy conservation equation \eqref{eq:discreteentropy} into linear jumps, we obtain the yet unknown components of the entropy conserving ideal GLM-MHD flux function:
\begin{subequations}
\begin{align}
	\jump{\rho} &: f_1^* \frac{\jump{\rho}}{\rholn} = \avg{u} \jump{\rho} \label{eq:jumprho}\\
	\jump{u} &: -2 f_1^* \avg{\beta} \avg{u} \jump{u} + 2 f_2^* \avg{\beta}\jump{u} = \avg{\rho}\jump{u} + \avg{\beta}\big(\avg{B_1^2} + \avg{B_2^2} + \avg{B_3^2}\big)\jump{u} -  \notag\\
	&\quad - 2 \avg{\beta}\avg{B_1}^2\jump{u} \label{eq:jumpu}\\
	\jump{v} &: -2 f_1^* \avg{\beta} \avg{v} \jump{v} + 2 f_3^* \avg{\beta}\jump{v} = - 2 \avg{\beta} \avg{B_1}\avg{B_2}\jump{v} \label{eq:jumpv}\\
	\jump{w} &: -2 f_1^* \avg{\beta} \avg{w} \jump{w} + 2 f_4^* \avg{\beta}\jump{w} = - 2 \avg{\beta} \avg{B_1}\avg{B_3}\jump{w} \label{eq:jumpw}\\
	\jump{B_1} &: 2 f_6^* \avg{\beta} \jump{B_1} = 2 \avg{\beta} \avg{B_1} \avg{u} \jump{B_1} - 2\avg{\beta}\avg{B_1} \avg{u}\jump{B_1} + 2 c_h \avg{\beta} \avg{\psi} \jump{B_1} \notag\\
	&\quad= 2\avg{\beta} c_h \avg{\psi} \jump{B_1} \label{eq:jumpB1}\\
	\jump{B_2} &: 2 f_7^* \avg{\beta} \jump{B_2} = 2 \avg{\beta} \avg{B_2} \avg{u} \jump{B_2} - 2 \avg{\beta} \avg{B_1} \avg{v} \jump{B_2} \label{eq:jumpB2}\\
	\jump{B_3} &: 2 f_8^* \avg{\beta} \jump{B_3} = 2 \avg{\beta} \avg{B_3} \avg{u} \jump{B_3} - 2 \avg{\beta} \avg{B_1} \avg{w} \jump{B_2} \label{eq:jumpB3}\\
	\jump{\psi} &: 2 f_9^* \avg{\beta} = 2 c_h \avg{\beta} \avg{B_1} \label{eq:jumppsi}\\
	\jump{\beta} &: \frac{f_1^*}{\betaln (\gamma-1)} \jump{\beta} - f_1^* \left(\avg{u^2} + \avg{v^2} + \avg{w^2}\right) \jump{\beta} + 2 f_2^* \avg{u} \jump{\beta} + 2 f_3^* \avg{v} \jump{\beta} \notag\\
	&\qquad + 2 f_4^* \avg{w} \jump{\beta} -2 f_5^* \jump{\beta} + 2 f_6^* \avg{B_1} \jump{\beta} + 2 f^*_7 \avg{B_2} \jump{\beta} + 2 f_8^* \avg{B_3} \jump{\beta} + 2 f_9^* \avg{\psi}\jump{\beta} \notag\\
	&\quad = \big(\avg{u B_1^2}+\avg{u B_2^2}+\avg{u B_3^2}\big)\jump{\beta} -2\avg{B_1} \big(\avg{u B_1}+\avg{v B_2}+\avg{w B_3}\big)\jump{\beta} \notag\\
	&\qquad + 2 c_h \avg{B_1 \psi} \jump{\beta} \label{eq:jumpbeta}
\end{align}
\end{subequations}

Solving \eqref{eq:jumprho} -- \eqref{eq:jumpbeta} gives the numerical entropy conserving flux function $\vec{f}^*$:
\begin{subequations}
\begin{align}
	f_1^* &= \rholn \avg{u} \label{eq:flux1}\\
	f_2^* &= f_1^* \avg{u} - \avg{B_1}^2  + \frac{\avg{\rho}}{2\avg{\beta}} + \frac{1}{2} \Big(\avg{B_1^2} + \avg{B_2^2} + \avg{B_3^2}\Big)\\
	f_3^* &= f_1^* \avg{v} - \avg{B_1}\avg{B_2}\\
	f_4^* &= f_1^* \avg{w} - \avg{B_1}\avg{B_3}\\
	f_5^* &= f_1^*\bigg[\frac{1}{2 (\gamma-1) \betaln} - \frac{1}{2} \left(\avg{u^2} + \avg{v^2} + \avg{w^2}\right) \bigg] + f_2^* \avg{u} + f_3^* \avg{v} + f_4^* \avg{w}  \notag\\
	&\quad + f_6^* \avg{B_1} + f_7^* \avg{B_2} + f_8^* \avg{B_3} + f_9^* \avg{\psi} - \frac{1}{2} \big(\avg{u B_1^2}+\avg{u B_2^2}+\avg{u B_3^2}\big) \notag\\
	&\quad + \avg{B_1}\left(\avg{u B_1}+\avg{v B_2}+\avg{w B_3}\right) - c_h \avg{B_1 \psi}. \label{eq:flux5}\\
	f_6^* &= c_h \avg{\psi}\\
	f_7^* &= \avg{u}\avg{B_2} - \avg{v}\avg{B_1}\\
	f_8^* &= \avg{u}\avg{B_3} - \avg{w}\avg{B_1}\\
	f_9^* &= c_h \avg{B_1} \label{eq:flux9}
\end{align}
\end{subequations}

\section{Discrete version of the entropy Jacobian}\label{app:entropyJacobian}
The entries of the matrix $\He{}$ are derived step-by-step through the solution of 81 equations in a similar fashion as done in \cite{Derigs2016_2} for the unmodified ideal MHD equations (64 equations):
\begin{equation}\label{eq:tobesolved}
	\jump{\vec{q}} = \jump{\begin{bmatrix}\rho\\\rho u\\\rho v\\\rho w\\E\\B_1\\B_2\\B_3\\\psi\end{bmatrix}} \stackrel{!}{=} \begin{bmatrix}
	\He_{1,1} & \He_{1,2} & \dots & \dots & \dots & \He_{1,8} & \He_{1,9} \\
	\He_{2,1} & \He_{2,2} & \dots & \dots & \dots & \He_{2,8} & \He_{2,9} \\
	\vdots  & \vdots & \ddots & \ddots & \ddots & \vdots & \vdots \\
	\vdots  & \vdots & \ddots & \ddots & \ddots & \vdots & \vdots \\
	\vdots  & \vdots & \ddots & \ddots & \ddots & \vdots & \vdots \\
	\He_{8,1} & \He_{8,2} & \dots & \dots & \dots & \He_{8,8} & \He_{9,9} \\
	\He_{9,1} & \He_{9,2} & \dots & \dots & \dots & \He_{9,8} & \He_{9,9} \\
	\end{bmatrix}
	\jump{\begin{bmatrix}\frac{\gamma - s}{\gamma - 1}-\beta \lVert\vec{u}\rVert^2\\2\beta  u\\2\beta  v\\2\beta  w\\-2\beta \\2\beta B_1\\2\beta B_2\\2\beta B_3\\2\beta \psi\end{bmatrix}}
	= \He{}\jump{\vec{v}}.
\end{equation}

First, we expand the jump in both the conservative and the entropy variables
\begin{align}
\jump{\vec{q}} &= \jump{\begin{bmatrix}\rho \\ \rho u \\ \rho v \\ \rho w \\ E \\ B_1 \\ B_2 \\ B_3 \\ \psi \\\end{bmatrix}} \notag\\
	&=\resizebox{.88\hsize}{!}{$
	\begin{bmatrix}
	\jump{\rho} \\ \avg{\rho} \jump{u} + \avg{u} \jump{\rho} \\ \avg{\rho} \jump{v} + \avg{v} \jump{\rho}\\ \avg{\rho} \jump{w} + \avg{w} \jump{\rho} \\ \big(\frac{1}{2(\gamma-1)\betaln} + \frac{1}{2}\uavg\big)\jump{\rho}+\avg{\rho}\big(\avg{u}\jump{u} + \avg{v}\jump{v} + \avg{w}\jump{w}\big) - \frac{\avg{\rho}}{2(\gamma-1)\betaavg}\jump{\beta} + \avg{B_1}\jump{B_1} + \avg{B_2}\jump{B_2} + \avg{B_3}\jump{B_3} + \avg{\psi} \jump{\psi}\\ \jump{B_1} \\ \jump{B_2} \\ \jump{B_3} \\ \jump{\psi} \\
	\end{bmatrix}$}
	\ ,\\
\jump{\vec{v}} &=
	\jump{\begin{bmatrix}\frac{\gamma - s}{\gamma - 1}-\beta \lVert\vec{u}\rVert^2\\2\beta  u\\2\beta  v\\2\beta  w\\-2\beta \\2\beta B_1\\2\beta B_2\\2\beta B_3\\2\beta \psi\end{bmatrix}} \notag\\
	&=\resizebox{.88\hsize}{!}{$
	\begin{bmatrix}
	\frac{\jump{\rho}}{\rholn}+\frac{\jump{\beta}}{\betaln(\gamma-1)}-\big(\avg{u^2}+\avg{v^2}+\avg{w^2}\big)\jump{\beta}-2\avg{\beta}\big( \avg{u}\jump{u} + \avg{v}\jump{v} + \avg{w}\jump{w} \big) \\
	2 \avg{\beta}\jump{u} + 2 \avg{u}\jump{\beta} \\
	2 \avg{\beta}\jump{v} + 2 \avg{v}\jump{\beta} \\
	2 \avg{\beta}\jump{w} + 2 \avg{w}\jump{\beta} \\
	-2 \jump{\beta} \\
	2 \avg{\beta}\jump{B_1} + 2 \avg{B_1}\jump{\beta} \\
	2 \avg{\beta}\jump{B_2} + 2 \avg{B_2}\jump{\beta} \\
	2 \avg{\beta}\jump{B_3} + 2 \avg{B_3}\jump{\beta} \\
	2 \avg{\beta}\jump{\psi} + 2 \avg{\psi}\jump{\beta} \\
	\end{bmatrix}$}
\ ,
\end{align}
with
\begin{equation}
	\betaavg = 2 \avg{\beta}^2 - \avg{\beta^2},\quad \pln = \frac{\rholn}{2 \betaln}, \quad\mbox{and} \quad \uavg = 2\left(\avg{u}^2 + \avg{v}^2 + \avg{w}^2\right)-\left(\avg{u^2} + \avg{v^2} + \avg{w^2}\right).
\end{equation}
Note that the jump in $E$ contains the specific modifications found in \cite{Derigs2016_2} to allow the derivation of a symmetric dissipation matrix.

\section{Discrete eigenvalues of the ideal GLM-MHD system}\label{App:discreteEigenvalues}
\newcommand{\dA}{\ensuremath{\hat{\mat{A}}}}
First, we transform the system into primitive variables as the analysis in conservative variables proved to be highly complicated.
This is straightforward because we can swap between variable spaces with the matrix
\begin{align}\label{eq:app:M}
	\mat{M} = \pderivative{\vec{q}}{\vec{\omega}} = 
	&\begin{bmatrix} 1 & 0 & 0 & 0 & 0 & 0 & 0 & 0 & 0
	\\[0.1cm] u & \rho & 0 & 0 & 0 & 0 & 0 & 0 & 0 \\[0.1cm] v & 0 & \rho & 0 & 0 & 0
	& 0 & 0 & 0 \\[0.1cm] w & 0 & 0 & \rho & 0 & 0 & 0 & 0 & 0 \\[0.1cm] \frac{1}{2}\|\vec{u}\|^2 & \rho\,u & \rho\,v & \rho\,w & \frac{1}{\gamma-1} & 
	{B_1} & {B_2} & {B_3} & \psi \\[0.1cm] 0 & 0 & 0 & 0 & 0 & 1 & 
	0 & 0 & 0 \\[0.1cm] 0 & 0 & 0 & 0 & 0 & 0 & 1 & 0 & 0 \\[0.1cm] 0 & 0 & 0 & 0 & 0
	& 0 & 0 & 1 & 0 \\[0.1cm] 0 & 0 & 0 & 0 & 0 & 0 & 0 & 0 & 1 \\[0.1cm] 
	\end{bmatrix},
	\shortintertext{and its inverse which can be used to go back to conservative variables}
	\mat{M}^{-1} = \pderivative{\vec{\omega}}{\vec{q}} =
	&\resizebox{.8\hsize}{!}{$ %
		\begin{bmatrix} 1 & 0 & 0 & 0 & 0 & 0 & 0 & 0 & 0
		\\ -\frac{u}{\rho} & \frac{1}{\rho} & 0 & 0 & 0 & 0 & 0 & 0 & 0 \\
		-\frac{v}{\rho} & 0 & \frac{1}{\rho} & 0 & 0 & 0 & 0 & 0 & 0 \\ -
		\frac{w}{\rho} & 0 & 0 & \frac{1}{\rho} & 0 & 0 & 0 & 0 & 0 \\ \frac{\gamma-1}{2}\|\vec{u}\|^2
		& u\left(1-\gamma\right) & v\left(1-\gamma\right) & w\left(1-
		\gamma\right) & \gamma-1 & { B_1}\left(1-\gamma\right) &
		{ B_2}\left(1-\gamma\right) & { B_3}\left(1-\gamma\right)
		& \psi\left(1-\gamma\right) \\ 0 & 0 & 0 & 0 & 0 & 1 & 0 & 0 & 0
		\\ 0 & 0 & 0 & 0 & 0 & 0 & 1 & 0 & 0 \\ 0 & 0 & 0 & 0 & 0 & 0 & 0
		& 1 & 0 \\ 0 & 0 & 0 & 0 & 0 & 0 & 0 & 0 & 1 \\ \end{bmatrix}$},
\end{align}
where $\vec{\xi}$ is the vector of primitive variables,
$\vec{\xi} := \begin{bmatrix}\rho & u & v & w & p & B_1 & B_2 & B_3 & \psi \end{bmatrix}^\intercal$.

The equation system
\begin{align}
	\pderivative{}{t}\vec{q} + \mat{A}_\mat{\Upsilon}\pderivative{}{x} \vec{q} &= \vec{0}\label{eq:app:dis:1}\\
	\shortintertext{can now be rewritten as}
	\mat{M}\pderivative{}{t} \vec{\xi} + \mat{A}_\mat{\Upsilon}\mat{M} \pderivative{}{x} \vec{\xi} &=0,\notag\\
	\pderivative{}{t} \vec{\xi} + \mat{M}^{-1}\mat{A}_\mat{\Upsilon}\mat{M} \pderivative{}{x} \vec{\xi} &= 0, \notag\\
	\pderivative{}{t} \vec{\xi} + \mat{C} \pderivative{}{x} \vec{\xi} &= 0.\label{eq:app:dis:2}
\end{align}

Comparing \eqref{eq:app:dis:1} and \eqref{eq:app:dis:2} we find that the matrix $\mat{A}_\mat{\Upsilon}$ can be transformed to primitive space to obtain the primitive matrix $\mat{C} :=  \mat{M}^{-1} \mat{A}_\mat{\Upsilon} \mat{M}$
\begin{align}
	\label{eq:matrixB}
	\mat{C} = 
	&\begin{bmatrix}
	u & \rho & 0 & 0 & 0 & 0 & 0 & 0 & 0\\[0.1cm]
	0 & u & 0 & 0 & \frac{1}{\rho} & 0 & \frac{{B_2}}{\rho} & \frac{{B_3}}{\rho} & 0 \\[0.1cm]
	0 & 0 & u & 0 & 0 & 0 & -\frac{{B_1}}{\rho} & 0 & 0 \\[0.1cm]
	0 & 0 & 0 & u & 0 & 0 & 0 & -\frac{{B_1}}{\rho} & 0 \\[0.1cm]
	0 & \gamma\,p & 0 & 0 & u & 0 & 0 & 0 & 0\\[0.1cm]
	0 & 0 & 0 & 0 & 0 & u & 0 & 0 & {c_h} \\[0.1cm]
	0 & {B_2} & - {B_1} & 0 & 0 & 0 & u & 0 & 0 \\[0.1cm]
	0 & {B_3} & 0 & -{B_1} & 0 & 0 & 0 & u & 0 \\[0.1cm]
	0 & 0 & 0 & 0 & 0 & {c_h} & 0 & 0 & u \\[0.1cm] 
	\end{bmatrix}.
\end{align}
The matrices $\mat{A}_\mat{\Upsilon}$ and $\mat{C}$ are \emph{similar}, \ie they have the same eigenvalues but not necessarily the same eigenvectors\footnote{The eigenvectors are transformed according to the base changing matrix $\mat{M}$.}.

From
\begin{align}
	\pderivative{}{t} \vec{\xi} + \mat{C} \pderivative{}{x} \vec{\xi} &= \pderivative{}{t} \vec{\xi} +
	\begin{bmatrix}
	u & \rho & 0 & 0 & 0 & 0 & 0 & 0 & 0\\
	0 & u & 0 & 0 & \frac{1}{\rho} & 0 & \frac{{B_2}}{\rho} & \frac{{B_3}}{\rho} & 0 \\
	0 & 0 & u & 0 & 0 & 0 & -\frac{{B_1}}{\rho} & 0 & 0 \\
	0 & 0 & 0 & u & 0 & 0 & 0 & -\frac{{B_1}}{\rho} & 0 \\
	0 & \gamma\,p & 0 & 0 & u & 0 & 0 & 0 & 0\\
	0 & 0 & 0 & 0 & 0 & u & 0 & 0 & {c_h} \\
	0 & {B_2} & - {B_1} & 0 & 0 & 0 & u & 0 & 0 \\
	0 & {B_3} & 0 & -{B_1} & 0 & 0 & 0 & u & 0 \\
	0 & 0 & 0 & 0 & 0 & {c_h} & 0 & 0 & u \\ 
	\end{bmatrix}
	\begin{bmatrix} \rho \\ u \vphantom{\frac{B_2}{\rho}} \\ v \vphantom{\frac{B_2}{\rho}} \\ w \vphantom{\frac{B_2}{\rho}} \\ p \\ B_1 \\ B_2 \\ B_3 \\ \psi \end{bmatrix}_x\\
	&= \pderivative{}{t} \vec{\xi} +
	\begin{bmatrix}
	u (\rho)_x + \rho (u)_x \\
	u (u)_x + \frac{1}{\rho}(p)_x + \frac{B_2}{\rho} (B_2)_x + \frac{B_3}{\rho} (B_3)_x \\[1mm]
	u v_x - \frac{B_1}{\rho} (B_2)_x\\[1mm]
	u w_x - \frac{B_1}{\rho} (B_3)_x\\[1mm]
	\gamma p (u)_x + u (p)_x\\
	u (B_1)_x + c_h (\psi)_x\\
	B_2 (u)_x - B_1 (v)_x + u (B_2)_x\\
	B_3 (u)_x - B_1 (w)_x + u (B_3)_x\\
	c_h (B_1)_x + u (\psi)_x
	\end{bmatrix} = \vec{0},
\end{align}
we see that we cannot bring the system into flux form for primitive variables. Hence, we refrain from calling $\mat{C}$ the primitive flux Jacobian, since it is not possible to bring the system described by
\begin{equation}
	\pderivative{}{t} \vec{\xi} + \mat{C} \pderivative{}{x} \vec{\xi} = \vec{0},
\end{equation}
into flux form for primitive variables. Therefore, unlike the matrix $\mat{A}$ in flux form, the coefficient matrix $\mat{C}$ in primitive variable form is not the flux Jacobian of \emph{any} flux function $\vec{f}(\vec{\xi})$. {We can easily see that, as the entries on the diagonal of $\mat{C}$ are all equal to $u$, and they are the only ones depending on $u$, the matrix $\mat{C}$ describes a Galiean invariant scheme \cite[Section 3]{Dedner2002}, as expected.}

In the next step, we make the discrete ansatz,  where we discretize the update of the primitive variables in the spatial dimension like
\newcommand{\dx}{\ensuremath{\,\mathrm{d}x}}
\newcommand{\dC}{\ensuremath{\hat{\mat{C}}}}
\begin{align}
	-\pderivative{}{t}\vec{\xi} &= \mat{C}\pderivative{}{x} \vec{\xi} \\
	&= \int_{L}^{R} \left( \mat{C}\pderivative{}{x} \vec{\xi}\right) \dx \label{eq:app:int}\\
	&\approx \frac{\Delta x}{2} \sum_{k=\{\L,\R\}}\left(\mat{C} \pderivative{}{x}\vec{\xi}\right)_k \label{eq:app:int1}\\
	&= \frac{\Delta x}{2} \left(\mat{C}_\L \frac{\vec{\xi}_\R - \vec{\xi}_\L}{\Delta x} + \mat{C}_\R \frac{\vec{\xi}_\R - \vec{\xi}_\L}{\Delta x}\right)\\
	&= \frac{1}{2} (\mat{C}_\L + \mat{C}_\R) (\vec{\xi}_\R - \vec{\omega}_\L)\\
	&= \avg{\mat{C}} \jump{\vec{\omega}},
\end{align}
where we used the trapezoidal rule for approximating the integral on the RHS of \eqref{eq:app:int} using the left and right states.
We immediately see that the discretized version of the coefficient matrix, $\mat{C}$, is the continuous coefficient matrix, $\mat{C}$, arithmetically averaged in each entry, $\dC := \avg{\mat{C}} = \frac{1}{2} \left(\mat{C}_\L + \mat{C}_\R\right)$:
\begin{equation}
	\dC =
	\begin{bmatrix}
	\avg{u} & \avg{\rho} & 0 & 0 & 0 & 0 & 0 & 0 & 0 \\[0.1cm]
	0 & \avg{u} & 0 & 0 & \avg{\rho^{-1}} & 0 & 
	\avg{\frac{B_2}{\rho}} & \avg{\frac{B_3}{\rho}} & 0 \\[0.175cm] 0 & 0 & 
	\avg{u} & 0 & 0 & 0 & -\avg{\frac{B_1}{\rho}}
	& 0 & 0 \\[0.175cm] 0 & 0 & 0 & \avg{u} & 0 & 0 & 0 & 
	-\avg{\frac{B_1}{\rho}} & 0 \\[0.1cm] 0 &\gamma \avg{p} & 0 & 0 & \avg{u} & 0 & 0 & 0 & 0 \\[0.1cm] 0 & 0 & 0 & 0 & 0 & 
	\avg{u} & 0 & 0 & \avg{c_h} \\[0.1cm] 0 & \avg{B_2} & -\avg{B_1} & 0 & 0 & 0 & 
	\avg{u} & 0 & 0 \\[0.1cm] 0 & \avg{B_3} & 0 & -\avg{B_1} & 0 & 0 & 0 & 
	\avg{u} & 0 \\[0.1cm] 0 & 0 & 0 & 0 & 0 & \avg{c_h} & 0 & 0 & \avg{u}
	\end{bmatrix}.
\end{equation}

The eigenvalues of $\dC$ are:
\begin{equation}\label{eq:supercomplicateddiscreteidealGLMMHDeigenvaluespressure}
\hat{\vec{\lambda}} = \resizebox{0.92\textwidth}{!}{$
	\begin{bmatrix}
	\avg{u}-\frac{\sqrt{2\,\sqrt{\avg{B_1}\,
				\avg{\frac{B_1}{\rho}}\,\avg{p}\,\avg{\rho^{-1}}\,\gamma}+\avg{p}
			\,\avg{\rho^{-1}}\,\gamma+\avg{B_3}\,\avg{\frac{B_3}{\rho}}+
			\avg{B_2}\,\avg{\frac{B_2}{\rho}}+\avg{B_1}\,\avg{\frac{B_1}{\rho}}}
		+\sqrt{-2\,\sqrt{\avg{B_1}\,\avg{\frac{B_1}{\rho}}\,\avg{p}\,
				\avg{\rho^{-1}}\,\gamma}+\avg{p}\,\avg{\rho^{-1}}\,\gamma+
			\avg{B_3}\,\avg{\frac{B_3}{\rho}}+\avg{B_2}\,\avg{\frac{B_2}{\rho}}+
			\avg{B_1}\,\avg{\frac{B_1}{\rho}}}}{2}\\
\avg{u}-\frac{\sqrt{2\,\sqrt{\avg{B_1}\,
			\avg{\frac{B_1}{\rho}}\,\avg{p}\,\avg{\rho^{-1}}\,\gamma}+\avg{p}
		\,\avg{\rho^{-1}}\,\gamma+\avg{B_3}\,\avg{\frac{B_3}{\rho}}+
		\avg{B_2}\,\avg{\frac{B_2}{\rho}}+\avg{B_1}\,\avg{\frac{B_1}{\rho}}}
	-\sqrt{-2\,\sqrt{\avg{B_1}\,\avg{\frac{B_1}{\rho}}\,\avg{p}\,
			\avg{\rho^{-1}}\,\gamma}+\avg{p}\,\avg{\rho^{-1}}\,\gamma+
		\avg{B_3}\,\avg{\frac{B_3}{\rho}}+\avg{B_2}\,\avg{\frac{B_2}{\rho}}+
		\avg{B_1}\,\avg{\frac{B_1}{\rho}}}}{2}\\
\avg{u}+\frac{\sqrt{2\,\sqrt{\avg{B_1}\,
				\avg{\frac{B_1}{\rho}}\,\avg{p}\,\avg{\rho^{-1}}\,\gamma}+\avg{p}
			\,\avg{\rho^{-1}}\,\gamma+\avg{B_3}\,\avg{\frac{B_3}{\rho}}+
			\avg{B_2}\,\avg{\frac{B_2}{\rho}}+\avg{B_1}\,\avg{\frac{B_1}{\rho}}}
		-\sqrt{-2\,\sqrt{\avg{B_1}\,\avg{\frac{B_1}{\rho}}\,\avg{p}\,
				\avg{\rho^{-1}}\,\gamma}+\avg{p}\,\avg{\rho^{-1}}\,\gamma+
			\avg{B_3}\,\avg{\frac{B_3}{\rho}}+\avg{B_2}\,\avg{\frac{B_2}{\rho}}+
			\avg{B_1}\,\avg{\frac{B_1}{\rho}}}}{2}\\
	\avg{u}+\frac{\sqrt{2\,\sqrt{\avg{B_1}\,
				\avg{\frac{B_1}{\rho}}\,\avg{p}\,\avg{\rho^{-1}}\,\gamma}+\avg{p}
			\,\avg{\rho^{-1}}\,\gamma+\avg{B_3}\,\avg{\frac{B_3}{\rho}}+
			\avg{B_2}\,\avg{\frac{B_2}{\rho}}+\avg{B_1}\,\avg{\frac{B_1}{\rho}}}
		+\sqrt{-2\,\sqrt{\avg{B_1}\,\avg{\frac{B_1}{\rho}}\,\avg{p}\,
				\avg{\rho^{-1}}\,\gamma}+\avg{p}\,\avg{\rho^{-1}}\,\gamma+
			\avg{B_3}\,\avg{\frac{B_3}{\rho}}+\avg{B_2}\,\avg{\frac{B_2}{\rho}}+
			\avg{B_1}\,\avg{\frac{B_1}{\rho}}}}{2}\\
	\avg{u}-\sqrt{\avg{B_1}\,\avg{\frac{B_1}{\rho}}}\\
	\avg{u}+\sqrt{\avg{B_1}\,\avg{\frac{B_1}{\rho}}}\\
	\avg{u}-\avg{c_h}\\
	\avg{u}+\avg{c_h}\\
	\avg{u}
	\end{bmatrix}$}
\end{equation}

After many manipulations, we find a greatly simplified form of the discrete eigenvalues using the discrete wave speeds
\begin{equation}\label{eq:app:characteristicSpeedsDiscrete}
	\hat{c}_\a = |\hat{b}_1|, \qquad
	\hat{c}_{\f,\s} = \frac{1}{2}\left( \sqrt{\hat{a}^2 + \hat{b}^2 + 2\sqrt{\hat{a}^2 \hat{b}_1^2}} \pm \sqrt{\hat{a}^2 + \hat{b}^2 - 2\sqrt{\hat{a}^2 \hat{b}_1^2}}\right),
\end{equation}
with the special discrete averages
\begin{equation}
	\hat{\vec{b}}^2 = \avg{\vec{B}} \cdot \avg{\frac{\vec{B}}{\rho}},\quad
	\hat{a}^2 = \gamma \avg{p}\avg{\rho^{-1}},\quad
	\hat{b}^2=\hat{b}_1^2+\hat{b}_2^2+\hat{b}_3^2.
\end{equation}
In \eqref{eq:app:characteristicSpeedsDiscrete}, the plus sign corresponds to the fast magnetoacoustic speed, $c_\f$, and the minus sign corresponds to the slow magnetoacoustic speed, $c_\s$.

The simplified eigenvalues of the ideal GLM-MHD system are:
\begin{equation}
	\hat{\vec{\lambda}} =
	\begin{bmatrix}
	\hat{\lambda}_{+\f}\vphantom{\hat{\lambda}_{+\f}}\\
	\hat{\lambda}_{+\a}\vphantom{\avg{u}+\hat{c}_\a}\\
	\hat{\lambda}_{+\s}\vphantom{\avg{u}+\hat{c}_\s}\\
	\hat{\lambda}_{+\psi}\vphantom{\avg{u}+\hat{c}_h}\\
	\hat{\lambda}_\E\vphantom{\avg{u}}\\
	\hat{\lambda}_{-\psi}\vphantom{\avg{u}+\hat{c}_h}\\
	\hat{\lambda}_{-\s}\vphantom{\avg{u}-\hat{c}_\s}\\
	\hat{\lambda}_{-\a}\vphantom{\avg{u}-\hat{c}_\a}\\
	\hat{\lambda}_{-\f}\vphantom{\avg{u}-\hat{c}_\f}
	\end{bmatrix} =
	\begin{bmatrix}
	\avg{u}+\hat{c}_\f\vphantom{\hat{\lambda}_{-\f}}\\
	\avg{u}+\hat{c}_\a\vphantom{\hat{\lambda}_{-\f}}\\
	\avg{u}+\hat{c}_\s\vphantom{\hat{\lambda}_{-\f}}\\
	\avg{u}+\avg{c_h}\\
	\avg{u}\\
	\avg{u}-\avg{c_h}\\
	\avg{u}-\hat{c}_\s\vphantom{\hat{\lambda}_{-\f}}\\
	\avg{u}-\hat{c}_\a\vphantom{\hat{\lambda}_{-\f}}\\
	\avg{u}-\hat{c}_\f\vphantom{\hat{\lambda}_{-\f}}
	\end{bmatrix}\quad
	\begin{matrix*}[l]
	\mbox{right going fast magnetoacoustic wave}\vphantom{\hat{\lambda}_{-\f}}\\
	\mbox{right going Alfv\'en wave}\vphantom{\hat{\lambda}_{-\f}}\\
	\mbox{right going slow magnetoacoustic wave}\vphantom{\hat{\lambda}_{-\f}}\\
	\mbox{right going GLM wave}\vphantom{\vphantom{\avg{u}+\hat{c}_h}}\\
	\mbox{entropy wave}\vphantom{\avg{u}}\\
	\mbox{left going GLM wave}\vphantom{\vphantom{\avg{u}+\hat{c}_h}}\\
	\mbox{left going slow magnetoacoustic wave}\vphantom{\hat{\lambda}_{-\f}}\\
	\mbox{left going Alfv\'en wave}\vphantom{\hat{\lambda}_{-\f}}\\
	\mbox{left going fast magnetoacoustic wave}\vphantom{\hat{\lambda}_{-\f}}
	\end{matrix*}
\end{equation}

\section{Eigenstructure}\label{sec:eigenstructure}
We outline the steps to obtain the eigenstructure of the flux Jacobian for the new ideal GLM-MHD system. For this one-dimensional analysis we forgo the addition of the matrix superscript $(\cdot)^x$ for the sake of convenience. We have already computed the flux Jacobian with the , $\mat{A}_\mat{\Upsilon}$ \eqref{eq:AP_matrix}, for the new system. For an entropy stable numerical flux with a matrix dissipation term we require a relationship between the entropy Jacobian, $\mat{H}$, and the right eigenvectors, $\mat{R}$.

The eigendecomposition of the matrix $\mat{A}_\mat{\Upsilon}$ supports nine propagating plane-wave solutions:
\begin{itemize}
	\item two fast magnetoacoustic waves ($\rm\pm f$),
	\item two slow magnetoacoustic waves ($\rm\pm s$),
	\item two Alfv\'{e}n waves ($\rm\pm a$),
	\item an entropy wave ($\rm E$),
	\item two GLM waves ($\pm \psi$).
\end{itemize}
It is known that the right eigenvectors may exhibit several forms of degeneracy that are carefully described by Roe and Balsara \cite{roe1996}. We follow the same rescaling procedure of Roe and Balsara to improve the numerical behavior of the fast/slow magnetoacoustic eigenvectors.

To compute the eigenvectors it is more convenient to work with primitive variables, $\vec{\omega}$, and then convert back to conservative space as noted in \ref{App:discreteEigenvalues}. Once we know the eigenvectors in primitive space, $\mat{R}_{\mat{C}}$ we return to conservative space by
\begin{equation}
       \mat{R} \mat{\Lambda} \mat{R}^{-1} = {\mat{A_P}} = \mat{M}{\mat{C}}\mat{M}^{-1} = (\mat{M}{\mat{R_{C}}}){\boldsymbol{\Lambda}}(\mat{M}{\mat{R_{C}}})^{-1}.
\end{equation}
with the eigenvalue matrix $\mat{\Lambda}$.
The matrix of right eigenvectors is given by
\begin{equation}\label{eq:RVs}
{\mat{R}}\coloneqq\mat{M}{\mat{R_{C}}} = \left[\,{\vec{r}}_{+{\f}} \,|\, {\vec{r}}_{+{\a}} \,|\, {\vec{r}}_{+{\s}} \,|\,\vec{r}_{+\psi} \,|\,{\vec{r}}_{\E} \,|\, {\vec{r}}_{-\psi} \,|\, {\vec{r}}_{-{\s}} \,|\, {\vec{r}}_{-{\a}} \,|\, {\vec{r}}_{-{\f}} \, \right],
\end{equation}
with the eigenvectors ${\vec{r}}$:
\begin{itemize}
\item[] \underline{GLM Waves}: $\lambda_{  \pm\psi} = {u\pm c_h}$, \underline{Entropy Wave}: $\lambda_{\E} = u$, and \underline{Alfv\'{e}n Waves}: $\lambda_{\pm \a} = u\pm c_\a$
\begin{equation}\label{DivAS1}
	{\vec{r}}_{\pm\psi} = \begin{bmatrix}
	 0 \\
	 0 \\
	 0 \\
	 0 \\
	 B_1\pm\psi \\[0.05cm]
	 1 \\
	 0 \\
	 0 \\
	 \pm 1\\
	 \end{bmatrix},
	\qquad
	{\vec{r}}_{\E} = \begin{bmatrix} 1 \\ u \\v \\w \\ \frac{\|\vec{u}\|^2}{2} \\[0.05cm]0 \\0 \\0 \\0\\\end{bmatrix}, \quad
	{\vec{r}}_{\pm \a} = \begin{bmatrix}
	0 \\
	0 \\
	\pm \rho^{\frac{3}{2}}\,\chi_3 \\
	\mp \rho^{\frac{3}{2}}\,\chi_2 \\
	\mp \rho^{\frac{3}{2}}(\chi_2 w - \chi_3 v) \\
	0 \\
	-\rho \chi_3 \\
	\rho \chi_2\\[0.1cm]
	0\\[0.1cm]
	\end{bmatrix},
\end{equation}
\item[] \underline{Magnetoacoustic Waves}: $\lambda_{\pm \f,\pm \s} = u\pm c_{\f,\s}$
\begin{equation}\label{MHDAS1}
	{\vec{r}}_{\pm \f} = \begin{bmatrix}
	\alpha_{\f}\rho \\[0.1cm]
	\alpha_{\f}\rho(u \pm c_{\f}) \\[0.1cm]
	\rho\left(\alpha_{\f} v \mp \alpha_{\s} c_{\s} \chi_2 \sigma(b_1) \right) \\[0.1cm]
	\rho\left(\alpha_{\f} w \mp \alpha_{\s} c_{\s} \chi_3 \sigma(b_1) \right) \\[0.1cm]
	\Psi_{\pm \f} \\[0.1cm]
	0 \\[0.1cm]
	\alpha_{\s} a \chi_2 \sqrt{\rho} \\[0.1cm]
	\alpha_{\s} a \chi_3 \sqrt{\rho} \\[0.1cm]
	0\\[0.1cm]
	\end{bmatrix},
	\qquad
	{\vec{r}}_{\pm\s} = \begin{bmatrix}
	\alpha_{\s}\rho \\[0.1cm]
	\alpha_{\s}\rho\left(u \pm c_{ s}\right) \\[0.1cm]
	\rho\left(\alpha_{\s} v \pm \alpha_{\f} c_{\f} \chi_2 \sigma(b_1)\right) \\[0.1cm]
	\rho\left(\alpha_{\s} w \pm \alpha_{\f} c_{\f} \chi_3 \sigma(b_1)\right) \\[0.1cm]
	\Psi_{\pm\s} \\[0.1cm]
	0 \\[0.1cm]
	-\alpha_{\f} a \chi_2 \sqrt{\rho} \\[0.1cm]
	-\alpha_{\f} a \chi_3 \sqrt{\rho}\\[0.1cm]
	0\\[0.1cm]
	\end{bmatrix},
\end{equation}
\end{itemize}
where we introduced several convenience variables
\begin{equation}\label{eq:alotofequations1}
	\begin{aligned}
	\Psi_{\pm{\s}} &= \frac{\alpha_{\s} \rho \lVert\vec{u}\rVert^2}{2} - a \alpha_{\f} \rho b_\perp + \frac{\alpha_{\s} \rho a^2}{\gamma-1} \pm \alpha_{\s} c_{\s} \rho u \pm \alpha_{\f} c_{\f} \rho \sigma(b_1) (v \chi_2 + w \chi_3),\quad b^2 = b_1^2 + b_2^2 + b_3^2, \\
	\Psi_{\pm{\f}} &= \frac{\alpha_{\f} \rho \lVert\vec{u}\rVert^2}{2} + a \alpha_{\s} \rho b_\perp + \frac{\alpha_{\f} \rho a^2}{\gamma-1} \pm \alpha_{\f} c_{\f} \rho u \mp \alpha_{\s} c_{\s} \rho \sigma(b_1) (v \chi_2 + w \chi_3),\quad b_\perp^2 = b_2^2 + b_3^2, \\
	c_{\a}^2& = b_1^2, \quad c_{\f,\s}^2 = \frac{1}{2}\left((a^2+b^2) \pm \sqrt{(a^2+b^2)^2 - 4a^2 b_1^2}\right), \quad a^2 = \gamma \, \frac{1}{2\beta}, \quad \vec{b}^2 = \frac{\vec{B}^2}{\rho},\\
	\alpha_{\f}^2 &= \frac{a^2 - c_{\s}^2}{c_{\f}^2 - c_{\s}^2}, \quad \alpha_{\s}^2 = \frac{c_{\f}^2 -a^2}{c_{\f}^2 - c_{\s}^2},\quad\chi_{1,2,3} = \frac{b_{1,2,3}}{b_\perp}, \quad 
	\sigma(\omega) = \begin{cases}
	+1 &\mbox{if } \omega \ge 0, \\
	-1 &\text{otherwise}
	\end{cases}.
	\end{aligned}
\end{equation}

\section{Derivation of the total energy equation}\label{app:thegorydetails}
The total energy equation \eqref{eq:Etresult} is obtained as described by \eqref{eq:Et}:
\begin{equation}
	\pderivative{E}{t} = \pderivative{}{t}\Bigg(\frac{1}{2}\rho\|\vec{u}\|^2 + \epsilon + \frac{1}{2}\|\vec{B}\|^2\Bigg).
\end{equation}

For now, we compute the contributions of the momentum and induction equation intentionally without the non-conservative terms to avoid confusion:

\begin{enumerate}
	\item Kinetic energy \emph{without} non-conservative term on the momentum equation
	\begin{align}
	\pderivative{}{t}\Bigg(\frac{1}{2}\rho\|\vec{u}\|^2\Bigg) &= \pderivative{}{t}\Bigg(\frac{1}{2}\frac{(\rho u)^2}{\rho} + \frac{1}{2}\frac{(\rho v)^2}{\rho} + \frac{1}{2}\frac{(\rho w)^2}{\rho} \Bigg) \notag\\
	&= u (\rho u)_t + v (\rho v)_t + w (\rho w)_t - \frac{1}{2} \|\vec{u}\|^2 (\rho)_t \notag\\
	&= \underbrace{-\frac{1}{2} u^2 \left( 3 \rho (u)_x + u (\rho)_x \right)}_{-\frac{1}{2}(\rho u^2)_x} - u(p)_x - \frac{1}{2} u \left(- (B_1)_x^2 + (B_2)_x^2 + (B_3)_x^2\right)\notag\\
	&\quad \underbrace{- \frac{1}{2} \left( 2\rho u v (v)_x +  \rho v^2 (u)_x + (\rho)_x u v^2 \right)}_{-\frac{1}{2}(\rho u v)_x} \underbrace{- \frac{1}{2}\left( 2\rho u w (w)_x +  \rho w^2 (u)_x + (\rho)_x u w^2 \right)}_{-\frac{1}{2}(\rho u w)_x}\notag\\
	&\quad + \underbrace{v B_1 (B_2)_x + v ( B_1)_x B_2}_{v (B_1 B_2)_x} + \underbrace{w B_1 (B_2)_x + w ( B_1)_x B_2}_{w (B_1 B_3)_x} \notag\\
	&= -\left\{u\left( \frac{1}{2} \rho \|\vec{u}\|^2\right)\right\}_x -u (p)_x - \frac{1}{2} u \left(- (B_1)_x^2 + (B_2)_x^2 + (B_3)_x^2\right) \notag\\
	&\quad+ v(B_1 B_2)_x + w (B_1 B_3)_x
	\end{align}
	\item Internal energy
	\begin{align}
	\pderivative{}{t} \epsilon = \frac{1}{\gamma-1}\pderivative{p}{t} = \frac{-1}{\gamma-1}\left(u (p)_x + \gamma p (u)_x\right)
	\end{align}
	\item Magnetic energy \emph{without} non-conservative term on the induction equation
	\begin{align}
	\pderivative{}{t}\bigg(\frac{1}{2}\|\vec{B}\|^2\bigg) &= \vec{B} \cdot \bigg(\pderivative{\vec{B}}{t}\bigg) = \begin{bmatrix}
	B_1\\B_2\\B_3
	\end{bmatrix}\cdot\left(-\pderivative{}{x}
	\begin{bmatrix}0 \\ u B_2 - v B_1 \\ u B_3 - w B_1\end{bmatrix}\right)\notag\\
	&= - B_2 (u B_2)_x + B_2 (v B_1)_x - B_3 (u B_3)_x + B_3 (w B_1)_x
	\end{align}
\end{enumerate}

Summing them all up, we obtain (we color code the individual contributions for the sake of readability)
\begin{align}
	&\mathcolor{ForestGreen}{\pderivative{}{t}\bigg(\frac{1}{2}\rho\|\vec{u}\|^2\bigg)} + \mathcolor{blue}{\pderivative{\epsilon}{t}} + \mathcolor{red}{\pderivative{}{t}\bigg(\frac{1}{2}\|\vec{B}\|^2\bigg)}\\
	&= \mathcolor{ForestGreen}{-\left\{u\left( \frac{1}{2} \rho \|\vec{u}\|^2\right)\right\}_x -u (p)_x- \frac{1}{2} u \left(- (B_1)_x^2 + (B_2)_x^2 + (B_3)_x^2\right) + v(B_1 B_2)_x + w (B_1 B_3)_x}\notag\\
	&\quad \mathcolor{blue}{- \frac{1}{\gamma-1}\left(u (p)_x + \gamma p (u)_x\right)}
	\mathcolor{red}{- B_2 (u B_2)_x + B_2 (v B_1)_x - B_3 (u B_3)_x + B_3 (w B_1)_x}\\
	&=-\left\{u\left( \frac{1}{2} \rho \|\vec{u}\|^2\right)\right\}_x \underbrace{-\frac{{u(p)_x + \gamma p (u)_x} + {\gamma u (p)_x - u (p)_x}}{\gamma-1}}_{-\frac{\gamma}{\gamma-1}(u p)_x}\notag\\
	&\quad + \mathcolor{orange}{u B_1 (B_1)_x} \underbrace{- B_2^2 (u)_x - u (B_2^2)_x}_{-(uB_2^2)_x} \underbrace{- B_3^2 (u)_x - u (B_3^2)_x}_{-(uB_3^2)_x}\notag\\
	&\quad + \underbrace{v B_1 (B_2)_x + 2 v B_2 (B_1)_x + B_1 B_2 (v)_x}_{(vB_1B_2)_x + \mathcolor{orange}{v B_2(B_1)_x}} + \underbrace{w B_1 (B_3)_x + 2 w B_3 (B_1)_x + B_1 B_3 (w)_x}_{(wB_1B_3)_x + \mathcolor{orange}{w B_3(B_1)_x}}
\end{align}
This means the total energy conservation law (using the induction equation and the given momentum conservation law \emph{without} the non-conservative terms) is
\begin{equation}
	\pderivative{E}{t} + \pderivative{}{x} \left( u\left(\frac{1}{2}\rho\|\vec{u}\|^2 + \frac{\gamma p}{\gamma-1} + \|\vec{B}\|^2\right) - \vec{B}(\vec{u}\cdot\vec{B}) \right) =\ \mathcolor{orange}{\stackrel{\downarrow}{+} (B_1)_x (\vec{u}\cdot\vec{B})}.
\end{equation}

As can be seen from these derivations, we obtained a non-conservative term contributing in the total energy equation \emph{although we started off from equations that do not contain non-conservative term}. It is this very specific (positive) contribution that cancels with one of the two (negative) $(B_1)_x (\vec{u} \cdot \vec{B})$ terms coming from the momentum and induction equations with non-conservative term:
\begin{enumerate}
	\item Kinetic energy \emph{with} non-conservative term
	\begin{align}
	\pderivative{}{t}\Bigg(\frac{1}{2}\rho\|\vec{u}\|^2\Bigg)
	&= u (\rho u)_t + v (\rho v)_t + w (\rho w)_t - \frac{1}{2} \|\vec{u}\|^2 (\rho)_t \notag\\
	&= -\left\{u\left( \frac{1}{2} \rho \|\vec{u}\|^2\right)\right\}_x -u (p)_x - \frac{1}{2} u \left(- (B_1)_x^2 + (B_2)_x^2 + (B_3)_x^2\right)\notag\\ &\quad+ v(B_1 B_2)_x + w (B_1 B_3)_x
	\mathcolor{orange}{-(B_1)_x(\vec{u}\cdot\vec{B})}
	\end{align}
	\item Magnetic energy \emph{with} non-conservative term
	\begin{align}
	\pderivative{}{t}\bigg(\frac{1}{2}\|\vec{B}\|^2\bigg) &= \vec{B} \cdot \bigg(\pderivative{\vec{B}}{t}\bigg) = \begin{bmatrix}B_1\\B_2\\B_3\end{bmatrix}
	\cdot\left(-\pderivative{}{x} \begin{bmatrix}0 \\ u B_2 - v B_1 \\ u B_3 - w B_1\end{bmatrix} \mathcolor{orange}{- (B_1)_x\begin{bmatrix}u\\v\\w\end{bmatrix}}\right)\notag\\
	&= - B_2 (u B_2)_x + B_2 (v B_1)_x - B_3 (u B_3)_x + B_3 (w B_1)_x \mathcolor{orange}{-(B_1)_x(\vec{B}\cdot\vec{u})}
	\end{align}
\end{enumerate}
Using the kinetic and magnetic energy contributions with non-conservative term as derived in Section~\ref{Sec:idealMHDdivB} results in the shown equation \eqref{eq:Etresult}:
\begin{equation}
\pderivative{E}{t} + \pderivative{}{x} \left( u\left(\frac{1}{2}\rho\|\vec{u}\|^2 + \frac{\gamma p}{\gamma-1} + \|\vec{B}\|^2\right) - \vec{B}(\vec{u}\cdot\vec{B}) \right) =\ \mathcolor{orange}{\stackrel{\downarrow}{-} (B_1)_x (\vec{u}\cdot\vec{B})}.
\end{equation}

\section{ideal GLM-MHD equations in $y$ and $z$-direction}\label{app:twoandthreedims}
For completeness, we summarize the ideal GLM-MHD equations in the two and three-dimensional case below:
\begin{align}
	\frac{\partial}{\partial t} \vec{q} &+ \frac{\partial}{\partial x} \vec{f}^x + \vec{\Upsilon} = \vec{0}, &&\text{in 1D}\\
	\frac{\partial}{\partial t} \vec{q} &+ \frac{\partial}{\partial x} \vec{f}^x + \frac{\partial}{\partial y} \vec{f}^y + \vec{\Upsilon} = \vec{0}, &&\text{in 2D}\\
	\frac{\partial}{\partial t} \vec{q} &+ \frac{\partial}{\partial x} \vec{f}^x + \frac{\partial}{\partial y} \vec{f}^y + \frac{\partial}{\partial z} \vec{f}^z + \vec{\Upsilon} = \vec{0}, &&\text{in 3D}
\end{align}
where $\vec{f}^{x,y,z}(\vec{q})$ are the flux vectors in $x$, $y$, and $z$-direction, and $\vec{\Upsilon}$ is the non-conservative term.
\begin{align}
	\vec{f}^x &=
	\begin{bmatrix}\rho \, u \\ \rho u^2 + p + \frac{1}{2} \lVert \vec{B} \rVert^2 -B_1 B_1 \\ \rho \, u \, v - B_1  B_2 \\ \rho \, u \, w - B_1  B_3 \\ u \hat{E} - B_1 \big(\vec{u} \cdot \vec{B}\big) + c_h \psi B_1 \\ c_h \psi \\ u\, B_2 - v \, B_1 \\ u \, B_3 - w\, B_1 \\ c_h B_1 \end{bmatrix},\qquad \vec{f}^y =
	\begin{bmatrix}\rho \, v \\ \rho \, v \, u - B_2  B_1 \\ \rho v^2 + p + \frac{1}{2} \lVert \vec{B} \rVert^2 - B_2 B_2 \\ \rho \, v \, w - B_2  B_3 \\ v \hat{E} - B_2 \big(\vec{u} \cdot \vec{B}\big) + c_h \psi B_2 \\ v\, B_1 - u \, B_2 \\ c_h \psi \\ v \, B_3 - w\, B_2 \\ c_h B_2 \end{bmatrix}, \notag\\
	\vec{f}^z &=
	\begin{bmatrix}\rho \, w \\ \rho \, w \, u - B_3  B_1 \\ \rho \, w \, v - B_3  B_2 \\ \rho w^2 + p + \frac{1}{2} \lVert \vec{B} \rVert^2 - B_3 B_3 \\ w \hat{E} - B_3 \big(\vec{u} \cdot \vec{B}\big) + c_h \psi B_3 \\ w\, B_1 - u \, B_3 \\ w \, B_2 - v\, B_3 \\ c_h \psi \\ c_h B_3 \end{bmatrix}.
\end{align}

\section{Dimensional ideal GLM-MHD equations}\label{App:DimidealMHD}
We used dimensionless quantities in this work for the sake of convenience as it allows us to hide some physical constants as they are set to one. However, as the authors have also shown in \cite[Appendix~D]{Derigs2016}, our EC/ES schemes trivially extend to dimensional units, where the magnetic permeability, $\mu_0$, has to explicitly be taken into account:
\begin{align}
E_\mathrm{mag} &= \frac{1}{2\mu_0} \|\vec{B}\|^2,\\
E_\psi &= \frac{1}{2\mu_0} \psi^2.
\end{align}

The dimensional ideal GLM-MHD equations are given by
\refstepcounter{equation}\label{eq:dim3DIDEALGLMMHDmixed}
\begin{equation}\tag{\theequation a-e}
	\pderivative{}{t}\vec{q} + \nabla\cdot\vec{f} =
	\pderivative{}{t}\begin{bmatrix} \rho \\ \vphantom{\big(}\rho\vec{u} \\ \vphantom{\big(}E \\ \vec{B}\\ \psi \end{bmatrix}
	+
	\nabla\cdot\begin{bmatrix} \rho\vec{u} \\
	\rho(\vec{u}\otimes\vec{u}) + \big(p+\frac{\|\vec{B}\|^2}{2\mu_0}\big)\mat{I}-\frac{\vec{B}\otimes\vec{B}}{\mu_0} \\
	\vec{u}\big(\frac{1}{2}\rho\|\vec{u}\|^2 + \frac{\gamma p}{\gamma - 1} + \frac{\|\vec{B}\|^2}{\mu_0} \big) - \frac{\vec{B}(\vec{u}\cdot\vec{B})}{\mu_0} + \frac{c_h}{\mu_0} \psi \vec{B} \\
	\vec{u}\otimes\vec{B} - \vec{B}\otimes\vec{u} + c_h \psi\mat{I} \\
	c_h \vec{B}
	\end{bmatrix}
	=
	-
	\vec{\Upsilon}_\mathrm{GLM},%
\end{equation}
with
\begin{equation}\label{eq:GLM-source-dim}
	\vec{\Upsilon}_\mathrm{GLM} := (\nabla \cdot \vec{B})
	\begin{bmatrix}
	0 \\ \vphantom{\big(}\mu_0^{-1}\vec{B} \\ \vphantom{\big(}\mu_0^{-1}\vec{u}\cdot\vec{B} \\ \vec{u} \\ 0
	\end{bmatrix}
	+ (\nabla\psi)\cdot
	\begin{bmatrix}
	\vec{0} \\ 0 \\ \mu_0^{-1}\vec{u}\psi \\ 0 \\ \vec{u}
	\end{bmatrix}
\end{equation}
where the thermal pressure is related to the conserved quantities through the dimensional ideal gas law:
\begin{equation}\label{eq:dimpressure}
	p = (\gamma-1)\left(E - \frac{\rho}{2}\|\vec{u}\|^2 - \frac{1}{2 \mu_0}\left(\|\vec{B}\|^2  + \psi^2\right) \right).
\end{equation}
The resulting units of the simulation quantities are can be determined by the chosen value for $\mu_0$. They are listed in Table~\ref{tab:units}. {In non-dimensional units, one typically uses $\mu_0=1$, where \eqref{eq:dim3DIDEALGLMMHDmixed} and \eqref{eq:dimpressure} are identical to \eqref{eq:3DIDEALGLMMHD} and \eqref{eq:pressure}.}

\begin{table}[h]
	\centering
	\begin{tabular}{lrcc}
		\toprule
		\multicolumn{2}{l}{Unit system:} & SI & CGS \\
		\midrule
		Length &$\ell$ 			& \si{m} & \si{cm}\\
		Time &$t$ 				& \si{s} & \si{s}\\
		Density &$\varrho$		& \si{kg.m^{-3}} & \si{g.cm^{-3}}\\
		Velocities &$\vec{u}$ 	& \si{m.s^{-1}} & \si{cm.s^{-1}}\\
		Specific energy &$E$ 	& \si{J.m^{-3}} & \si{erg.cm^{-3}}\\
		Pressure &$p$			& \si{N.m^{-2}} & \si{dyn.cm^{-2}}\\
		Magnetic field &$B$ 	& \si{T} & \si{G}\\
		Damping coefficient & $\alpha$	& \si{s^{-1}} & \si{s^{-1}}\\
		\midrule
		\multicolumn{2}{r}{with $\mu_0 := $}	& \SI{4\pi e-7}{T^2.m^3.J^{-1}} & \SI{4\pi}{G^2.cm^3.erg^{-1}}\\
		\bottomrule
	\end{tabular}
	\caption{Simulation units determined by different values for $\mu_0$ \cite{Derigs2016}.}
	\label{tab:units}
\end{table}

\end{document}